\documentstyle[psfig,morefloats,url]{mn}

\footnotesize
\newdimen\digitwidth    
\setbox0=\hbox{\rm0}
\digitwidth=\wd0
\catcode`!=\active
\def!{\kern\digitwidth}
\normalsize

\title[The Parkes multibeam pulsar survey: IV.] {The Parkes multibeam
pulsar survey: IV. Discovery of 180 pulsars and parameters for 281 previously known pulsars}
\author[G. Hobbs et al.]
{G. Hobbs$^1$,\thanks{Email: george.hobbs@csiro.au}
A. Faulkner$^2$,
I.~H. Stairs$^3$, 
F. Camilo$^4$,
R. N. Manchester$^1$,
\newauthor
A. G. Lyne$^2$,
M. Kramer$^2$,
N. D'Amico$^{5,6}$, 
V.~M. Kaspi$^7$,
A. Possenti$^5$,
\newauthor
M.~A. McLaughlin$^2$,
D.~R. Lorimer$^2$,
M. Burgay$^5$,
B.~C. Joshi$^{2,8}$,
F. Crawford$^9$
\\
$^1$ Australia Telescope National Facility, CSIRO, PO~Box~76, Epping
NSW~1710, Australia \\
$^2$ University of Manchester, Jodrell Bank Observatory, Macclesfield,
Cheshire SK11~9DL \\
$^3$ Department of Physics \& Astronomy, University of British Columbia,
6224 Agricultural Road, Vancouver, B.C. V6T 1Z1, Canada \\
$^4$ Columbia Astrophysics Laboratory, Columbia University, 550 West 
120th Street, New York, NY 10027, USA \\
$^5$ INAF - Osservatorio Astronomico di Cagliari, Loc. Poggio dei Pini,
Strada 54, 09012, Capoterra (CA), Italy \\
$^6$ Universita' degli Studi di Cagliari, Dipartimento di Fisica, SP
   Monserrato-Sestu km 0,7, 90042, Monserrato (CA), Italy \\
$^7$ Physics Department, McGill University, Montreal, Quebec H3W 2C4,
Canada \\
$^8$ National Center for Radio Astrophysics, PO Bag No. 3,
Ganeshkhind, Pune, India\\
$^9$ Department of Physics, Haverford College, Haverford, PA 19041,
USA \\}
\date{}
\begin{document}

\maketitle
\newcommand{\setthebls}{
}

\setthebls

\begin{abstract} 

 The Parkes multibeam pulsar survey has led to the discovery of more
 than 700 pulsars. In this paper, we provide timing solutions, flux
 densities and pulse profiles for 180 of these new discoveries.  Two
 pulsars, PSRs J1736$-$2843 and J1847$-$0130 have rotational periods
 $P > 6$\,s and are therefore among the slowest rotating radio pulsars
 known. Conversely, with $P = 1.8$\,ms, PSR~J1843$-$1113 has the third
 shortest period of pulsars currently known.  This pulsar and
 PSR~J1905$+$0400 ($P = 3.8$\,ms) are both solitary. We also provide
 orbital parameters for a new binary system, PSR~J1420$-$5625, which
 has $P = 34$\,ms, an orbital period of 40\,days and a minimum
 companion mass of 0.4 solar masses. The 10$^\circ$-wide strip along
 the Galactic plane that was surveyed is known to contain 264 radio
 pulsars that were discovered prior to the multibeam pulsar survey.
 We have redetected almost all of these pulsars and provide new
 dispersion measure values and flux densities at 20\,cm for the
 redetected pulsars.

\end{abstract}


\section{INTRODUCTION}

 \begin{table*} 
\caption{Positions, flux densities and pulse widths for 180 pulsars discovered 
in the Parkes multibeam pulsar survey. Radial angular distances are given in
units of beam radii.  Timing solutions indicated by a $J$ or an $A$ have a significant
number of observations from the Jodrell and Arecibo telescopes respectively. Pulse widths at 10 percent of the peak are given only for high signal--to--noise profiles.}\label{tb:posn}
\begin{center}\begin{footnotesize}
\begin{tabular}{lllrrccrlrr}
\hline
PSR J & R.A. (J2000) & Dec. (J2000) & \multicolumn{1}{c}{$l$} & \multicolumn{1}{c}{$b$} & Beam & Radial & S/N & $S_{1400}$ & W$_{50}$ & W$_{10}$ \\
 & (h~~~m~~~s) & (\degr ~~~\arcmin ~~~\arcsec) & \multicolumn{1}{c}{(\degr)} &\multicolumn{1}{c}{(\degr)} & & Distance & & (mJy) & (ms) & (ms) \\ 
\hline 
0843$-$5022	& 08:43:09.884(8) & $-$50:22:43.10(8) & 268.50 & $-$4.90 & 7 & 0.46 & 31.6 & 0.31(4) & 6.1 & 29\\
1016$-$5857	& 10:16:21.16(1) & $-$58:57:12.1(1) & 284.08 & $-$1.88 & 3 & 0.74 & 22.3 & 0.46(5) & 7.8 & -- \\
1021$-$5601	& 10:21:24.82(15) & $-$56:01:50.9(11) & 283.04 & !0.94 & 7 & 0.77 & 21.8 & 0.37(5) & 52.0 & -- \\
1032$-$5206	& 10:32:27.69(7) & $-$52:06:08.5(6) & 282.35 & !5.13 & 11 & 0.64 & 34.7 & 0.19(3) & 21.0 & 71\\
1052$-$6348	& 10:52:53.39(6) & $-$63:48:16.6(3) & 290.29 & $-$3.88 & 3 & 0.49 & 9.5 & 0.11(2) & 10.0 & -- \\ \\ 
1054$-$6452	& 10:54:08.84(16) & $-$64:52:37.5(8) & 290.89 & $-$4.78 & 8 & 1.00 & 22.2 & 0.25(4) & 22.0 & 42\\
1055$-$6022	& 10:55:48.5(4) & $-$60:22:52(3) & 289.11 & $-$0.65 & 13 & 0.90 & 10.9 & 0.16(3) & 23.0 & -- \\
1106$-$6438	& 11:06:28.44(14) & $-$64:38:60.0(6) & 291.99 & $-$4.03 & 10 & 0.63 & 23.4 & 0.19(3) & 27.0 & 46\\
1152$-$5800	& 11:52:10.0(4) & $-$58:00:34(4) & 295.13 & !3.96 & 5 & 0.71 & 10.2 & 0.12(2) & 16.0 & -- \\
1156$-$5707	& 11:56:07.45(4) & $-$57:07:01.9(5) & 295.45 & !4.95 & 7 & 0.76 & 14.1 & 0.19(3) & 4.8 & 17\\ \\ 
1210$-$6550	& 12:10:42.0(3) & $-$65:50:04.6(19) & 298.77 & $-$3.29 & 11 & 0.91 & 15.5 & 0.17(3) & 39.0 & -- \\
1337$-$6306	& 13:37:20.35(5) & $-$63:06:23.3(3) & 308.10 & $-$0.70 & 1 & 0.82 & 8.7 & 0.11(2) & 12.0 & -- \\
1352$-$6803	& 13:52:34.44(3) & $-$68:03:36.79(19) & 308.61 & $-$5.87 & 12 & 1.77 & 17.4 & 0.68(8) & 28.0 & 45\\
1415$-$6621	& 14:15:31.27(3) & $-$66:21:12.2(3) & 311.23 & $-$4.85 & 6 & 0.18 & 97.8 & 0.71(8) & 8.9 & 17\\
1420$-$5625	& 14:20:03.062(3) & $-$56:25:55.00(3) & 315.00 & !4.35 & 8 & 0.86 & 12.3 & 0.13(2) & 1.3 & 12\\ \\ 
1424$-$5556	& 14:24:12.76(3) & $-$55:56:13.9(3) & 315.72 & !4.61 & 6 & 0.65 & 40.3 & 0.38(5) & 22.0 & 36\\
1435$-$5954	& 14:35:00.36(5) & $-$59:54:49.2(3) & 315.58 & !0.39 & 1 & 0.75 & 96.1 & 3.6(4) & 19.0 & 33\\
1437$-$6146	& 14:37:15.31(9) & $-$61:46:02.0(8) & 315.10 & $-$1.42 & 13 & 0.43 & 19.4 & 0.24(3) & 17.0 & -- \\
1502$-$5653	& 15:02:57.389(11) & $-$56:53:39.21(12) & 320.19 & !1.51 & 3 & 0.74 & 46.3 & 0.39(5) & 7.2 & 14\\
1519$-$6308	& 15:19:09.56(18) & $-$63:08:19.5(10) & 318.74 & $-$4.90 & 1 & 0.55 & 45.0 & 0.32(4) & 22.0 & -- \\ \\ 
1538$-$5551	& 15:38:45.03(4) & $-$55:51:36.9(6) & 324.91 & $-$0.30 & 1 & 0.87 & 10.6 & 0.25(4) & 11.0 & -- \\
1542$-$5133	& 15:42:19.93(13) & $-$51:33:35(3) & 327.91 & !2.83 & 7 & 0.97 & 19.7 & 0.27(4) & 35.0 & 59\\
1547$-$5750	& 15:47:30.60(10) & $-$57:50:29.4(15) & 324.66 & $-$2.60 & 11 & 0.20 & 27.3 & 0.23(3) & 39.0 & -- \\
1551$-$5310	& 15:51:41.25(6) & $-$53:10:59.6(8) & 328.03 & !0.67 & 5 & 0.97 & 18.2 & 0.54(6) & 62.0 & -- \\
1609$-$4616	& 16:09:41.13(4) & $-$46:16:22.5(4) & 334.76 & !3.99 & 12 & 0.28 & 36.6 & 0.38(5) & 5.6 & 9\\ \\ 
1620$-$5414	& 16:20:14.44(10) & $-$54:14:51.7(16) & 330.47 & $-$2.94 & 4 & 0.82 & 11.4 & 0.13(2) & 26.0 & -- \\
1632$-$4509	& 16:32:14.00(17) & $-$45:09:09(9) & 338.34 & !2.00 & 11 & 0.06 & 14.4 & 0.16(3) & 18.0 & -- \\
1632$-$4757	& 16:32:16.72(6) & $-$47:57:34.3(14) & 336.30 & !0.08 & 8 & 0.97 & 9.5 & 0.30(4) & 21.0 & -- \\
1638$-$4417	& 16:38:46.221(14) & $-$44:17:03.6(4) & 339.77 & !1.73 & 12 & 0.88 & 12.2 & 0.21(3) & 5.6 & -- \\
1658$-$4958	& 16:58:54.92(6) & $-$49:58:58.4(6) & 337.60 & $-$4.55 & 9 & 0.94 & 64.7 & 0.87(10) & 13.0 & 26\\ \\ 
1700$-$3919	& 17:00:22.27(3) & $-$39:19:00.02(142) & 346.16 & !1.83 & 11 & 0.26 & 23.2 & 0.23(3) & 14.0 & -- \\
1702$-$4217	& 17:02:36.44(6) & $-$42:17:01.2(22) & 344.08 & $-$0.33 & 13 & 0.48 & 18.4 & 0.50(6) & 41.0 & -- \\
1708$-$4522	& 17:08:12.92(6) & $-$45:22:51(3) & 342.22 & $-$3.01 & 1 & 0.63 & 25.1 & 0.22(3) & 17.0 & 37\\
1715$-$4254	& 17:15:10.54(9) & $-$42:54:54(4) & 344.95 & $-$2.56 & 6 & 0.71 & 9.5 & 0.07(2) & 26.0 & -- \\
1718$-$3714	& 17:18:18.59(16) & $-$37:14:16(9) & 349.93 & !0.24 & 11 & 0.42 & 9.5 & 0.23(3) & 96.0 & -- \\ \\ 
1718$-$3718	& 17:18:10.0(3) & $-$37:18:53(11) & 349.85 & !0.22 & 11 & 0.27 & 13.3 & 0.18(3) & 130.0 & -- \\
1724$-$3149	& 17:24:44.87(7) & $-$31:49:04(4) & 355.14 & !2.23 & 11 & 0.42 & 24.9 & 0.36(5) & 32.0 & -- \\
1726$-$4006	& 17:26:33.37(7) & $-$40:06:02(4) & 348.48 & $-$2.71 & 4 & 0.86 & 17.7 & 0.21(3) & 20.0 & -- \\
1727$-$2739	& 17:27:30.99(12) & $-$27:39:00.5(169) & 358.94 & !4.05 & 11 & 0.80 & 79.5 & 1.60(17) & 90.0 & -- \\
1730$-$3353	& 17:30:55.58(10) & $-$33:53:38(12) & 354.14 & $-$0.00 & 3 & 0.38 & 43.9 & 0.38(5) & 54.0 & -- \\ \\ 
1731$-$3123	& 17:31:00.53(3) & $-$31:23:43(4) & 356.23 & !1.35 & 6 & 1.06 & 14.7 & 0.29(4) & 19.0 & -- \\
1732$-$4156	& 17:32:48.86(4) & $-$41:56:29.6(16) & 347.59 & $-$4.71 & 5 & 0.67 & 17.9 & 0.22(3) & 21.0 & -- \\
1733$-$3030	& 17:33:58.89(6) & $-$30:30:49(7) & 357.32 & !1.30 & 13 & 0.64 & 10.2 & 0.20(3) & 16.0 & -- \\
1733$-$4005	& 17:33:58.64(3) & $-$40:05:39.7(15) & 349.27 & $-$3.89 & 13 & 0.71 & 40.0 & 0.49(6) & 14.0 & 23\\
1736$-$2819	& 17:36:24.73(9) & $-$28:19:42(16) & 359.45 & !2.04 & 7 & 0.78 & 10.3 & 0.16(3) & 21.0 & -- \\ \\ 
1736$-$2843	& 17:36:42.59(16) & $-$28:43:51(22) & 359.14 & !1.76 & 1 & 0.63 & 47.4 & 0.43(5) & 145.0 & -- \\
1737$-$3320	& 17:37:10.51(5) & $-$33:20:20(5) & 355.31 & $-$0.79 & 5 & 0.40 & 22.2 & 0.35(4) & 42.0 & -- \\
1738$-$2647	& 17:38:05.03(4) & $-$26:47:46(26) & 0.94 & !2.54 & 12 & 0.93 & 20.6 & 0.44(5) & 12.0 & -- \\
1738$-$3107	& 17:38:47.4(3) & $-$31:07:44(14) & 357.36 & !0.10 & 12 & 1.04 & 12.0 & 0.26(4) & 34.0 & -- \\
1738$-$3316	& 17:38:34.45(6) & $-$33:16:01.6(49) & 355.53 & $-$1.00 & 9 & 0.70 & 21.0 & 0.55(7) & 91.0 & -- \\\hline 
\end{tabular}\end{footnotesize}\end{center}\end{table*}
\addtocounter{table}{-1}\begin{table*} 
\begin{center}\begin{footnotesize}
\caption{-- {\it continued}}
\begin{tabular}{lllrrccrlrr}
\hline
PSR J & R.A. (J2000) & Dec. (J2000) & \multicolumn{1}{c}{$l$} & \multicolumn{1}{c}{$b$} & Beam & Radial & S/N & $S_{1400}$ & W$_{50}$ & W$_{10}$ \\
 & (h~~~m~~~s) & (\degr ~~~\arcmin ~~~\arcsec) & \multicolumn{1}{c}{(\degr)} &\multicolumn{1}{c}{(\degr)} & & Distance & & (mJy) & (ms) & (ms) \\ 
\hline 
1740$-$2540	& 17:40:45.32(6) & $-$25:40:19(18) & 2.21 & !2.63 & 6 & 0.21 & 17.6 & 0.16(3) & 30.0 & -- \\
1740$-$3327	& 17:40:25.72(3) & $-$33:27:53.5(17) & 355.56 & $-$1.44 & 3 & 1.08 & 15.8 & 0.30(4) & 11.0 & 23\\
1743$-$2442	& 17:43:20.12(10) & $-$24:42:55(41) & 3.33 & !2.64 & 4 & 0.38 & 14.8 & 0.14(2) & 75.0 & -- \\
1745$-$2229	& 17:45:16.71(7) & $-$22:29:14(25) & 5.47 & !3.42 & 12 & 0.33 & 18.9 & 0.13(2) & 14.0 & -- \\
1749$-$2347	& 17:49:15.61(12) & $-$23:47:17(82) & 4.83 & !1.97 & 5 & 1.40 & 11.5 & 0.13(2) & 9.1 & -- \\ \\ 
1750$-$2444	& 17:50:22.96(8) & $-$24:44:47(42) & 4.14 & !1.26 & 13 & 0.50 & 15.7 & 0.27(4) & 32.0 & -- \\
1752$-$2410	& 17:52:58.742(15) & $-$24:10:26(15) & 4.93 & !1.04 & 10 & 2.71 & 30.2 & 0.47(6) & 7.1 & -- \\
1754$-$3443	& 17:54:37.372(14) & $-$34:43:53.9(10) & 355.99 & $-$4.61 & 8 & 0.91 & 33.4 & 0.49(6) & 12.0 & 21\\
1755$-$25211	& 17:55:19.31(5) & $-$25:21:09(18) & 4.18 & $-$0.02 & 6 & 0.26 & 15.1 & 0.17(3) & 20.0 & -- \\
1755$-$2534	& 17:55:49.82(3) & $-$25:34:39(11) & 4.05 & $-$0.23 & 10 & 0.29 & 9.2 & 0.17(3) & 15.0 & -- \\ \\ 
1756$-$2225	& 17:56:25.56(8) & $-$22:25:48(66) & 6.84 & !1.24 & 1 & 0.67 & 17.0 & 0.25(4) & 19.0 & -- \\
1758$-$1931	& 17:58:05.60(7) & $-$19:31:41(11) & 9.54 & !2.35 & 6 & 0.79 & 20.4 & 0.38(5) & 21.0 & -- \\
1759$-$1903	& 17:59:41.76(16) & $-$19:03:19(27) & 10.14 & !2.26 & 2 & 0.39 & 16.4 & 0.16(3) & 34.0 & -- \\
1759$-$3107	& 17:59:22.056(14) & $-$31:07:21.5(20) & 359.63 & $-$3.67 & 10 & 0.99 & 97.8 & 0.91(10) & 16.0 & 27\\
1800$-$2114	& 18:00:12.3(4) & $-$21:14:19(75) & 8.31 & !1.07 & 1 & 0.56 & 18.7 & 0.30(4) & 90.0 & -- \\ \\ 
1801$-$2115	& 18:01:32.49(12) & $-$21:15:18(53) & 8.45 & !0.79 & 1 & 0.46 & 12.0 & 0.19(3) & 47.0 & -- \\
1801$-$2154	& 18:01:08.33(3) & $-$21:54:32(12) & 7.83 & !0.55 & 11 & 0.49 & 15.0 & 0.18(3) & 9.6 & -- \\
1803$-$1616	& 18:03:34.68(3) & $-$16:16:30(4) & 13.02 & !2.83 & 5 & 0.55 & 16.9 & 0.16(3) & 19.0 & -- \\
1803$-$1920	& 18:03:29.44(3) & $-$19:20:41(7) & 10.33 & !1.34 & 9 & 0.49 & 18.6 & 0.27(4) & 12.0 & -- \\
1805$-$2447	& 18:05:25.93(3) & $-$24:47:30(14) & 5.81 & $-$1.72 & 3 & 0.88 & 13.6 & 0.27(4) & 11.0 & 22\\ \\ 
1806$-$1618	& 18:06:25.78(6) & $-$16:18:38(8) & 13.32 & !2.21 & 11 & 0.53 & 11.9 & 0.22(3) & 34.0 & -- \\
1809$-$1850	& 18:09:37.21(12) & $-$18:50:55(21) & 11.47 & !0.32 & 6 & 0.93 & 13.0 & 0.20(3) & 55.0 & -- \\
1810$-$1441	& 18:10:59.162(14) & $-$14:41:33.6(13) & 15.27 & !2.03 & 5 & 0.96 & 10.6 & 0.21(3) & 14.0 & -- \\
1812$-$1910	& 18:12:34.89(10) & $-$19:10:39(10) & 11.52 & $-$0.46 & 4 & 0.88 & 10.2 & 0.22(3) & 28.0 & -- \\
1813$-$2242	& 18:13:29.16(8) & $-$22:42:06(42) & 8.53 & $-$2.33 & 8 & 0.75 & 11.9 & 0.21(3) & 21.0 & -- \\ \\ 
1815$-$1738	& 18:15:14.672(9) & $-$17:38:03.0(12) & 13.18 & $-$0.27 & 1 & 0.58 & 13.9 & 0.25(4) & 29.0 & -- \\
1816$-$1446	& 18:16:29.19(3) & $-$14:46:30(3) & 15.84 & !0.83 & 3 & 0.16 & 17.5 & 0.23(3) & 25.0 & -- \\
1817$-$1511$^J$	& 18:17:36.20(6) & $-$15:11:39(6) & 15.59 & !0.39 & 2 & 0.24 & 18.9 & 0.43(5) & 0.0 & -- \\
1818$-$1116$^J$	& 18:18:26.45(6) & $-$11:16:29(7) & 19.14 & !2.07 & 11 & 0.11 & 23.2 & 0.50(6) & 25.0 & -- \\
1819$-$0925$^J$	& 18:19:50.542(19) & $-$09:25:49.9(13) & 20.93 & !2.63 & 9 & 0.68 & 60.0 & 0.72(8) & 20.0 & 34\\ \\ 
1819$-$1008$^J$	& 18:19:39.986(17) & $-$10:08:28(4) & 20.29 & !2.34 & 1 & 0.73 & 21.4 & 0.35(4) & 9.9 & -- \\
1819$-$1131	& 18:19:58.15(10) & $-$11:31:29(9) & 19.10 & !1.62 & 7 & 0.37 & 10.8 & 0.15(3) & 64.0 & -- \\
1820$-$1529$^J$	& 18:20:40.82(9) & $-$15:29:50(10) & 15.68 & $-$0.41 & 11 & 0.24 & 19.6 & 0.61(7) & 0.0 & -- \\
1821$-$1419	& 18:21:34.3(4) & $-$14:19:26(32) & 16.82 & $-$0.04 & 12 & 0.20 & 11.1 & 0.20(3) & 99.0 & -- \\
1822$-$0907	& 18:22:39.80(5) & $-$09:07:36(3) & 21.53 & !2.16 & 1 & 0.71 & 11.4 & 0.12(2) & 28.0 & -- \\ \\ 
1822$-$1252	& 18:22:41.7(3) & $-$12:52:49(29) & 18.22 & !0.39 & 7 & 0.27 & 15.0 & 0.25(4) & 105.0 & -- \\
1822$-$1617	& 18:22:36.6(3) & $-$16:17:35(25) & 15.19 & $-$1.19 & 6 & 0.34 & 9.8 & 0.20(3) & 115.0 & -- \\
1823$-$1126	& 18:23:19.86(10) & $-$11:26:04(5) & 19.57 & !0.93 & 9 & 0.68 & 36.4 & 0.51(6) & 23.0 & 44\\
1823$-$1526	& 18:23:21.42(6) & $-$15:26:22(7) & 16.03 & $-$0.95 & 5 & 0.75 & 26.8 & 0.47(6) & 41.0 & -- \\
1824$-$1500	& 18:24:14.10(7) & $-$15:00:33(8) & 16.51 & $-$0.93 & 1 & 0.86 & 10.9 & 0.16(3) & 18.0 & -- \\ \\ 
1828$-$0611	& 18:28:20.715(6) & $-$06:11:51.5(4) & 24.78 & !2.28 & 10 & 0.60 & 68.3 & 1.20(13) & 10.0 & 19\\
1828$-$1007	& 18:28:30.356(15) & $-$10:07:10.1(10) & 21.32 & !0.42 & 12 & 0.53 & 11.2 & 0.21(3) & 8.7 & -- \\
1828$-$1057	& 18:28:33.21(4) & $-$10:57:26(3) & 20.59 & !0.02 & 13 & 0.67 & 11.7 & 0.23(3) & 15.0 & -- \\
1831$-$0823$^J$	& 18:31:36.334(8) & $-$08:23:23.9(5) & 23.21 & !0.55 & 12 & 0.66 & 78.8 & 0.97(11) & 14.0 & 23\\
1831$-$1423	& 18:31:29.10(3) & $-$14:23:46(4) & 17.87 & $-$2.20 & 11 & 0.11 & 13.3 & 0.19(3) & 23.0 & -- \\ \\ 
1833$-$0556	& 18:33:38.88(16) & $-$05:56:05(9) & 25.62 & !1.23 & 7 & 0.57 & 11.4 & 0.20(3) & 69.0 & -- \\
1834$-$0633	& 18:34:29.25(15) & $-$06:33:01.1(63) & 25.17 & !0.76 & 6 & 0.43 & 17.8 & 0.28(4) & 101.6 & -- \\
1834$-$0731	& 18:34:16.00(7) & $-$07:31:07(4) & 24.29 & !0.37 & 4 & 0.44 & 46.3 & 1.00(11) & 30.0 & -- \\
1834$-$0742$^J$	& 18:34:31.32(3) & $-$07:42:20.6(14) & 24.15 & !0.22 & 2 & 0.22 & 27.4 & 0.35(4) & 18.0 & -- \\
1834$-$1202	& 18:34:23.12(3) & $-$12:02:26.4(13) & 20.29 & $-$1.74 & 2 & 1.01 & 26.4 & 0.70(8) & 63.0 & 87\\\hline 
\end{tabular}\end{footnotesize}\end{center}\end{table*}
\addtocounter{table}{-1}\begin{table*} 
\begin{center}\begin{footnotesize}
\caption{-- {\it continued}}
\begin{tabular}{lllrrccrlrr}
\hline
PSR J & R.A. (J2000) & Dec. (J2000) & \multicolumn{1}{c}{$l$} & \multicolumn{1}{c}{$b$} & Beam & Radial & S/N & $S_{1400}$ & W$_{50}$ & W$_{10}$ \\
 & (h~~~m~~~s) & (\degr ~~~\arcmin ~~~\arcsec) & \multicolumn{1}{c}{(\degr)} &\multicolumn{1}{c}{(\degr)} & & Distance & & (mJy) & (ms) & (ms) \\ 
\hline 
1835$-$0522	& 18:35:08.12(5) & $-$05:22:08(3) & 26.30 & !1.16 & 6 & 0.76 & 18.4 & 0.23(3) & 23.0 & -- \\
1836$-$0517	& 18:36:25.20(11) & $-$05:17:35(4) & 26.51 & !0.91 & 1 & 0.89 & 9.5 & 0.15(3) & 13.0 & -- \\
1838$-$0549	& 18:38:38.09(3) & $-$05:49:12(3) & 26.30 & !0.18 & 11 & 0.37 & 18.9 & 0.29(4) & 7.0 & -- \\
1838$-$0624	& 18:38:51.78(11) & $-$06:24:54(4) & 25.79 & $-$0.14 & 5 & 0.54 & 10.9 & 0.16(3) & 36.0 & -- \\
1839$-$0905	& 18:39:53.458(20) & $-$09:05:14.8(14) & 23.53 & $-$1.59 & 13 & 0.69 & 12.1 & 0.16(3) & 17.0 & -- \\ \\ 
1840$-$0559	& 18:40:23.18(4) & $-$05:59:16.2(18) & 26.35 & $-$0.28 & 12 & 1.16 & 12.7 & 0.31(4) & 16.0 & 34\\
1840$-$0809$^J$	& 18:40:33.364(6) & $-$08:09:03.3(4) & 24.44 & $-$1.31 & 1 & 1.02 & 134.2 & 2.3(2) & 10.0 & 35\\
1840$-$0815$^J$	& 18:40:13.775(13) & $-$08:15:10.6(7) & 24.31 & $-$1.28 & 1 & 1.79 & 19.6 & 1.40(15) & 22.0 & 35\\
1840$-$1122	& 18:40:24.066(19) & $-$11:22:10.7(16) & 21.56 & $-$2.74 & 3 & 0.87 & 14.5 & 0.13(2) & 11.0 & -- \\
1841$-$0157$^J$	& 18:41:56.207(20) & $-$01:57:54.6(8) & 30.10 & !1.22 & 9 & 0.35 & 97.7 & 0.81(9) & 21.0 & -- \\ \\ 
1841$-$0310	& 18:41:25.89(14) & $-$03:10:21(9) & 28.97 & !0.78 & 13 & 0.39 & 14.1 & 0.15(3) & 59.0 & -- \\
1841$-$0524	& 18:41:49.32(5) & $-$05:24:29.5(12) & 27.02 & $-$0.33 & 3 & 0.59 & 13.1 & 0.20(3) & 13.0 & -- \\
1842$-$0309	& 18:42:19.02(5) & $-$03:09:46(3) & 29.08 & !0.58 & 9 & 0.64 & 11.5 & 0.25(4) & 43.0 & -- \\
1842$-$0612	& 18:42:43.05(10) & $-$06:12:36(5) & 26.41 & $-$0.90 & 12 & 0.61 & 21.7 & 0.54(6) & 53.0 & -- \\
1843$-$0000$^J$	& 18:43:27.962(9) & $-$00:00:40.9(6) & 32.01 & !1.77 & 10 & 0.62 & 250.4 & 2.9(3) & 26.0 & 40\\ \\ 
1843$-$0137	& 18:43:12.63(3) & $-$01:37:46.3(12) & 30.54 & !1.09 & 4 & 1.04 & 13.5 & 0.26(4) & 17.0 & -- \\
1843$-$0211	& 18:43:30.328(20) & $-$02:11:02.8(7) & 30.08 & !0.77 & 10 & 0.65 & 69.8 & 0.93(10) & 24.0 & 105\\
1843$-$0408	& 18:43:43.44(7) & $-$04:08:04(3) & 28.37 & $-$0.17 & 9 & 0.60 & 17.1 & 0.17(3) & 12.0 & -- \\
1843$-$0702	& 18:43:22.441(10) & $-$07:02:54.6(7) & 25.74 & $-$1.43 & 3 & 0.40 & 20.1 & 0.17(3) & 4.7 & -- \\
1843$-$0806	& 18:43:28.715(11) & $-$08:06:44.9(7) & 24.81 & $-$1.93 & 12 & 0.97 & 13.5 & 0.36(5) & 21.0 & -- \\ \\ 
1843$-$1113	& 18:43:41.26225(20) & $-$11:13:31.052(16) & 22.05 & $-$3.40 & 7 & 0.18 & 17.6 & 0.10(2) & 0.2 & -- \\
1844$-$0030$^A$	& 18:44:41.099(19) & $-$00:30:25.8(13) & 31.71 & !1.27 & 5 & 0.61 & 33.5 & 0.42(5) & 16.0 & -- \\
1844$-$0452	& 18:44:01.54(4) & $-$04:52:20.9(19) & 27.75 & $-$0.58 & 9 & 0.84 & 8.7 & 0.19(3) & 16.0 & -- \\
1844$-$0502	& 18:44:33.96(7) & $-$05:02:00.5(23) & 27.67 & $-$0.77 & 5 & 1.06 & 10.2 & 0.40(5) & 24.0 & -- \\
1845$-$0545	& 18:45:38.49(4) & $-$05:45:18.2(10) & 27.15 & $-$1.34 & 10 & 0.97 & 20.9 & 0.47(6) & 18.0 & 32\\ \\ 
1846+0051$^A$	& 18:46:43.821(20) & +00:51:39.0(7) & 33.16 & !1.44 & 11 & 1.02 & 16.9 & 0.34(9) & 28.2 & 60\\
1847$-$0130	& 18:47:35.21(9) & $-$01:30:46(3) & 31.15 & !0.17 & 3 & 0.41 & 15.6 & 0.33(4) & 205.0 & -- \\
1847$-$0443	& 18:47:51.85(3) & $-$04:43:36.2(8) & 28.32 & $-$1.36 & 7 & 0.62 & 11.1 & 0.16(3) & 9.0 & -- \\
1848$-$0023$^A$	& 18:48:37.89(9) & $-$00:23:17(4) & 32.27 & !0.45 & 7 & 1.23 & 13.1 & 0.6(3) & 17.7 & 38\\
1848$-$0055	& 18:48:45.50(19) & $-$00:55:53(4) & 31.80 & !0.17 & 6 & 0.65 & 9.5 & 0.19(3) & 27.9 & -- \\ \\ 
1848$-$0511	& 18:48:15.01(14) & $-$05:11:38(5) & 27.95 & $-$1.66 & 8 & 0.25 & 21.4 & 0.40(5) & 99.0 & -- \\
1849$-$0040$^A$	& 18:49:10.25(8) & $-$00:40:20(6) & 32.08 & !0.20 & 11 & 0.34 & 10.9 & 0.20(3) & 64.6 & -- \\
1849$-$0614	& 18:49:45.157(19) & $-$06:14:31.5(8) & 27.18 & $-$2.47 & 3 & 1.18 & 36.8 & 0.59(7) & 14.0 & 32\\
1850$-$0031	& 18:50:33.39(9) & $-$00:31:09(4) & 32.37 & $-$0.04 & 11 & 0.93 & 9.3 & 0.23(3) & 32.0 & -- \\
1851+0118	& 18:51:52.18(13) & +01:18:59(5) & 34.15 & !0.50 & 1 & 0.95 & 10.9 & 0.10(2) & 24.0 & -- \\ \\ 
1851$-$0053$^A$	& 18:51:03.17(8) & $-$00:53:07.3(19) & 32.10 & $-$0.32 & 5 & 1.24 & 35.9 & 1.00(11) & 19.0 & -- \\
1851$-$0241	& 18:51:15.26(10) & $-$02:41:31(3) & 30.51 & $-$1.19 & 4 & 0.56 & 10.9 & 0.20(3) & 35.0 & -- \\
1852+0008$^A$	& 18:52:42.78(3) & +00:08:09.6(8) & 33.20 & $-$0.22 & 5 & 0.58 & 15.8 & 0.31(4) & 16.0 & -- \\
1852+0013$^A$	& 18:52:41.779(20) & +00:13:57.1(12) & 33.28 & $-$0.17 & 5 & 0.22 & 14.8 & 0.30(4) & 19.0 & -- \\
1852$-$0118	& 18:52:17.15(3) & $-$01:18:14.8(17) & 31.87 & $-$0.78 & 6 & 0.98 & 22.1 & 0.35(4) & 24.0 & -- \\ \\ 
1852$-$0127	& 18:52:03.60(4) & $-$01:27:23.4(14) & 31.71 & $-$0.80 & 6 & 0.34 & 36.1 & 0.58(7) & 18.0 & -- \\
1852$-$0635	& 18:52:57.38(14) & $-$06:35:57(8) & 27.22 & $-$3.34 & 12 & 0.86 & 160.4 & 5.9(6) & 90.0 & -- \\
1853+0011$^A$	& 18:53:29.968(14) & +00:11:29.7(5) & 33.34 & $-$0.37 & 6 & 0.34 & 8.3 & 0.30(8) & 11.8 & 21\\
1853+0505	& 18:53:04.36(7) & +05:05:26.1(18) & 37.65 & !1.96 & 9 & 1.03 & 67.4 & 1.50(16) & 92.0 & 195\\
1853$-$0004$^A$	& 18:53:23.018(6) & $-$00:04:32.3(4) & 33.09 & $-$0.47 & 3 & 0.76 & 22.5 & 0.87(10) & 2.2 & 6\\ \\ 
1855+0307$^A$	& 18:55:26.63(3) & +03:07:20.2(9) & 36.17 & !0.53 & 4 & 0.94 & 31.1 & 0.97(11) & 12.0 & -- \\
1855+0700	& 18:55:17.72(4) & +07:00:37.1(9) & 39.61 & !2.34 & 4 & 0.95 & 11.0 & 0.10(2) & 5.4 & -- \\
1856+0102$^A$	& 18:56:28.503(13) & +01:02:10.6(5) & 34.43 & $-$0.65 & 3 & 0.54 & 11.4 & 0.38(11) & 16.8 & 28\\
1857+0143$^A$	& 18:57:33.008(15) & +01:43:47.0(9) & 35.17 & $-$0.57 & 4 & 0.41 & 29.9 & 0.74(18) & 15.7 & 43\\
1857+0809	& 18:57:09.31(3) & +08:09:04.3(8) & 40.84 & !2.45 & 12 & 0.23 & 14.4 & 0.14(2) & 12.0 & -- \\\hline 
\end{tabular}\end{footnotesize}\end{center}\end{table*}
\addtocounter{table}{-1}\begin{table*} 
\begin{center}\begin{footnotesize}
\caption{-- {\it continued}}
\begin{tabular}{lllrrccrlrr}
\hline
PSR J & R.A. (J2000) & Dec. (J2000) & \multicolumn{1}{c}{$l$} & \multicolumn{1}{c}{$b$} & Beam & Radial & S/N & $S_{1400}$ & W$_{50}$ & W$_{10}$ \\
 & (h~~~m~~~s) & (\degr ~~~\arcmin ~~~\arcsec) & \multicolumn{1}{c}{(\degr)} &\multicolumn{1}{c}{(\degr)} & & Distance & & (mJy) & (ms) & (ms) \\ 
\hline 
1858+0241	& 18:58:53.81(14) & +02:41:38(6) & 36.18 & $-$0.43 & 6 & 0.09 & 11.9 & 0.10(2) & 80.0 & -- \\
1859+0601$^A$	& 18:59:45.76(5) & +06:01:46.1(18) & 39.25 & !0.90 & 6 & 0.83 & 12.1 & 0.30(4) & 24.3 & -- \\
1900+0634$^A$	& 19:00:28.034(20) & +06:34:20.9(6) & 39.81 & !1.00 & 12 & 0.54 & 11.7 & 0.24(9) & 11.1 & 20\\
1900$-$0051$^J$	& 19:00:46.644(7) & $-$00:51:08.4(5) & 33.24 & $-$2.47 & 5 & 1.00 & 30.1 & 0.45(6) & 5.6 & 17\\
1901+0124$^A$	& 19:01:52.545(14) & +01:24:49.3(8) & 35.38 & $-$1.68 & 12 & 1.06 & 13.2 & 0.30(4) & 8.3 & -- \\ \\ 
1901+0254	& 19:01:15.67(7) & +02:54:41(5) & 36.64 & $-$0.86 & 11 & 0.74 & 42.8 & 0.58(7) & 70.0 & -- \\
1901+0320$^A$	& 19:01:03.01(9) & +03:20:18(4) & 37.00 & $-$0.61 & 8 & 0.42 & 10.4 & 0.89(10) & 47.2 & -- \\
1901+0355	& 19:01:30.81(4) & +03:55:58.9(9) & 37.58 & $-$0.44 & 1 & 0.33 & 15.9 & 0.15(3) & 12.0 & -- \\
1901+0510$^A$	& 19:01:57.85(11) & +05:10:34(4) & 38.74 & !0.03 & 10 & 1.51 & 8.6 & 0.66(8) & 56.9 & -- \\
1901$-$0312	& 19:01:15.675(18) & $-$03:12:29.5(9) & 31.19 & $-$3.65 & 7 & 0.74 & 16.3 & 0.23(3) & 7.9 & -- \\ \\ 
1902+0248	& 19:02:50.26(7) & +02:48:56(3) & 36.74 & $-$1.25 & 5 & 1.24 & 11.7 & 0.17(3) & 35.4 & -- \\
1902$-$0340	& 19:02:50.70(19) & $-$03:40:18(4) & 30.96 & $-$4.21 & 6 & 0.48 & 16.5 & 0.22(3) & 25.0 & -- \\
1903+0601$^J$	& 19:03:20.874(16) & +06:01:34.0(6) & 39.65 & !0.11 & 11 & 1.03 & 12.7 & 0.26(4) & 12.0 & -- \\
1905+0400	& 19:05:28.2736(3) & +04:00:10.922(13) & 38.09 & $-$1.29 & 4 & 1.10 & 26.5 & 0.050(10) & 0.6 & -- \\
1905+0600	& 19:05:04.35(5) & +06:00:59.9(14) & 39.84 & $-$0.28 & 8 & 0.83 & 17.4 & 0.42(5) & 13.0 & -- \\ \\ 
1906+0649$^A$	& 19:06:11.97(3) & +06:49:48.1(10) & 40.69 & $-$0.15 & 11 & 0.25 & 36.4 & 0.30(4) & 37.0 & -- \\
1907+0249$^A$	& 19:07:42.03(4) & +02:49:41(3) & 37.31 & $-$2.32 & 11 & 1.03 & 10.7 & 0.46(12) & 21.7 & 32\\
1907+0345	& 19:07:14.543(19) & +03:45:10.6(4) & 38.08 & $-$1.80 & 1 & 0.62 & 13.3 & 0.17(3) & 6.3 & -- \\
1907+0731	& 19:07:54.79(3) & +07:31:21.9(5) & 41.50 & $-$0.21 & 7 & 1.21 & 20.9 & 0.35(4) & 11.0 & -- \\
1907+0918	& 19:07:22.441(4) & +09:18:30.76(4) & 43.02 & !0.73 & 3 & 0.97 & 12.8 & 0.29(4) & 4.4 & -- \\ \\ 
1910+0225$^A$	& 19:10:10.359(17) & +02:25:23.6(5) & 37.23 & $-$3.06 & 7 & 0.25 & 39.2 & 0.6(2) & 16.4 & 29\\
1910+0728	& 19:10:22.079(6) & +07:28:37.09(15) & 41.74 & $-$0.77 & 11 & 0.52 & 78.2 & 0.87(10) & 15.0 & 22\\
1913+1000	& 19:13:03.59(5) & +10:00:02.4(14) & 44.29 & $-$0.19 & 2 & 0.48 & 42.5 & 0.53(6) & 32.0 & 52\\
1914+0631$^A$	& 19:14:17.24(4) & +06:31:56.3(10) & 41.35 & $-$2.07 & 11 & 0.83 & 15.6 & 0.26(10) & 12.6 & 27\\
1915+0838$^J$	& 19:15:13.87(3) & +08:38:59.7(13) & 43.34 & $-$1.30 & 13 & 1.17 & 13.5 & 0.29(4) & 14.0 & -- \\ \\ 
1916+0844$^J$	& 19:16:19.081(9) & +08:44:07.0(4) & 43.54 & $-$1.49 & 10 & 1.03 & 30.1 & 0.44(5) & 8.3 & 19\\
1916+0852	& 19:16:24.6(3) & +08:52:36(5) & 43.67 & $-$1.45 & 4 & 0.75 & 16.4 & 0.13(2) & 48.0 & -- \\
1916+1023$^A$	& 19:16:36.91(15) & +10:23:03(6) & 45.03 & $-$0.79 & 4 & 0.20 & 29.7 & 0.36(5) & 72.0 & -- \\
1920+1040$^A$	& 19:20:55.38(8) & +10:40:31(3) & 45.78 & $-$1.59 & 11 & 0.39 & 47.8 & 0.57(7) & 45.0 & -- \\
1937+1505$^J$	& 19:37:16.31(14) & +15:05:19(4) & 51.57 & $-$2.98 & 8 & 0.76 & 11.3 & 0.13(2) & 46.0 & -- \\ \\ 
\hline 
\end{tabular}\end{footnotesize}\end{center}\end{table*}

  Observing for the Parkes Multibeam pulsar survey (hereafter referred
  to as the `multibeam survey') has been completed. Full details of
  the telescope, hardware and software used were provided in
  Manchester et al. (2001)\nocite{mlc+01} along with the rotational,
  astrometric and derived parameters for 100 pulsar discoveries.
  Morris et al. (2002)\nocite{mhl+02} and Kramer et
  al. (2003)\nocite{kbm+03} provided parameters for a further 320
  discoveries timed for at least one year at the Parkes, Jodrell Bank
  and/or Arecibo observatories. Here, we provide timing solutions for
  180 newly discovered pulsars.  With this paper a total of 600 timing
  solutions for the multibeam discoveries have now been published.

  We have successfully redetected 249 of the 264 previously known
  radio pulsars that lie within the survey region (defined by the
  Galactic coordinates, $260^\circ < l < 50^\circ$ and $|b| <
  5^\circ$) and have redetected a further 32 pulsars that lie
  outside this nominal survey region.  For many of these
  redetected pulsars, the 35\,minute observation used during the
  survey is longer than any previous observation. Here, we analyse
  these long observations to obtain, for each pulsar, a flux density
  at 20\,cm, pulse widths and dispersion measure.  The flux densities
  are compared to other flux density measurements, at different
  observing frequencies, existing in the literature to obtain new
  spectral indices for 38 pulsars.

  This paper is divided into three major parts.  In the first, we
  describe the observing systems used for the timing of the new
  multibeam survey discoveries and provide timing solutions for these
  pulsars.  In the second, we highlight some particularly interesting
  discoveries such as the long-period pulsars PSRs~J1736$-$2843 and
  J1847$-$0130, the solitary millisecond pulsars PSRs J1843$-$1113 and
  J1905$+$0400 and the binary system PSR~J1420$-$5625.  The third
  section contains a discussion on previously known pulsars redetected
  during the multibeam survey. We conclude by mentioning how these new
  results will be used to improve upon earlier studies of the Galactic
  pulsar population.

\section{DISCOVERY AND TIMING OF 180 PULSARS}

 The pulsars detailed in this paper have been observed multiple
 (typically around 25) times for at least one year using the Parkes
 64-m, the 305-m Arecibo and/or the 76-m Lovell telescopes. The
 observing and analysis methods used at Parkes and at Jodrell Bank
 were described by Manchester et al. (2001)\nocite{mlc+01} and Morris
 et al. (2002)\nocite{mhl+02} respectively.  The Arecibo timing
 observations were taken with 100\,MHz bandwidth centred at 1400\,MHz,
 using the ``L-narrow'' receiver with a system temperature of
 25--30\,K\footnote{see \url{http://www.naic.edu/~astro/RXstatus} for
 further technical details of this receiver}.  The incoming telescope
 voltages were sampled by the Wideband Arecibo Pulsar Processor
 \cite{dsh00}, a digital correlator with three--level sampling,
 producing 128 lags across 100\,MHz.  Dual circular polarizations were
 summed in hardware, and 16-bit data samples written to disk every 64
 $\mu$sec.  Typical integration times were 60 or 120 seconds.  The
 data were then dedispersed and folded off-line modulo the predicted
 topocentric pulse period.

\begin{table*} 
\caption{Periods, period derivatives and dispersion measures for 180 pulsars 
discovered in the Parkes multibeam pulsar survey. We also give the MJD of the epoch
used for period determination, the number of TOAs included in the timing solution, the MJD range covered
and the RMS of the post-fit timing residuals.  Asterisks indicate those pulsars
which exhibit significant timing noise that has been removed, to first order, by
the fitting of a frequency second derivative (even higher derivatives were included in the timing model if necessary). A superscript `G' indicates pulsars that have glitched. A $^\dagger$ symbol indicates timing solutions already published; see text.}\label{tb:prd}
\begin{center}\begin{footnotesize}
\begin{tabular}{llllllll}
\hline 
PSR J & Period, $P$ & $\dot{P}$ & Epoch & N$_{toa}$ & Data Span & Residual & DM \\ 
      & (s)       & (10$^{-15}$) & (MJD) &         & (MJD)     & (ms)     & (cm$^{-3}$pc) \\ 
\hline 
0843$-$5022$^\dagger$ & 0.2089556931527(14)	& 0.17238(14)	& 51500.0	& --- 	&51060$-$51888	& !--- & 178.47(9) \\
1016$-$5857$^G$$^*$ & 0.1073864584588(15)	& 80.8342(6)	& 52717.0	& 20	&52571$-$52862	& !2.87 & 394.2(2) \\
1021$-$5601 & 0.67002629761(14)	& 0.053(6)	& 51875.0	& 26	&51632$-$52465	& !3.73 & 212(8) \\
1032$-$5206 & 2.40762231135(12)	& 17.891(11)	& 52161.0	& 28	&51744$-$52577	& !1.97 & 139(4) \\
1052$-$6348 & 0.383830423532(15)	& 0.387(6)	& 52477.0	& 30	&52260$-$52692	& !0.65 & 167.5(14) \\ \\ 
1054$-$6452 & 1.8400035414(3)	& 3.14(5)	& 51875.0	& 23	&51632$-$52118	& !1.97 & 234(4) \\
1055$-$6022 & 0.9475584093(3)	& 92.39(12)	& 52746.0	& 23	&52569$-$52923	& !2.83 & 590(5) \\
1106$-$6438 & 2.7179341360(3)	& 2.33(7)	& 51875.0	& 23	&51632$-$52118	& !1.42 & 203(3) \\
1152$-$5800 & 1.7898294664(5)	& 1.3(3)	& 52746.0	& 21	&52569$-$52923	& !3.46 & 191(7) \\
1156$-$5707 & 0.288409420325(8)	& 26.451(5)	& 52149.0	& 23	&51944$-$52352	& !0.57 & 243.5(6) \\ \\ 
1210$-$6550 & 4.2370102164(18)	& 0.43(11)	& 52499.0	& 20	&52305$-$52893	& !2.89 & 37(6) \\
1337$-$6306 & 0.207953012189(4)	& 0.3558(3)	& 51900.0	& 95	&51220$-$52578	& !1.86 & 777.7(17) \\
1352$-$6803 & 0.628902611443(12)	& 1.2380(16)	& 52260.0	& 28	&51944$-$52574	& !0.56 & 215.1(11) \\
1415$-$6621$^\dagger$ & 0.39247897310(5)	& 0.5800(10)	& 51396.2	& --- 	&51569$-$52151	& !--- & 260.17(7) \\
1420$-$5625 & 0.03411713084786(12)	& 0.000068(17)	& 52558.0	& 57	&52293$-$52853	& !0.07 & 64.56(9) \\ \\ 
1424$-$5556 & 0.770374845874(18)	& 0.780(3)	& 52260.0	& 31	&51945$-$52574	& !0.75 & 198.7(20) \\
1435$-$5954 & 0.472995435514(16)	& 1.5433(7)	& 50545.0	& 82	&49955$-$51134	& !0.10 & 44.26(11) \\
1437$-$6146$^G$$^*$ & 0.46761632345(7)	& 6.3304(9)	& 51614.0	& 56	&51710$-$52883	& !2.42 & 200.5(13) \\
1502$-$5653 & 0.535504397790(3)	& 1.8286(3)	& 51780.0	& 30	&51214$-$52345	& !0.29 & 194.0(4) \\
1519$-$6308 & 1.25405163056(10)	& 5.96(5)	& 52144.0	& 19	&51945$-$52342	& !1.49 & 250(3) \\ \\ 
1538$-$5551 & 0.104674920844(4)	& 3.2082(4)	& 51886.0	& 42	&51466$-$52305	& !1.38 & 603(3) \\
1542$-$5133 & 1.7838649877(3)	& 0.59(6)	& 51911.0	& 26	&51634$-$52187	& !3.17 & 186(4) \\
1547$-$5750 & 0.64719774666(6)	& 0.026(5)	& 52044.0	& 17	&51744$-$52342	& !1.26 & 148(4) \\
1551$-$5310$^*$ & 0.453394164640(11)	& 195.1299(9)	& 52204.0	& 71	&51555$-$52852	& !2.68 & 493(2) \\
1609$-$4616 & 0.249608954839(5)	& 0.501(4)	& 52133.0	& 29	&51945$-$52319	& !0.40 & 150.1(6) \\ \\ 
1620$-$5414 & 1.15636028782(9)	& 0.067(3)	& 51830.0	& 16	&51466$-$52923	& !1.98 & 71(8) \\
1632$-$4509 & 1.0468098769(4)	& 14.88(9)	& 51973.0	& 13	&51744$-$52201	& !1.59 & 412(4) \\
1632$-$4757 & 0.228564097816(10)	& 15.0749(10)	& 51909.0	& 42	&51467$-$52350	& !2.39 & 578(3) \\
1638$-$4417 & 0.1178015507206(9)	& 1.60526(12)	& 52142.0	& 40	&51712$-$52572	& !0.53 & 436.0(7) \\
1658$-$4958 & 0.416873770089(11)	& 3.856(7)	& 52146.0	& 18	&51946$-$52344	& !0.57 & 193.4(13) \\ \\ 
1700$-$3919 & 0.560503533065(16)	& 0.0050(13)	& 52025.0	& 18	&51495$-$52596	& !0.79 & 354.3(18) \\
1702$-$4217 & 0.227564958458(7)	& 0.0114(6)	& 51895.0	& 29	&51216$-$52572	& !1.93 & 629(5) \\
1708$-$4522 & 1.29783683782(7)	& 2.612(7)	& 52158.0	& 28	&51744$-$52572	& !2.19 & 454(3) \\
1715$-$4254 & 0.57374535892(7)	& 0.876(10)	& 52278.0	& 31	&51984$-$52571	& !4.16 & 407(9) \\
1718$-$3714 & 1.2893787775(3)	& 26.21(3)	& 51799.0	& 28	&51463$-$52135	& !6.74 & 833(21) \\ \\ 
1718$-$3718 & 3.3782065287(14)	& 1598.15(8)	& 51776.0	& 41	&51245$-$52926	& 27.33 & 373(13) \\
1724$-$3149 & 0.94823697410(9)	& 7.25(4)	& 51995.0	& 16	&51803$-$52186	& !0.99 & 409(4) \\
1726$-$4006 & 0.88277826565(10)	& 3.33(4)	& 52475.0	& 26	&52262$-$52686	& !1.96 & 277(3) \\
1727$-$2739 & 1.29309994235(16)	& 1.10(3)	& 52263.0	& 22	&51946$-$52579	& !2.21 & 147(4) \\
1730$-$3353 & 3.2702418032(5)	& 21.96(4)	& 51672.0	& 16	&51157$-$52186	& !3.36 & 256(10) \\ \\ 
1731$-$3123 & 0.75304798900(3)	& 1.679(3)	& 51702.0	& 18	&51217$-$52186	& !0.89 & 354(3) \\
1732$-$4156 & 0.323434055969(13)	& 0.661(6)	& 51911.0	& 21	&51687$-$52133	& !0.91 & 228.7(15) \\
1733$-$3030 & 0.362051962860(14)	& 1.6485(11)	& 51914.0	& 26	&51300$-$52526	& !2.03 & 636(3) \\
1733$-$4005 & 0.561778329077(13)	& 3.624(3)	& 52385.0	& 27	&51946$-$52823	& !0.94 & 317.8(11) \\
1736$-$2819 & 1.59241948019(13)	& 14.921(13)	& 51716.0	& 19	&51244$-$52186	& !2.24 & 261(11) \\ \\ 
1736$-$2843 & 6.4450360861(17)	& 29.99(17)	& 52344.0	& 28	&52001$-$52686	& !7.48 & 331(10) \\
1737$-$3320 & 0.81627308018(6)	& 2.251(6)	& 51887.0	& 21	&51581$-$52192	& !1.70 & 804(5) \\
1738$-$2647 & 0.34959099098(5)	& 3.153(15)	& 52359.0	& 13	&52146$-$52572	& !0.47 & 182.2(16) \\
1738$-$3107 & 0.54949769829(13)	& 0.30(7)	& 52333.0	& 28	&52134$-$52532	& !5.49 & 735(10) \\
1738$-$3316 & 0.73037251102(5)	& 0.089(4)	& 51702.0	& 24	&51217$-$52186	& !1.99 & 273(4) \\\hline 
\end{tabular}\end{footnotesize}\end{center}\end{table*}
\addtocounter{table}{-1}\begin{table*} 
\begin{center}\begin{footnotesize}
\caption{-- {\it continued}}\begin{tabular}{llllllll}
\hline 
PSR J & Period, $P$ & $\dot{P}$ & Epoch & N$_{toa}$ & Data Span & Residual & DM \\ 
      & (s)       & (10$^{-15}$) & (MJD) &         & (MJD)     & (ms)     & (cm$^{-3}$pc) \\ 
\hline 
1740$-$2540 & 1.69265634040(9)	& 1.851(12)	& 52162.0	& 27	&51746$-$52578	& !2.11 & 333(5) \\
1740$-$3327 & 0.515000656402(14)	& 3.8978(14)	& 52189.0	& 26	&51805$-$52572	& !0.91 & 274.1(15) \\
1743$-$2442 & 1.24250830909(11)	& 0.472(10)	& 52133.0	& 30	&51687$-$52579	& !3.42 & 239(5) \\
1745$-$2229 & 1.160592538166(19)	& 2.857(3)	& 52242.0	& 17	&51632$-$52851	& !0.80 & 299(3) \\
1749$-$2347 & 0.87448588170(4)	& 2.4236(10)	& 52055.0	& 16	&51217$-$52926	& !1.74 & 344(4) \\ \\ 
1750$-$2444 & 0.89937672140(5)	& 0.264(3)	& 51835.0	& 27	&51091$-$52579	& !3.24 & 331(4) \\
1752$-$2410 & 0.1910366935341(19)	& 0.61773(8)	& 51865.0	& 33	&51159$-$52571	& !0.59 & 508.3(9) \\
1754$-$3443 & 0.361690595821(10)	& 0.5712(7)	& 52260.0	& 21	&51947$-$52572	& !0.40 & 187.7(9) \\
1755$-$25211 & 1.00451255339(7)	& 31.2081(13)	& 51908.0	& 22	&51244$-$52926	& !2.09 & 835(5) \\
1755$-$2534 & 0.233540625445(9)	& 11.2061(9)	& 52188.0	& 31	&51804$-$52571	& !1.45 & 590(3) \\ \\ 
1756$-$2225 & 0.40498030096(9)	& 52.691(5)	& 52360.0	& 20	&52149$-$52926	& !3.68 & 326(4) \\
1758$-$1931 & 0.69255157815(7)	& 16.92(3)	& 52353.0	& 24	&52134$-$52571	& !1.69 & 207(5) \\
1759$-$1903 & 0.73150547250(17)	& 3.07(7)	& 52003.0	& 26	&51804$-$52201	& !2.67 & 467(4) \\
1759$-$3107 & 1.07895331831(4)	& 3.769(9)	& 52608.0	& 19	&52391$-$52823	& !0.27 & 128.6(11) \\
1800$-$2114 & 1.7992724923(7)	& 0.55(26)	& 52360.0	& 21	&52149$-$52571	& !7.56 & 641(18) \\ \\ 
1801$-$2115 & 0.4381131572(4)	& 0.016(6)	& 51887.0	& 22	&52134$-$53032	& !5.06 & 778.8(1) \\
1801$-$2154 & 0.375296924221(10)	& 15.9982(5)	& 51895.0	& 25	&51218$-$52571	& !1.02 & 387.9(14) \\
1803$-$1616 & 0.536595822673(18)	& 1.7714(12)	& 51991.0	& 26	&51410$-$52571	& !1.58 & 388.1(20) \\
1803$-$1920 & 0.443648922251(16)	& 0.3297(8)	& 51843.0	& 24	&51159$-$52527	& !1.23 & 436.1(17) \\
1805$-$2447 & 0.661401759025(14)	& 0.0058(7)	& 51936.0	& 28	&51300$-$52571	& !1.06 & 269(3) \\ \\ 
1806$-$1618 & 0.66830920992(5)	& 0.862(3)	& 51843.0	& 24	&51159$-$52527	& !2.51 & 319(5) \\
1809$-$1850 & 1.12448108944(17)	& 10.575(8)	& 51871.0	& 26	&51214$-$52527	& !6.01 & 598(13) \\
1810$-$1441 & 0.217213500357(4)	& 0.02387(19)	& 51874.0	& 28	&51220$-$52527	& !0.74 & 304.9(13) \\
1812$-$1910 & 0.43099098078(6)	& 37.74(3)	& 51997.0	& 12	&51804$-$52190	& !0.94 & 892(5) \\
1813$-$2242 & 0.328514216183(15)	& 0.0478(6)	& 51843.0	& 29	&51159$-$52527	& !1.67 & 333(3) \\ \\ 
1815$-$1738 & 0.1984357331818(16)	& 77.85293(12)	& 52204.0	& 39	&51584$-$52823	& 14.31 & 728(3) \\
1816$-$1446 & 0.594499815819(18)	& 1.3261(11)	& 51969.0	& 27	&51410$-$52527	& !1.29 & 629(4) \\
1817$-$1511$^J$ & 0.224603837185(16)	& 1.4313(9)	& 52160.0	& 31	&51643$-$52676	& !2.79 & 970(5) \\
1818$-$1116$^J$ & 0.54479952470(5)	& 3.824(6)	& 52158.0	& 29	&51669$-$52647	& !2.58 & 422(7) \\
1819$-$0925$^J$ & 0.852047483505(15)	& 3.1302(11)	& 52105.0	& 34	&51563$-$52647	& !1.03 & 378(3) \\ \\ 
1819$-$1008$^J$ & 0.301489849278(7)	& 1.3209(5)	& 52182.0	& 22	&51743$-$52619	& !0.68 & 404(3) \\
1819$-$1131 & 1.38813712856(16)	& 0.761(10)	& 51874.0	& 18	&51220$-$52527	& !3.91 & 578(13) \\
1820$-$1529$^*$$^J$ & 0.333242850831(18)	& 37.9072(10)	& 52051.0	& 21	&51643$-$52857	& !1.17 & 772(28) \\
1821$-$1419 & 1.6560095697(9)	& 894.5(3)	& 52303.0	& 19	&52077$-$52527	& !8.85 & 1123(16) \\
1822$-$0907 & 0.97470037336(6)	& 0.355(3)	& 52267.0	& 22	&51681$-$52852	& !2.15 & 467(5) \\ \\ 
1822$-$1252 & 2.0710402725(5)	& 84.76(3)	& 51809.0	& 31	&51091$-$52527	& 15.48 & 925(25) \\
1822$-$1617 & 0.8311557274(3)	& 1.885(12)	& 51932.0	& 30	&51335$-$52527	& 11.69 & 647(19) \\
1823$-$1126 & 1.8465342015(3)	& 36.52(11)	& 52331.0	& 17	&52135$-$52527	& !1.58 & 607(3) \\
1823$-$1526 & 1.62540547240(16)	& 4.52(3)	& 52267.0	& 21	&52001$-$52532	& !2.16 & 611(5) \\
1824$-$1500 & 0.41222995857(6)	& 0.758(11)	& 52265.0	& 22	&52001$-$52527	& !3.13 & 571(6) \\ \\ 
1828$-$0611 & 0.269414732063(3)	& 1.4595(3)	& 52045.0	& 21	&51563$-$52527	& !0.28 & 363.2(5) \\
1828$-$1007 & 0.1531969792414(15)	& 0.61339(12)	& 51837.0	& 17	&51091$-$52582	& !0.49 & 302.6(12) \\
1828$-$1057 & 0.246327587410(12)	& 20.7007(19)	& 52166.0	& 29	&51805$-$52527	& !1.96 & 245(3) \\
1831$-$0823$^J$ & 0.612132963973(15)	& 0.3091(12)	& 51763.0	& 25	&51410$-$52115	& !0.35 & 245.9(17) \\
1831$-$1423 & 0.507945195243(13)	& 1.0947(11)	& 52109.0	& 11	&51689$-$52527	& !0.43 & 352(3) \\ \\ 
1833$-$0556 & 1.5215451697(4)	& 1.28(5)	& 52166.0	& 32	&51805$-$52527	& !8.11 & 461(13) \\
1834$-$0633 & 0.31731488279(7)	& 0.60(4)	& 52333.0	& 21	&52138$-$52528	& !2.86 & 707(9) \\
1834$-$0731 & 0.51297990872(4)	& 58.20(3)	& 52323.0	& 18	&52117$-$52527	& !1.37 & 295(3) \\
1834$-$0742$^J$ & 0.788353566535(19)	& 32.4702(9)	& 52145.0	& 32	&51466$-$52823	& !1.47 & 552(18) \\
1834$-$1202 & 0.610258781957(12)	& 0.0067(3)	& 52366.0	& 14	&51026$-$52582	& !0.30 & 342.4(13) \\\hline 
\end{tabular}\end{footnotesize}\end{center}\end{table*}
\addtocounter{table}{-1}\begin{table*} 
\begin{center}\begin{footnotesize}
\caption{-- {\it continued}}\begin{tabular}{llllllll}
\hline 
PSR J & Period, $P$ & $\dot{P}$ & Epoch & N$_{toa}$ & Data Span & Residual & DM \\ 
      & (s)       & (10$^{-15}$) & (MJD) &         & (MJD)     & (ms)     & (cm$^{-3}$pc) \\ 
\hline 
1835$-$0522 & 1.08774918841(8)	& 0.47(4)	& 52331.0	& 20	&52134$-$52528	& !1.08 & 456(4) \\
1836$-$0517 & 0.45724503044(3)	& 1.3019(11)	& 52156.0	& 27	&51460$-$52852	& !2.51 & 564(5) \\
1838$-$0549 & 0.235303200419(10)	& 33.429(6)	& 52642.0	& 14	&52459$-$52824	& !0.47 & 274(7) \\
1838$-$0624 & 0.92717774151(5)	& 0.077(4)	& 51875.0	& 20	&51090$-$52660	& !3.31 & 424(9) \\
1839$-$0905 & 0.418968843399(16)	& 26.033(3)	& 52287.0	& 23	&52002$-$52570	& !0.99 & 348(4) \\ \\ 
1840$-$0559 & 0.85936846681(5)	& 9.602(9)	& 52418.0	& 18	&52149$-$52687	& !0.76 & 321.7(14) \\
1840$-$0809$^J$ & 0.955672135909(8)	& 2.3510(5)	& 52091.0	& 30	&51562$-$52620	& !0.27 & 349.8(8) \\
1840$-$0815$^J$ & 1.096439972228(14)	& 2.4151(9)	& 52091.0	& 29	&51562$-$52620	& !0.61 & 233.2(12) \\
1840$-$1122 & 0.94096161661(4)	& 6.409(6)	& 52287.0	& 25	&52002$-$52570	& !0.96 & 311(3) \\
1841$-$0157$^J$ & 0.663321097981(9)	& 18.0768(5)	& 52163.0	& 39	&51467$-$52858	& !1.26 & 475(3) \\ \\ 
1841$-$0310 & 1.6576564582(4)	& 0.335(9)	& 52353.0	& 16	&51096$-$52918	& !4.47 & 216(12) \\
1841$-$0524 & 0.44574893108(3)	& 233.724(10)	& 52360.0	& 16	&52149$-$52570	& !0.79 & 289(15) \\
1842$-$0309 & 0.40491964372(3)	& 4.518(4)	& 52493.0	& 30	&52134$-$52852	& !2.92 & 955(7) \\
1842$-$0612 & 0.56447537999(16)	& 0.022(9)	& 52402.0	& 28	&52233$-$52929	& !4.90 & 485(10) \\
1843$-$0000$^J$ & 0.88033048188(3)	& 7.792(4)	& 51910.0	& 21	&51633$-$52186	& !0.47 & 101.5(8) \\ \\ 
1843$-$0137 & 0.66987238665(3)	& 2.468(14)	& 52490.0	& 21	&52293$-$52687	& !0.62 & 486(3) \\
1843$-$0211 & 2.02752438872(5)	& 14.44(3)	& 52608.0	& 17	&52391$-$52824	& !0.39 & 441.7(9) \\
1843$-$0408 & 0.78193369973(7)	& 2.39(3)	& 52353.0	& 21	&52135$-$52570	& !1.99 & 246(3) \\
1843$-$0702 & 0.191614063108(3)	& 2.1402(15)	& 52608.0	& 19	&52391$-$52824	& !0.32 & 228.1(7) \\
1843$-$0806 & 0.536413665506(9)	& 17.359(4)	& 52608.0	& 17	&52390$-$52824	& !0.29 & 215.8(9) \\ \\ 
1843$-$1113 & 0.0018456662924246(5)	& 0.00000959(5)	& 52374.0	& 36	&52041$-$52852	& !0.01 & 59.96(3) \\
1844$-$0030$^A$ & 0.64109785122(3)	& 6.078(8)	& 52632.0	& 25	&52407$-$52857	& !0.86 & 605(3) \\
1844$-$0452 & 0.26944332141(3)	& 0.680(6)	& 52471.0	& 14	&52254$-$52687	& !0.89 & 626(4) \\
1844$-$0502 & 0.33516252825(4)	& 0.062(12)	& 52608.0	& 20	&52391$-$52824	& !1.34 & 318(5) \\
1845$-$0545 & 1.09234815284(5)	& 13.43(3)	& 52608.0	& 16	&52391$-$52824	& !0.51 & 315.9(12) \\ \\ 
1846+0051$^A$ & 0.434372879784(16)	& 11.226(3)	& 52554.0	& 31	&52279$-$52828	& !0.86 & 140(3) \\
1847$-$0130 & 6.7070457241(9)	& 1274.9(3)	& 52353.0	& 22	&52135$-$52571	& !2.08 & 667(6) \\
1847$-$0443 & 0.340832130441(14)	& 0.0283(4)	& 52403.0	& 24	&51090$-$52660	& !0.91 & 454.9(20) \\
1848$-$0023$^A$ & 0.53762373255(9)	& 1.610(18)	& 52522.0	& 18	&52279$-$52763	& !1.89 & 30.6(1) \\
1848$-$0055 & 0.27455668472(7)	& 1.35(3)	& 52353.0	& 22	&52134$-$52571	& !4.07 & 1166(7) \\ \\ 
1848$-$0511 & 1.6371290072(6)	& 8.863(11)	& 52777.0	& 23	&51470$-$53033	& !5.12 & 418(7) \\
1849$-$0040$^A$ & 0.67248060717(14)	& 11.14(4)	& 52633.0	& 25	&52409$-$52857	& !3.18 & 1234.9(1) \\
1849$-$0614 & 0.95338418817(3)	& 53.889(6)	& 52417.0	& 20	&52146$-$52687	& !0.63 & 119.6(12) \\
1850$-$0031 & 0.73418485978(10)	& 1.263(6)	& 52015.0	& 25	&51460$-$52569	& !4.91 & 895(8) \\
1851+0118 & 0.90697686069(14)	& 136.705(9)	& 51936.0	& 23	&51301$-$52569	& !5.71 & 418(7) \\ \\ 
1851$-$0053$^A$ & 1.40906524128(16)	& 0.87(8)	& 52585.0	& 28	&52407$-$52763	& !1.21 & 24(4) \\
1851$-$0241 & 0.43519385185(4)	& 7.963(3)	& 52300.0	& 25	&51747$-$52852	& !2.85 & 515(5) \\
1852+0008$^A$ & 0.467894113075(20)	& 5.679(7)	& 52584.0	& 32	&52310$-$52857	& !0.72 & 254.9(18) \\
1852+0013$^A$ & 0.95775094505(5)	& 14.034(12)	& 52633.0	& 38	&52408$-$52857	& !0.99 & 545(3) \\
1852$-$0118 & 0.45147285265(3)	& 1.757(11)	& 52611.0	& 18	&52396$-$52824	& !0.99 & 286(3) \\ \\ 
1852$-$0127 & 0.42897892562(3)	& 5.149(8)	& 52608.0	& 21	&52390$-$52824	& !1.12 & 431(3) \\
1852$-$0635 & 0.52415088472(14)	& 14.64(5)	& 52477.0	& 18	&52265$-$52687	& !3.31 & 171(6) \\
1853+0011$^A$ & 0.397881893633(12)	& 33.5381(16)	& 52554.0	& 28	&52279$-$52828	& !0.53 & 568.8(16) \\
1853+0505 & 0.90513715648(5)	& 1.288(5)	& 52321.0	& 21	&51817$-$52825	& !1.93 & 279(3) \\
1853$-$0004$^A$ & 0.1014357461981(14)	& 5.5745(5)	& 52633.0	& 19	&52409$-$52857	& !0.19 & 438.2(8) \\ \\ 
1855+0307$^A$ & 0.84534757461(4)	& 18.110(10)	& 52632.0	& 29	&52407$-$52857	& !0.61 & 402.5(19) \\
1855+0700 & 0.258684648071(7)	& 0.7516(6)	& 51991.0	& 21	&51413$-$52569	& !1.35 & 244(4) \\
1856+0102$^A$ & 0.620217115135(20)	& 1.222(3)	& 52568.0	& 40	&52279$-$52857	& !0.65 & 554(3) \\
1857+0143$^A$ & 0.139760064515(4)	& 31.1674(12)	& 52632.0	& 30	&52407$-$52857	& !0.65 & 249(3) \\
1857+0809 & 0.502923870516(13)	& 4.7374(8)	& 51991.0	& 26	&51413$-$52569	& !1.33 & 282(3) \\\hline 
\end{tabular}\end{footnotesize}\end{center}\end{table*}
\addtocounter{table}{-1}\begin{table*} 
\begin{center}\begin{footnotesize}
\caption{-- {\it continued}}\begin{tabular}{llllllll}
\hline 
PSR J & Period, $P$ & $\dot{P}$ & Epoch & N$_{toa}$ & Data Span & Residual & DM \\ 
      & (s)       & (10$^{-15}$) & (MJD) &         & (MJD)     & (ms)     & (cm$^{-3}$pc) \\ 
\hline 
1858+0241 & 4.6932329333(12)	& 24.32(9)	& 52111.0	& 22	&51688$-$52532	& !5.64 & 336(15) \\
1859+0601$^A$ & 1.04431270179(15)	& 25.51(4)	& 52503.0	& 56	&52279$-$52726	& !1.83 & 276(7) \\
1900+0634$^A$ & 0.389869101178(20)	& 5.125(5)	& 52554.0	& 54	&52279$-$52828	& !1.16 & 323.4(18) \\
1900$-$0051$^J$ & 0.385194094862(6)	& 0.1421(9)	& 51912.0	& 24	&51634$-$52188	& !0.26 & 136.8(7) \\
1901+0124$^A$ & 0.318817259335(11)	& 3.241(4)	& 52632.0	& 29	&52407$-$52857	& !0.56 & 314.4(13) \\ \\ 
1901+0254 & 1.2996934495(3)	& 0.46(11)	& 52626.0	& 15	&52426$-$52824	& !2.09 & 185(5) \\
1901+0320$^A$ & 0.63658447822(8)	& 0.52(3)	& 52503.0	& 26	&52279$-$52726	& !2.48 & 393(7) \\
1901+0355 & 0.55475646483(3)	& 12.741(10)	& 52352.0	& 21	&52134$-$52569	& !0.92 & 547(3) \\
1901+0510$^A$ & 0.61475669408(12)	& 31.10(4)	& 52618.0	& 30	&52407$-$52828	& !2.89 & 429(7) \\
1901$-$0312 & 0.355725186569(14)	& 2.292(6)	& 52608.0	& 18	&52390$-$52824	& !0.53 & 106.4(11) \\ \\ 
1902+0248 & 1.22377745359(17)	& 2.41(3)	& 52554.0	& 35	&52279$-$52828	& !2.99 & 272.0(1) \\
1902$-$0340 & 1.5246721060(4)	& 2.00(17)	& 52724.0	& 18	&52532$-$52916	& !2.43 & 114(6) \\
1903+0601$^J$ & 0.374117028251(5)	& 19.2039(3)	& 52146.0	& 33	&51467$-$52824	& !0.94 & 388(3) \\
1905+0400 & 0.0037844047875897(12)	& 0.00000486(6)	& 52173.0	& 107	&51492$-$52853	& !0.04 & 25.71(6) \\
1905+0600 & 0.441209731966(18)	& 1.1123(10)	& 52048.0	& 42	&51469$-$52626	& !0.86 & 730.1(19) \\ \\ 
1906+0649$^A$ & 1.28656437956(10)	& 0.152(5)	& 52317.0	& 31	&51805$-$52828	& !1.37 & 249(4) \\
1907+0249$^*$$^A$ & 0.351879439822(20)	& 1.135(4)	& 52554.0	& 36	&52279$-$52828	& !3.59 & 261(6) \\
1907+0345 & 0.240153263208(5)	& 8.222(3)	& 51999.0	& 19	&51805$-$52193	& !0.35 & 311.7(9) \\
1907+0731 & 0.363676330005(12)	& 18.416(4)	& 52352.0	& 22	&52134$-$52569	& !0.69 & 239.8(13) \\
1907+0918$^\dagger$ & 0.2261071099878(6)	& 94.2955(4)	& 51319.0	& --- 	&51257$-$51540	& !--- & 357.9(1) \\ \\ 
1910+0225$^A$ & 0.337854845269(11)	& 0.2623(14)	& 52586.0	& 33	&52315$-$52857	& !0.75 & 209(3) \\
1910+0728 & 0.325415321974(3)	& 8.3062(3)	& 52187.0	& 23	&51805$-$52569	& !0.22 & 283.7(4) \\
1913+1000 & 0.83714819649(5)	& 16.737(6)	& 52187.0	& 17	&51805$-$52569	& !1.21 & 422(3) \\
1914+0631$^A$ & 0.69381120574(5)	& 0.033(13)	& 52582.0	& 39	&52335$-$52828	& !1.16 & 58(3) \\
1915+0838$^J$ & 0.34277679653(4)	& 1.571(4)	& 52025.0	& 24	&51743$-$52306	& !0.99 & 358(3) \\ \\ 
1916+0844$^J$ & 0.439995272067(8)	& 2.9009(4)	& 52018.0	& 30	&51467$-$52569	& !0.54 & 339.4(8) \\
1916+0852 & 2.1827459895(6)	& 13.1(3)	& 52352.0	& 15	&52134$-$52569	& !5.21 & 295(10) \\
1916+1023$^A$ & 1.6183389208(8)	& 0.68(14)	& 52529.0	& 24	&52294$-$52763	& !4.65 & 329.8(1) \\
1920+1040$^*$$^A$ & 2.21580173889(20)	& 6.48(3)	& 52596.0	& 38	&52335$-$52857	& !4.46 & 304(9) \\
1937+1505$^J$ & 2.8727736506(7)	& 5.6(3)	& 51969.0	& 19	&51743$-$52194	& !4.11 & 237(11) \\ \\ 
\hline 
\end{tabular}\end{footnotesize}\end{center}\end{table*}

 For every observation of each pulsar we obtained a pulse topocentric
 arrival time (TOA).  Using the \textsc{TEMPO} program\footnote{See
 \url{http://www.atnf.csiro.au/research/pulsar/tempo/}} we
 fitted a timing model, which contained the pulsar's position,
 rotational period and its derivative, to the TOAs of each pulsar.  In
 Table~\ref{tb:posn} we provide these positions in equatorial and
 Galactic coordinates. Subsequent columns contain information on the
 discovery of each pulsar: the beam number (corresponding to the 13
 beams of the multibeam receiver) for the highest S/N discovery
 observation of this pulsar, the radial distance between the centre of
 this beam and the position of the pulsar (beam radii greater than one
 beam width occur if the pulsar scintillates or nulls or the closest
 pointing was contaminated by interference) and the S/N of the profile
 during this observation.  The observations used to form TOAs were
 added together to provide a characteristic pulse profile for each
 pulsar at 1374\,MHz (Figure~\ref{fg:prf}). The final three columns in
 Table~\ref{tb:posn} contain the flux densities measured from these
 mean profiles and the pulse widths at 50\% and 10\% of the pulse
 height. The 10\% width is not measurable for pulsars with mean
 profiles that have poor signal--to--noise ratios. For profiles
 containing multiple components the widths are measured across the
 entire profile. 

 Some of the profiles observed at Arecibo had significant dips at the
 start of the observed pulse shape.  These are due to instrumental
 data quantization problems.  For cases where dips remained even after
 applying a correction scheme (van Vleck \& Middleton
 1966\nocite{vm66}), we used Parkes observations to determine the flux
 density and pulse widths (and provide the profile obtained at Parkes
 in Figure~\ref{fg:prf}) if high S/N profiles were available. For the
 remaining pulsars that were observed at the Arecibo Observatory, flux
 densities were obtained using estimates of the gain and system
 temperatures (these vary significantly with zenith angle and were
 therefore found separately for each observation\footnote{see
 \url{http://www.naic.edu/~astro/RXstatus/Lnarrow/ln_gain_postaug01.shtml}
 and
 \url{http://www.naic.edu/~astro/RXstatus/Lnarrow/ln_tsys_2001.shtml}}). The
 sky temperature at each position was estimated from Haslam et
 al. (1982)\nocite{hssw82} assuming a spectral index of $-$2.5
 \cite{rr88b}.  The scale was determined by comparing the baseline
 noise with the predictions of the radiometer equation and flux
 densities were found by integrating under the peak of each profile.
 Individual observations were averaged to find the mean flux density
 for each pulsar.  This was then corrected for off-centre pointing by
 assuming a Gaussian beam shape with a beam width of
 3.6\,arcminutes. Estimated uncertainties for all parameters are given
 in parentheses where relevant and refer to the last quoted digit.

 The pulsars' rotational parameters are given in Table~\ref{tb:prd}.
 In column order, this table provides each pulsar's name,
 solar--system barycentric pulse period, period derivative, epoch of
 the period, the number of TOAs used in the timing solution, the MJD
 range covered by the timing observations, the final
 root--mean--square values for the timing residuals and the dispersion
 measure.  The data have been folded at two and three times the
 tabulated periods in order to confirm that they represent the
 fundamental periods of the pulsars. Pulsars timed primarily at
 Arecibo or Jodrell Bank are indicated by a superscript `$A$' or `$J$'
 respectively; all other pulsars were timed using the Parkes
 telescope.  PSRs~J1016$-$5857 and J1437$-$6146 have both glitched.
 Table~\ref{tb:prd} contains their post--glitch solution; full details
 of the glitches will be provided in a later paper. A pre-glitch
 timing solution for PSR~J1016$-$5857 has also been published by
 Camilo et al. (2001a)\nocite{cbm+01}. A timing solution for
 PSR~J1847$-$0130 was previously published by McLaughlin et al.
 (2003a)\nocite{msk+03a}. 

 Five pulsars in our sample were independently discovered by other
 surveys. We define pulsars as independent discoveries if our
 confirmation of the pulsar candidate occurred prior to the pulsar's
 parameters being published elsewhere. Three, PSRs~J0843$-$5022,
 J1352$-$6803 and J1415$-$6621, were detected in the Swinburne
 multibeam pulsar survey \cite{ebvb01}.  The timing solution given in
 Tables~\ref{tb:posn} and \ref{tb:prd} for PSR~J0843$-$5022 was
 obtained from Edwards et al. (2001).  Bailes (2003; private
 communication) provided a timing solution for PSR~J1415$-$6621 and
 the solution for PSR~J1352$-$6803 was obtained from our observations.
 PSR~J1907$+$0918 was discovered by Lorimer \& Xilouris
 (2000)\nocite{lx00} during a search for radio emission from
 SGR~1900+14.  The timing solution given in Tables~\ref{tb:posn} and
 \ref{tb:prd} is obtained from Lorimer \& Xilouris (2000). The flux
 density and pulse widths were obtained from the Parkes multibeam
 data.  The flux density tabulated of $0.29(4)$\,mJy agrees well with
 the earlier measurement of $0.3(1)$\,mJy.  PSR~J1435$-$5954, was
 independently discovered at Parkes in an unpublished pulsar search
 during the year 1995.  We provide a timing solution from observations
 between 1995 and 1998.

 PSR~J1420$-$5625 is a 34\,ms pulsar in a 40-day binary system. The
 orbital parameters for this intermediate--mass binary pulsar system
 are given in Table~\ref{tb:binary}\nocite{bt76} and discussed in
 Section~\ref{sec:1420}.  All published parameters may also be
 obtained online using the ATNF pulsar
 catalogue\footnote{\url{http://www.atnf.csiro.au/research/pulsar/psrcat}}
 (Manchester et al., in preparation).

\begin{figure*} 
\centerline{\psfig{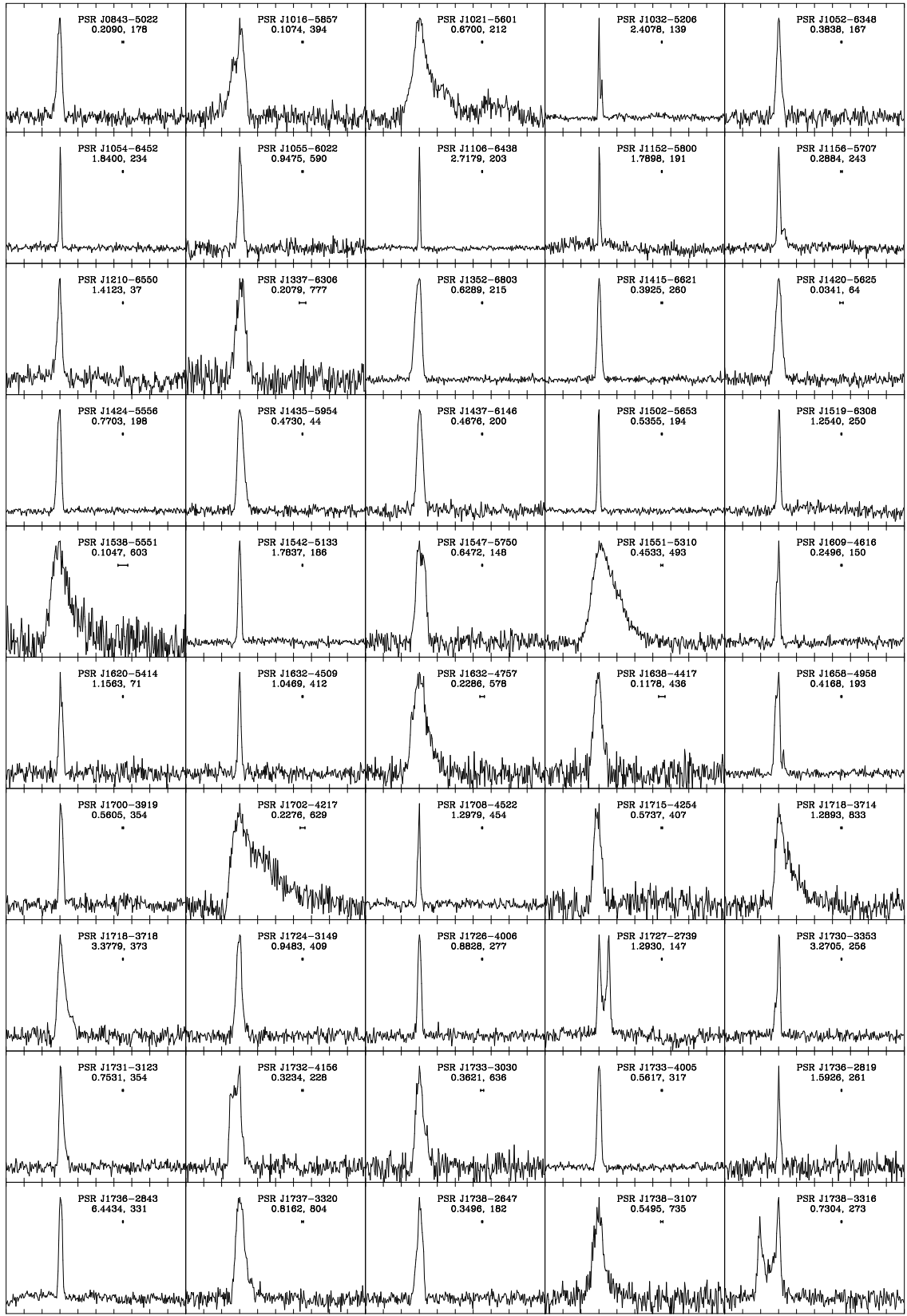}} 
\caption{Mean 1374\,MHz pulse profiles for 180 pulsars discovered in the
multibeam survey. The highest point in the profile is placed at phase
0.3. For each profile, the pulsar Jname, pulse period (s) and dispersion
measure (cm$^{-3}$pc) are given. The
small horizontal bar under the period indicates the effective resolution of
the profile by adding the bin size to the effects of interstellar dispersion
in quadrature.}
\label{fg:prf}
\end{figure*}
\addtocounter{figure}{-1}
\begin{figure*} 
\centerline{\psfig{file=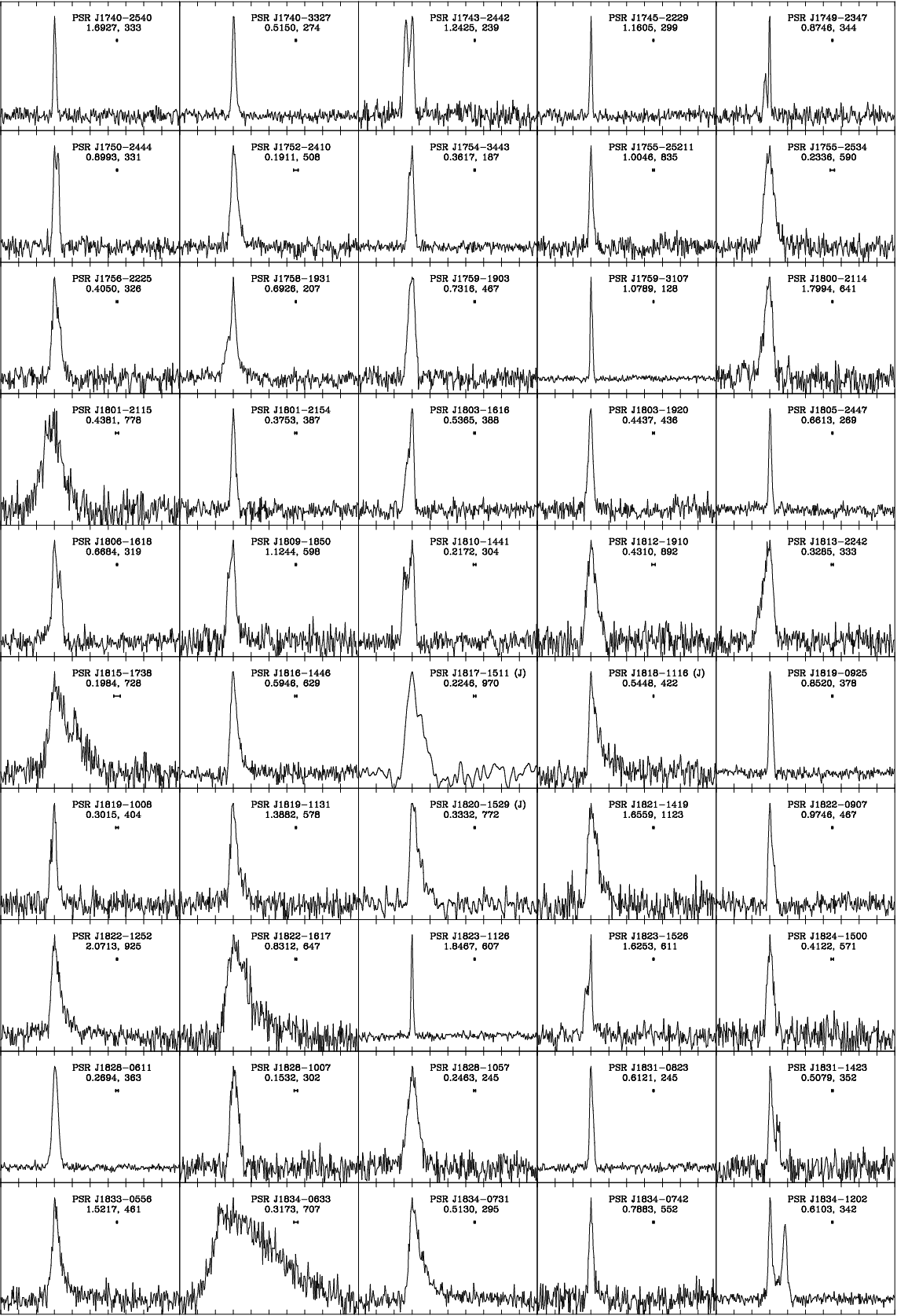,width=150mm}} 
\caption{-- {\it continued}}
\end{figure*}
\addtocounter{figure}{-1}
\begin{figure*} 
\centerline{\psfig{file=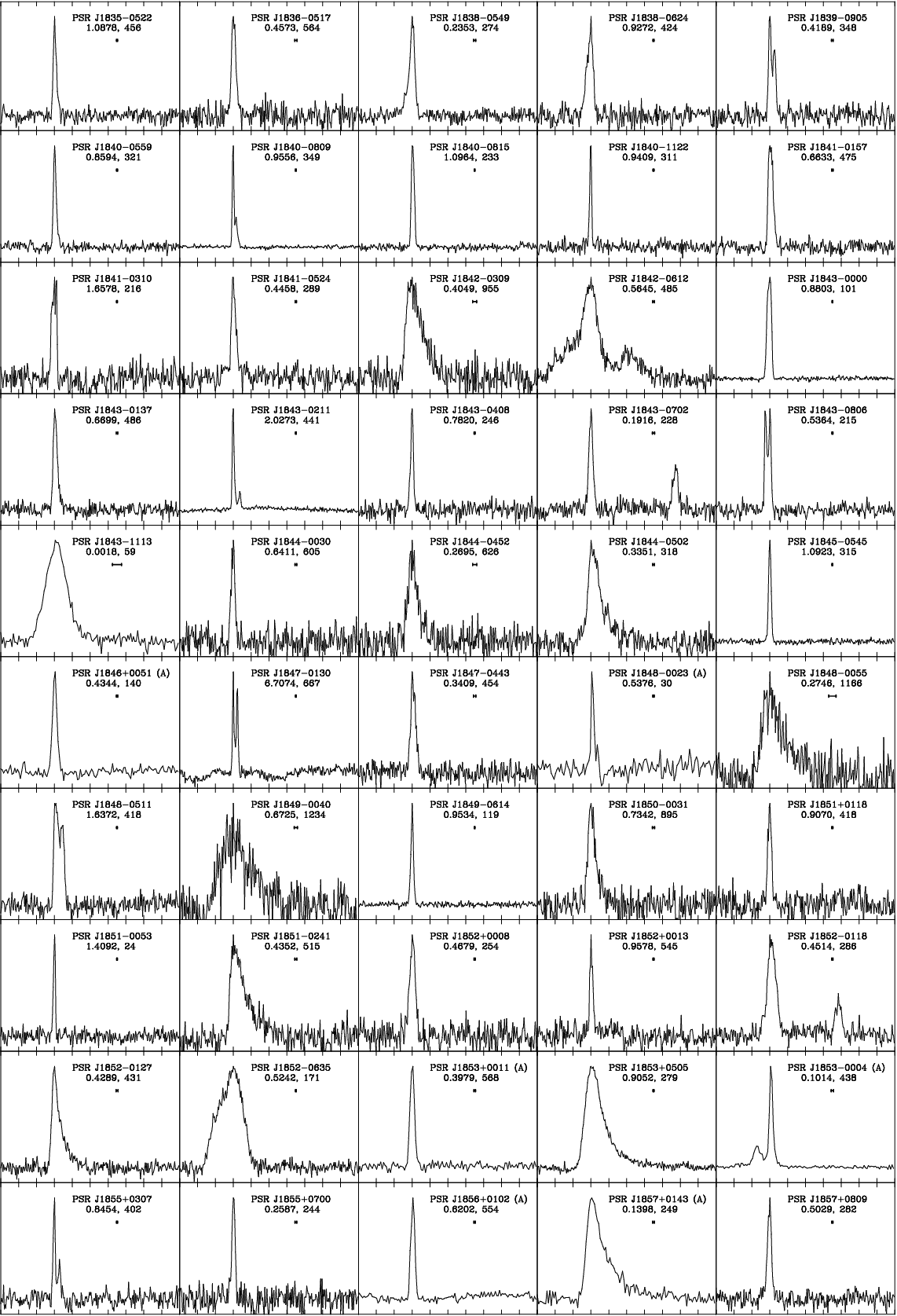,width=150mm}} 
\caption{-- {\it continued}}
\end{figure*}
\addtocounter{figure}{-1}
\begin{figure*}
\centerline{\psfig{file=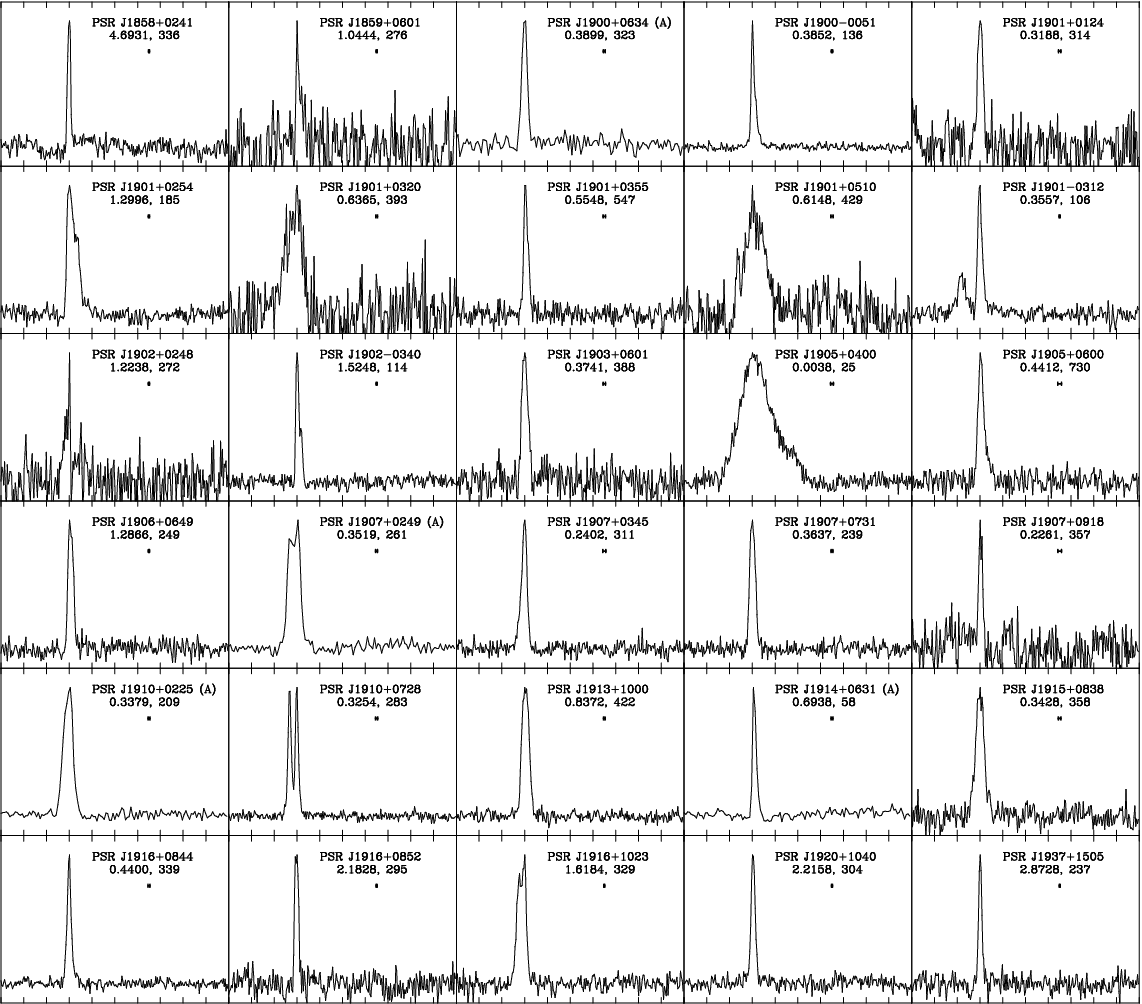,width=150mm}} 
\caption{-- {\it continued}}
\end{figure*}

\section{DISCUSSION}

 This discussion section is in two parts: we first describe the
 newly discovered pulsars and second discuss those that were detected, but
 not discovered, during the multibeam survey.

\begin{table}
\caption{Orbital parameters for PSR~J1420$-$5625 obtained using the
Blandford \& Teukolsky (1976) binary model.  The minimum companion
mass is calculated by assuming an inclination angle of $90^\circ$ and
a neutron star mass of 1.35\,M$_\odot$.}
\begin{tabular}{ll} \hline
Orbital period (d) & 40.294523(4) \\
Projected semi-major axis of orbit (lt sec) & 29.53977(4) \\
Eccentricity & 0.003500(3) \\
Epoch of periastron (MJD) & 52388.945(6) \\
Longitude of periastron (degrees) & 337.30(5) \\  \\
Minimum companion mass (M$_\odot$) & 0.37 \\ \hline
\end{tabular} 
\label{tb:binary}
\end{table}

\subsection{New discoveries}

 In Table~\ref{tb:deriv} we list the pulsars' derived parameters: the
 logarithm of the characteristic age in years, the surface dipole
 magnetic field strength, $B_s = 3.2\times10^{19}(P\dot{P})^{1/2}$ in
 Gauss, and the rate of loss of rotational energy in erg\,s$^{-1}$
 where a neutron star with moment of inertia of $10^{45}$\,g\,cm$^2$
 is assumed.  The final columns contain the pulsar distances and
 luminosities.  The distances are computed from their dispersion
 measures assuming the Taylor \& Cordes (1993)\nocite{tc93} model for
 the Galactic distribution of free electrons. This model is used,
 rather than the more recent Cordes \& Lazio (2002)\nocite{cl02} or
 Gomez, Benjamin \& Cox (2002)\nocite{gbc02} models for consistency
 with the distance and luminosity values provided in earlier papers of
 this series.  The implications of using the Cordes \& Lazio (2002)
 model for the determination of the distances and luminosities of the
 multibeam pulsars was described in Kramer et
 al. (2003)\nocite{kbm+03}.  In general, the distances are less
 accurate than the 0.1\,kpc quoted because of uncertainties in the
 electron density model.

 \subsubsection{Rotational properties}

\begin{figure}
 \centerline{\psfig{file=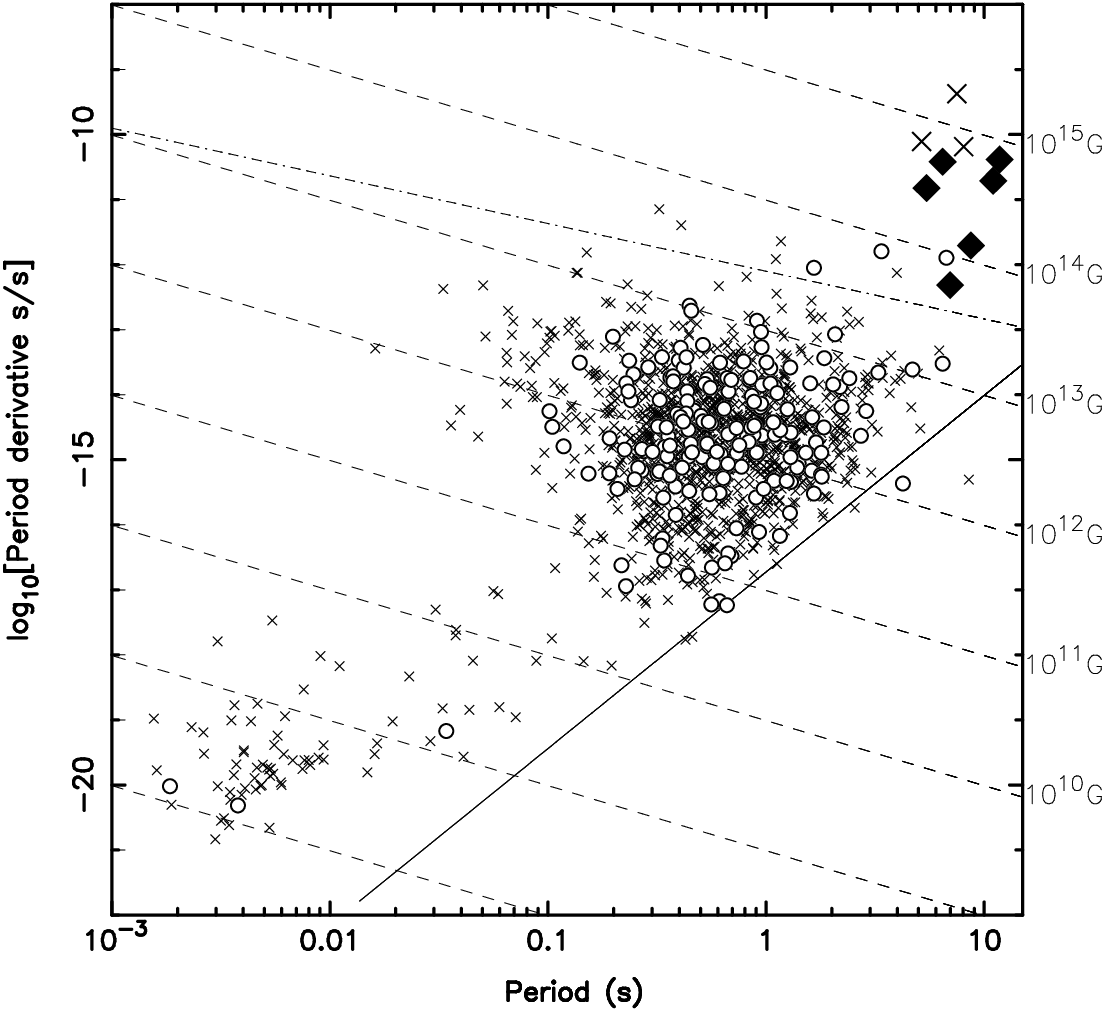,width=8cm,angle=-90}} 
 \caption{P--$\dot{\rm{P}}$ diagram containing the multibeam pulsars 
 listed in this paper (circles) overlaid on the previously known
 population.  Diamonds indicate the anomalous X-ray pulsars (AXPs) and 
 large crosses the soft $\gamma$-ray repeaters (SGRs) that are listed in the
 ATNF pulsar catalogue (version 1.13; Manchester et al., in preparation).  Lines of
 constant magnetic field are shown as dashed lines and assume that
 pulsars, AXPs and SGRs spin down due to magnetic dipole radiation. The AXPs
 and SGRs lie in a region of the diagram that is predicted to be radio
 quiet (indicated using a dot-dashed line; and defined by 
 Equation 10 in Baring \& Harding 2001). The solid line is a `death
 line' defined by $7 \log B_s - 13 \log P = 78$
 (Chen \& Ruderman 1993).}\label{fg:ppdot}
 \end{figure}

 A $P$--$\dot{P}$ diagram is shown in
 Figure~\ref{fg:ppdot}\nocite{cr93} with the 180 new discoveries
 presented in this paper highlighted (open circles).  Three of these
 pulsars (PSRs~J1821$-$1419, J1718$-$3718 and J1847$-$0130) lie just
 inside a region of the diagram that is expected to be radio quiet
 \cite{bh01}.  However, the exact position of the boundary that
 defines this region is not well determined and, for instance, depends
 upon the poorly known height of the radio emission above the neutron
 star's surface.  Two of these pulsars, PSRs~J1718$-$3718 and
 J1847$-$0130, have rotational parameters similar to the anomalous
 X-ray pulsars (diamond symbols in Figure~\ref{fg:ppdot}) and have
 already been discussed by McLaughlin et al.
 (2003a,b)\nocite{msk+03a}\nocite{msk+03b}.  PSRs~J1847$-$0130 and
 J1736$-$2843 have rotational periods greater than six seconds and are
 the second and third slowest rotating radio pulsars known.  However,
 even with its long rotational period, PSR~J1736$-$2843 lies in the
 $P$--$\dot{P}$ diagram below the `radio-quiet' boundary and above the
 death-line.

 Our sample also includes the solitary millisecond pulsars,
 PSRs~J1843$-$1113 ($P=1.85$\,ms) and J1905$+$0400 ($P=3.78$\,ms).
 PSR~J1843$-$1113 is the third fastest rotating pulsar known (after
 PSRs~B1937$+$21 and B1957$+$20 which have spin periods of 1.56 and
 1.61\,ms respectively). We reported in earlier multibeam papers that
 the multibeam survey had discovered fewer recycled pulsars than
 expected (see also Toscano et al. 1998\nocite{tbms98}).  This lack of
 recycled pulsars was partly due to a poor choice of software filters
 that were applied to remove known interference before searching
 begins for pulsar candidates.  Hobbs (2002)\nocite{hob02} showed that
 significant increases in the detection rates of millisecond pulsars
 could be made by improving these filters.  The multibeam data are
 currently being reanalysed with updated search code.  A full
 description of this reprocessing will be published by Faulkner et
 al. (in preparation).

 Bailes et al. (1997)\nocite{bjb+97} reported on the discovery of four
 isolated millisecond pulsars. Three of these have very low
 luminosities while the other had a more intermediate luminosity. They
 concluded that the solitary millisecond pulsars are less luminous
 than those in binary systems. This result was confirmed by Kramer et
 al. (1998)\nocite{kxl+98} using a sample of seven isolated
 millisecond pulsars. The two solitary pulsars, PSRs J1843$-$1113 and
 J1905$+$0400, reported in this paper also have low luminosities at
 1400\,MHz of 0.39 and 0.09\,mJy\,kpc$^2$ respectively.  The median
 luminosity at 1400\,MHz for the 16 solitary millisecond pulsars known
 is 0.4\,mJy\,kpc$^2$ and the luminosity range is from
 0.03\,mJy\,kpc$^2$ for PSR~J0030+0451 to 207\,mJy\,kpc$^2$ for
 PSR~B1937+21.  The corresponding luminosities for millisecond pulsars
 in binary systems range between 0.39\,mJy\,kpc$^2$ to
 126\,mJy\,kpc$^2$ and have a median value of 3.7\,mJy\,kpc$^2$.  We
 must, however, emphasise that (i) the fastest rotating and solitary
 pulsar, PSR~B1937$+$21, has the highest observed luminosity of all
 the recycled pulsars, (ii) derived luminosities are highly dependent
 upon the distance to the pulsar and therefore have large
 uncertainties and (iii) many millisecond pulsars scintillate and
 therefore published flux densities may not give a true representation
 of the pulsar's intrinsic luminosity.

 \begin{table*} 
\caption{Derived parameters for 180 pulsars discovered in the Parkes multibeam pulsar
survey. We list the characteristic age, the surface dipole magnetic field strength,
the loss in rotational energy, the distance derived from the DM and the Taylor \&
Cordes (1993) model, the inferred $z$-height and the corresponding radio luminosity
at 1400 MHz.}\label{tb:deriv}
\begin{center}\begin{footnotesize}
\begin{tabular}{lllllll}
\hline 
PSR J & $\log[\tau_c({\rm yr})]$ & $\log[B_s({\rm G})]$ & $\log[\dot{E}({\rm erg\,s}^{-1})]$ & Dist. & z & Luminosity \\ 
 & & & & (kpc) & (kpc) & (mJy kpc$^2$) \\ 
\hline 
0843$-$5022 	& 7.28 & 11.28 & 32.88 & !7.7 & $-$0.66 & !18.3 \\
1016$-$5857 	& 4.32 & 12.47 & 36.41 & !9.3 & $-$0.31 & !39.9 \\
1021$-$5601 	& 8.30 & 11.28 & 30.85 & !4.2 & !0.07 & !!6.5 \\
1032$-$5206 	& 6.33 & 12.82 & 31.71 & !4.3 & !0.39 & !!3.6 \\
1052$-$6348 	& 7.20 & 11.59 & 32.43 & !5.3 & $-$0.36 & !!3.1 \\ \\ 
1054$-$6452 	& 6.97 & 12.39 & 31.30 & 13.5 & $-$1.13 & !45.8 \\
1055$-$6022 	& 5.21 & 12.98 & 33.63 & 25.6 & $-$0.29 & 105.1 \\
1106$-$6438 	& 7.27 & 12.40 & 30.66 & !7.9 & $-$0.55 & !11.8 \\
1152$-$5800 	& 7.34 & 12.18 & 30.94 & !7.9 & !0.54 & !!7.4 \\
1156$-$5707 	& 5.24 & 12.45 & 34.64 & 20.4 & !1.76 & !79.1 \\ \\ 
1210$-$6550 	& 8.19 & 12.14 & 29.34 & !1.6 & $-$0.09 & !!0.4 \\
1337$-$6306 	& 6.97 & 11.44 & 33.20 & 16.0 & $-$0.19 & !28.2 \\
1352$-$6803 	& 6.91 & 11.95 & 32.30 & 14.4 & $-$1.47 & 140.4 \\
1415$-$6621 	& 7.03 & 11.68 & 32.58 & 14.4 & $-$1.22 & 147.2 \\
1420$-$5625 	& 9.90 & !9.19 & 31.83 & !1.7 & !0.13 & !!0.4 \\ \\ 
1424$-$5556 	& 7.19 & 11.89 & 31.83 & !7.7 & !0.62 & !22.6 \\
1435$-$5954 	& 6.69 & 11.94 & 32.76 & !1.5 & !0.01 & !!7.7 \\
1437$-$6146 	& 6.07 & 12.24 & 33.38 & !4.8 & $-$0.12 & !!5.5 \\
1502$-$5653 	& 6.67 & 12.00 & 32.67 & !4.5 & !0.12 & !!7.9 \\
1519$-$6308 	& 6.52 & 12.44 & 32.08 & 17.8 & $-$1.52 & 101.5 \\ \\ 
1538$-$5551 	& 5.71 & 11.77 & 35.04 & 10.4 & $-$0.05 & !27.0 \\
1542$-$5133 	& 7.68 & 12.02 & 30.61 & !4.9 & !0.24 & !!6.4 \\
1547$-$5750 	& 8.60 & 11.11 & 30.57 & !3.9 & $-$0.18 & !!3.5 \\
1551$-$5310 	& 4.57 & 12.98 & 34.92 & !7.5 & !0.09 & !30.5 \\
1609$-$4616 	& 6.90 & 11.55 & 33.11 & !4.1 & !0.29 & !!6.5 \\ \\ 
1620$-$5414 	& 8.43 & 11.45 & 30.23 & !1.7 & $-$0.09 & !!0.4 \\
1632$-$4509 	& 6.05 & 12.60 & 32.71 & !9.1 & !0.32 & !13.3 \\
1632$-$4757 	& 5.38 & 12.27 & 34.70 & !7.0 & !0.01 & !14.5 \\
1638$-$4417 	& 6.06 & 11.64 & 34.59 & !8.5 & !0.26 & !15.0 \\
1658$-$4958 	& 6.23 & 12.11 & 33.32 & !6.3 & $-$0.50 & !34.1 \\ \\ 
1700$-$3919 	& 9.25 & 10.73 & 30.04 & !6.3 & !0.20 & !!9.2 \\
1702$-$4217 	& 8.50 & 10.71 & 31.58 & !7.5 & $-$0.04 & !28.1 \\
1708$-$4522 	& 6.90 & 12.27 & 31.67 & 30.0 & $-$1.57 & 198.0 \\
1715$-$4254 	& 7.02 & 11.86 & 32.26 & 12.6 & $-$0.56 & !11.1 \\
1718$-$3714 	& 5.89 & 12.77 & 32.68 & 10.9 & !0.04 & !27.2 \\ \\ 
1718$-$3718 	& 4.53 & 13.87 & 33.20 & !5.1 & !0.02 & !!4.7 \\
1724$-$3149 	& 6.32 & 12.42 & 32.53 & 10.5 & !0.41 & !39.7 \\
1726$-$4006 	& 6.62 & 12.24 & 32.28 & !6.2 & $-$0.29 & !!8.0 \\
1727$-$2739 	& 7.27 & 12.08 & 31.30 & !3.8 & !0.26 & !22.5 \\
1730$-$3353 	& 6.37 & 12.93 & 31.40 & !4.2 & $-$0.00 & !!6.8 \\ \\ 
1731$-$3123 	& 6.85 & 12.06 & 32.20 & !5.3 & !0.13 & !!8.2 \\
1732$-$4156 	& 6.89 & 11.67 & 32.89 & !8.8 & $-$0.72 & !17.0 \\
1733$-$3030 	& 6.54 & 11.89 & 33.15 & 14.2 & !0.32 & !40.2 \\
1733$-$4005 	& 6.39 & 12.16 & 32.91 & 13.9 & $-$0.95 & !95.4 \\
1736$-$2819 	& 6.23 & 12.69 & 32.18 & !4.9 & !0.17 & !!3.8 \\ \\ 
1736$-$2843 	& 6.53 & 13.15 & 30.64 & !5.5 & !0.17 & !12.9 \\
1737$-$3320 	& 6.76 & 12.14 & 32.20 & 14.1 & $-$0.20 & !69.4 \\
1738$-$2647 	& 6.25 & 12.03 & 33.46 & !3.9 & !0.17 & !!6.6 \\
1738$-$3107 	& 7.47 & 11.61 & 31.85 & 10.8 & !0.02 & !30.1 \\
1738$-$3316 	& 8.12 & 11.41 & 30.95 & !4.6 & $-$0.08 & !11.4 \\\hline 
\end{tabular}\end{footnotesize}\end{center}\end{table*}
\addtocounter{table}{-1}\begin{table*} 
\begin{center}\begin{footnotesize}
\caption{-- {\it continued}}\begin{tabular}{lllllll}
\hline 
PSR J & $\log[\tau_c({\rm yr})]$ & $\log[B_s({\rm G})]$ & $\log[\dot{E}({\rm erg\,s}^{-1})]$ & Dist. & z & Luminosity \\ 
 & & & & (kpc) & (kpc) & (mJy kpc$^2$) \\ 
\hline 
1740$-$2540 	& 7.16 & 12.25 & 31.18 & !8.3 & !0.38 & !10.9 \\
1740$-$3327 	& 6.32 & 12.16 & 33.04 & !4.7 & $-$0.12 & !!6.7 \\
1743$-$2442 	& 7.62 & 11.89 & 30.99 & !5.0 & !0.23 & !!3.4 \\
1745$-$2229 	& 6.81 & 12.26 & 31.86 & !9.3 & !0.56 & !11.2 \\
1749$-$2347 	& 6.76 & 12.17 & 32.15 & !6.1 & !0.21 & !!4.8 \\ \\ 
1750$-$2444 	& 7.73 & 11.69 & 31.15 & !5.0 & !0.11 & !!6.8 \\
1752$-$2410 	& 6.69 & 11.54 & 33.54 & !7.4 & !0.13 & !25.9 \\
1754$-$3443 	& 7.00 & 11.66 & 32.68 & !5.6 & $-$0.45 & !15.4 \\
1755$-$25211 	& 5.71 & 12.75 & 33.08 & 11.5 & $-$0.00 & !22.6 \\
1755$-$2534 	& 5.52 & 12.21 & 34.54 & !7.2 & $-$0.03 & !!8.9 \\ \\ 
1756$-$2225 	& 5.09 & 12.67 & 34.49 & !5.0 & !0.11 & !!6.3 \\
1758$-$1931 	& 5.81 & 12.54 & 33.30 & !4.2 & !0.17 & !!6.8 \\
1759$-$1903 	& 6.58 & 12.18 & 32.49 & 13.3 & !0.52 & !28.3 \\
1759$-$3107 	& 6.66 & 12.31 & 32.08 & !3.4 & $-$0.22 & !10.3 \\
1800$-$2114 	& 7.71 & 12.00 & 30.57 & 11.2 & !0.21 & !37.5 \\ \\ 
1801$-$2115 	& 8.65 & 10.92 & 30.86 & !-1.0 & $-$0.00 & !!0.2 \\
1801$-$2154 	& 5.57 & 12.39 & 34.08 & !5.2 & !0.05 & !!4.8 \\
1803$-$1616 	& 6.68 & 11.99 & 32.65 & 12.9 & !0.64 & !26.6 \\
1803$-$1920 	& 7.33 & 11.59 & 32.18 & !6.5 & !0.15 & !11.3 \\
1805$-$2447 	& 9.26 & 10.80 & 29.90 & !4.8 & $-$0.14 & !!6.2 \\ \\ 
1806$-$1618 	& 7.09 & 11.89 & 32.04 & !6.4 & !0.25 & !!8.9 \\
1809$-$1850 	& 6.23 & 12.54 & 32.46 & !7.1 & !0.04 & !!10.0 \\
1810$-$1441 	& 8.16 & 10.86 & 31.96 & !5.9 & !0.21 & !!7.3 \\
1812$-$1910 	& 5.26 & 12.61 & 34.28 & 11.5 & $-$0.09 & !29.3 \\
1813$-$2242 	& 8.04 & 11.10 & 31.72 & !6.8 & $-$0.28 & !!9.7 \\ \\ 
1815$-$1738 	& 4.61 & 12.60 & 35.59 & !9.0 & $-$0.04 & !20.3 \\
1816$-$1446 	& 6.85 & 11.95 & 32.40 & !9.0 & !0.13 & !18.5 \\
1817$-$1511 	& 6.40 & 11.76 & 33.70 & 11.6 & !0.08 & !57.7 \\
1818$-$1116 	& 6.35 & 12.16 & 32.97 & 10.3 & !0.37 & !52.9 \\
1819$-$0925 	& 6.63 & 12.22 & 32.30 & 11.1 & !0.51 & !88.7 \\ \\ 
1819$-$1008 	& 6.56 & 11.81 & 33.28 & 10.8 & !0.44 & !40.4 \\
1819$-$1131 	& 7.46 & 12.02 & 31.04 & 12.2 & !0.35 & !22.5 \\
1820$-$1529 	& 5.14 & 12.56 & 34.60 & !9.6 & $-$0.07 & !56.0 \\
1821$-$1419 	& 4.47 & 13.59 & 33.89 & 11.9 & $-$0.01 & !28.5 \\
1822$-$0907 	& 7.64 & 11.78 & 31.18 & 12.3 & !0.46 & !18.2 \\ \\ 
1822$-$1252 	& 5.59 & 13.13 & 32.58 & 10.6 & !0.07 & !28.1 \\
1822$-$1617 	& 6.84 & 12.10 & 32.11 & 11.6 & $-$0.24 & !27.1 \\
1823$-$1126 	& 5.90 & 12.92 & 32.36 & !8.7 & !0.14 & !38.8 \\
1823$-$1526 	& 6.76 & 12.44 & 31.62 & !9.2 & $-$0.15 & !39.5 \\
1824$-$1500 	& 6.94 & 11.75 & 32.63 & !8.3 & $-$0.14 & !11.0 \\ \\ 
1828$-$0611 	& 6.47 & 11.80 & 33.46 & !8.8 & !0.35 & !92.1 \\
1828$-$1007 	& 6.60 & 11.49 & 33.83 & !4.7 & !0.03 & !!4.7 \\
1828$-$1057 	& 5.28 & 12.36 & 34.74 & !4.3 & $-$0.00 & !!4.2 \\
1831$-$0823 	& 7.50 & 11.64 & 31.72 & !4.4 & !0.04 & !18.9 \\
1831$-$1423 	& 6.87 & 11.88 & 32.52 & !8.0 & $-$0.31 & !12.2 \\ \\ 
1833$-$0556 	& 7.27 & 12.15 & 31.15 & !7.4 & !0.16 & !10.9 \\
1834$-$0633 	& 6.92 & 11.65 & 32.88 & !9.4 & !0.12 & !24.5 \\
1834$-$0731 	& 5.15 & 12.74 & 34.23 & !4.8 & !0.03 & !22.7 \\
1834$-$0742 	& 5.59 & 12.71 & 33.41 & !6.9 & !0.03 & !16.7 \\
1834$-$1202 	& 9.16 & 10.81 & 30.08 & !6.3 & $-$0.19 & !27.6 \\\hline 
\end{tabular}\end{footnotesize}\end{center}\end{table*}
\addtocounter{table}{-1}\begin{table*} 
\begin{center}\begin{footnotesize}
\caption{-- {\it continued}}\begin{tabular}{lllllll}
\hline 
PSR J & $\log[\tau_c({\rm yr})]$ & $\log[B_s({\rm G})]$ & $\log[\dot{E}({\rm erg\,s}^{-1})]$ & Dist. & z & Luminosity \\ 
 & & & & (kpc) & (kpc) & (mJy kpc$^2$) \\ 
\hline 
1835$-$0522 	& 7.56 & 11.86 & 31.18 & !7.1 & !0.14 & !11.7 \\
1836$-$0517 	& 6.75 & 11.89 & 32.73 & !8.1 & !0.13 & !!9.8 \\
1838$-$0549 	& 5.05 & 12.45 & 35.00 & !4.7 & !0.02 & !!6.5 \\
1838$-$0624 	& 8.28 & 11.43 & 30.58 & !5.8 & $-$0.01 & !!5.4 \\
1839$-$0905 	& 5.41 & 12.52 & 34.15 & !6.1 & $-$0.17 & !!6.0 \\ \\ 
1840$-$0559 	& 6.15 & 12.46 & 32.78 & !5.0 & $-$0.02 & !!7.9 \\
1840$-$0809 	& 6.81 & 12.18 & 32.04 & !5.8 & $-$0.13 & !76.6 \\
1840$-$0815 	& 6.86 & 12.22 & 31.86 & !4.5 & $-$0.10 & !28.3 \\
1840$-$1122 	& 6.37 & 12.39 & 32.48 & !8.3 & $-$0.40 & !!9.0 \\
1841$-$0157 	& 5.76 & 12.54 & 33.38 & !7.3 & !0.15 & !42.8 \\ \\ 
1841$-$0310 	& 7.89 & 11.88 & 30.46 & !4.5 & !0.06 & !!3.0 \\
1841$-$0524 	& 4.48 & 13.01 & 35.00 & !4.9 & $-$0.03 & !!4.8 \\
1842$-$0309 	& 6.15 & 12.14 & 33.43 & 11.4 & !0.12 & !32.5 \\
1842$-$0612 	& 8.61 & 11.05 & 30.68 & !7.0 & $-$0.11 & !26.5 \\
1843$-$0000 	& 6.25 & 12.42 & 32.65 & !2.7 & !0.08 & !21.6 \\ \\ 
1843$-$0137 	& 6.63 & 12.11 & 32.51 & !7.1 & !0.14 & !13.2 \\
1843$-$0211 	& 6.35 & 12.74 & 31.83 & !6.3 & !0.08 & !37.4 \\
1843$-$0408 	& 6.71 & 12.14 & 32.30 & !4.7 & $-$0.01 & !!3.7 \\
1843$-$0702 	& 6.15 & 11.81 & 34.08 & !4.6 & $-$0.11 & !!3.6 \\
1843$-$0806 	& 5.69 & 12.49 & 33.64 & !4.6 & $-$0.15 & !!7.5 \\ \\ 
1843$-$1113 	& 9.48 & !8.13 & 34.78 & !2.0 & $-$0.12 & !!0.4 \\
1844$-$0030 	& 6.22 & 12.30 & 32.96 & !8.7 & !0.19 & !31.5 \\
1844$-$0452 	& 6.80 & 11.64 & 33.15 & !8.1 & $-$0.08 & !12.3 \\
1844$-$0502 	& 7.93 & 11.16 & 31.81 & !5.2 & $-$0.07 & !10.8 \\
1845$-$0545 	& 6.11 & 12.59 & 32.61 & !5.4 & $-$0.13 & !13.7 \\ \\ 
1846+0051 	& 5.79 & 12.35 & 33.73 & !3.2 & !0.08 & !!3.5 \\
1847$-$0130 	& 4.92 & 13.97 & 32.23 & !7.6 & !0.02 & !19.3 \\
1847$-$0443 	& 8.28 & 11.00 & 31.45 & !7.6 & $-$0.18 & !!9.2 \\
1848$-$0023 	& 6.72 & 11.97 & 32.61 & !1.5 & !0.01 & !!1.4 \\
1848$-$0055 	& 6.51 & 11.79 & 33.41 & 15.1 & !0.05 & !43.6 \\ \\ 
1848$-$0511 	& 6.47 & 12.59 & 31.90 & !7.8 & $-$0.23 & !24.2 \\
1849$-$0040 	& 5.98 & 12.44 & 33.15 & 20.9 & !0.07 & !87.0 \\
1849$-$0614 	& 5.45 & 12.86 & 33.40 & !2.8 & $-$0.12 & !!4.8 \\
1850$-$0031 	& 6.96 & 11.99 & 32.11 & !10.0 & $-$0.01 & !22.9 \\
1851+0118 	& 5.02 & 13.05 & 33.86 & !6.8 & !0.06 & !!4.6 \\ \\ 
1851$-$0053 	& 7.41 & 12.05 & 31.08 & !1.2 & $-$0.01 & !!1.5 \\
1851$-$0241 	& 5.94 & 12.27 & 33.58 & !7.6 & $-$0.16 & !11.6 \\
1852+0008 	& 6.12 & 12.22 & 33.34 & !5.1 & $-$0.02 & !!8.1 \\
1852+0013 	& 6.03 & 12.57 & 32.80 & !7.2 & $-$0.02 & !15.6 \\
1852$-$0118 	& 6.61 & 11.95 & 32.88 & !5.4 & $-$0.07 & !10.2 \\ \\ 
1852$-$0127 	& 6.12 & 12.18 & 33.41 & !6.4 & $-$0.09 & !24.0 \\
1852$-$0635 	& 5.75 & 12.45 & 33.60 & !4.6 & $-$0.27 & 124.8 \\
1853+0011 	& 5.27 & 12.57 & 34.32 & !7.5 & $-$0.05 & !16.8 \\
1853+0505 	& 7.05 & 12.04 & 31.84 & !7.5 & !0.26 & !85.3 \\
1853$-$0004 	& 5.46 & 11.88 & 35.32 & !6.6 & $-$0.05 & !37.7 \\ \\ 
1855+0307 	& 5.87 & 12.60 & 33.08 & !7.4 & !0.07 & !53.8 \\
1855+0700 	& 6.74 & 11.65 & 33.23 & !6.8 & !0.28 & !!4.7 \\
1856+0102 	& 6.91 & 11.94 & 32.30 & !8.6 & $-$0.10 & !28.2 \\
1857+0143 	& 4.85 & 12.32 & 35.65 & !5.2 & $-$0.05 & !19.9 \\
1857+0809 	& 6.23 & 12.19 & 33.18 & !8.7 & !0.37 & !10.6 \\\hline 
\end{tabular}\end{footnotesize}\end{center}\end{table*}
\addtocounter{table}{-1}\begin{table*} 
\begin{center}\begin{footnotesize}
\caption{-- {\it continued}}\begin{tabular}{lllllll}
\hline 
PSR J & $\log[\tau_c({\rm yr})]$ & $\log[B_s({\rm G})]$ & $\log[\dot{E}({\rm erg\,s}^{-1})]$ & Dist. & z & Luminosity \\ 
 & & & & (kpc) & (kpc) & (mJy kpc$^2$) \\ 
\hline 
1858+0241 	& 6.49 & 13.03 & 30.97 & !6.5 & $-$0.05 & !!4.2 \\
1859+0601 	& 5.81 & 12.72 & 32.94 & !6.0 & !0.09 & !10.7 \\
1900+0634 	& 6.08 & 12.16 & 33.53 & !7.3 & !0.13 & !12.8 \\
1900$-$0051 	& 7.63 & 11.37 & 31.99 & !3.3 & $-$0.14 & !!4.9 \\
1901+0124 	& 6.19 & 12.01 & 33.59 & !7.2 & $-$0.21 & !15.6 \\ \\ 
1901+0254 	& 7.66 & 11.89 & 30.91 & !4.0 & $-$0.06 & !!9.0 \\
1901+0320 	& 7.29 & 11.77 & 31.90 & !7.7 & $-$0.08 & !53.3 \\
1901+0355 	& 5.84 & 12.43 & 33.46 & 10.2 & $-$0.08 & !15.5 \\
1901+0510 	& 5.50 & 12.65 & 33.72 & !8.5 & $-$0.00 & !47.3 \\
1901$-$0312 	& 6.39 & 11.96 & 33.30 & !2.9 & $-$0.18 & !!1.9 \\ \\ 
1902+0248 	& 6.91 & 12.24 & 31.72 & !6.1 & $-$0.13 & !!6.3 \\
1902$-$0340 	& 7.08 & 12.25 & 31.34 & !3.1 & $-$0.23 & !!2.2 \\
1903+0601 	& 5.49 & 12.43 & 34.15 & !7.8 & !0.01 & !16.0 \\
1905+0400 	& 10.09 & !8.14 & 33.54 & !1.3 & $-$0.03 & !!0.1 \\
1905+0600 	& 6.80 & 11.85 & 32.71 & 18.1 & $-$0.09 & 137.0 \\ \\ 
1906+0649 	& 8.13 & 11.65 & 30.45 & !5.1 & $-$0.01 & !!7.7 \\
1907+0249 	& 6.69 & 11.81 & 33.00 & !7.5 & $-$0.30 & !25.8 \\
1907+0345 	& 5.67 & 12.15 & 34.36 & !8.6 & $-$0.27 & !12.6 \\
1907+0731 	& 5.50 & 12.42 & 34.18 & !4.9 & $-$0.02 & !!8.4 \\
1907+0918 	& 4.58 & 12.67 & 35.51 & !7.7 & !0.10 & !17.1 \\ \\ 
1910+0225 	& 7.31 & 11.48 & 32.43 & !6.3 & $-$0.34 & !24.0 \\
1910+0728 	& 5.79 & 12.22 & 33.98 & !6.0 & $-$0.08 & !31.7 \\
1913+1000 	& 5.90 & 12.58 & 33.04 & !7.9 & $-$0.03 & !33.0 \\
1914+0631 	& 8.52 & 11.19 & 30.59 & !2.7 & $-$0.10 & !!1.9 \\
1915+0838 	& 6.54 & 11.87 & 33.18 & !8.1 & $-$0.18 & !19.1 \\ \\ 
1916+0844 	& 6.38 & 12.06 & 33.11 & !8.0 & $-$0.21 & !28.0 \\
1916+0852 	& 6.42 & 12.73 & 31.70 & !7.0 & $-$0.18 & !!6.3 \\
1916+1023 	& 7.58 & 12.03 & 30.80 & !6.9 & $-$0.10 & !17.3 \\
1920+1040 	& 6.73 & 12.58 & 31.38 & !7.2 & $-$0.20 & !29.5 \\
1937+1505 	& 6.91 & 12.61 & 30.97 & 14.1 & $-$0.73 & !25.7 \\ \\ 
\hline 
\end{tabular}\end{footnotesize}\end{center}\end{table*}

 The three millisecond pulsar discoveries all have implications for
 high precision pulsar timing and its applications.  For instance,
 observations of PSR~B1937$+$21 have been used to place limits on the
 gravitational wave background \cite{lom02}.  The three discoveries
 are all well out of the ecliptic plane and have moderate dispersion
 measures (from 26\,cm$^{-3}$pc for PSR~J1905$+$0400 to
 64.9\,cm$^{-3}$pc for the binary system PSR~J1420$-$5625).  A figure
 of merit for precision timing measurement is the ratio $S/W^{3/2}$
 where $W$ is the pulse width and $S$ the flux density.  However,
 because of their low flux densities, these three discoveries have
 lower figures of merit than some other millisecond pulsars such as
 PSRs~B1937$+$21 and J0437$-$4715, but they still may be
 useful as part of a millisecond pulsar timing array. PSR~J1843$-$1113
 has been observed at the Parkes observatory using a filterbank with a
 channel bandwidth of 0.5\,MHz and a sampling time of 80\,$\mu$s.  In
 a typical ten minute observation, signal--to--noise ratios of 10 and
 uncertainties in the arrival times between 3 and 8\,$\mu$s have been
 achieved.  It should also be possible to decrease the uncertainties
 in the arrival times with improved instrumentation.

 \subsubsection{PSR~J1420$-$5625}\label{sec:1420}

  The binary system PSR~J1420$-$5625 has a rotational period of
  34\,ms, a companion mass $>$0.4M$_\odot$, a relatively large orbital
  eccentricity of $e = 0.0035$ and an orbital period of $\sim40$\,days
  (Table~\ref{tb:binary}). We note that the periastron advance,
  $\dot{\omega}$, is likely to be measurable soon which would provide
  a value for the total system mass. As PSR~J1420$-$5625 is a recycled
  pulsar and the system's orbital eccentricity is significantly less
  than that measured for double--neutron--star systems, the companion
  is almost certainly a white dwarf (WD) star. As reviewed in Tauris
  \& van den Heuvel (2003)\nocite{th03} a pulsar is expected to have
  1) a He--WD companion if $M_{WD} \stackrel{<}{_\sim}
  0.45$\,M$_\odot$, 2) a CO--WD companion if $0.45 \stackrel{<}{_\sim}
  M_{WD} \stackrel{<}{_\sim} 0.8$\,M$_\odot$ or 3) an O--Ne--Mg WD
  companion if $0.8 \stackrel{<}{_\sim} M_{WD} \stackrel{<}{_\sim}
  1.4$\,M$_\odot$. Due to the relatively high lower limit on the
  companion mass for PSR~J1420$-$5625, it is unlikely that the
  companion is a He--WD; most likely it is a CO--WD.

  Similar binary systems have already been discovered during the
  multibeam survey and were described by Camilo et
  al. (2001b
)\nocite{clm+01}.  Due to their relatively long millisecond
  spin periods and/or large orbital eccentricities these pulsars are
  unlike the more common low--mass binary pulsars (LMBPs) and were
  thus catagorised as being intermediate--mass binary pulsars
  (IMBPs)\footnote{Camilo et al. (2001) defined IMBPs as objects that
  once had intermediate--mass donor stars.  This applies to pulsar
  systems with spin periods between 10 and 200\,ms and orbital
  eccentricities less than 0.01.}.  PSR~J1420$-$5625 has a similarly
  large spin period and orbital eccentricity and is therefore also an
  IMBP making 14 such systems known (Table~\ref{tb:imbp}).  

  PSR~J1420$-$5625 has the longest orbital period and largest orbital
  eccentricity of the known IMBPs.  Edwards \& Bailes
  (2001)\nocite{eb01b} reviewed two plausible scenarios for such
  binary systems.  The first, consisting of massive late case A/early
  case B mass transfer \cite{ths00} where the Roche lobe overflow
  started before or soon after the termination of hydrogen core
  burning, is limited to systems with orbital periods greater than a
  few days and up to $\sim70$\,days and companions lighter than
  $\sim0.9$\,M$_\odot$. The second, common envelope evolution on the
  first or second ascent of the red giant branch, is able to account
  for the remaining systems.  PSR~J1420$-$5625 is well modelled by the
  first scenario. Edwards \& Bailes (2001)\nocite{eb01b} noted that
  there seemed an underdensity of pulsars with orbital periods between
  12 and 56 days.  PSR~J1420$-$5625 lies within this range and this
  apparent underdensity may have been due only to the small number of
  such systems known.

  \begin{table}\begin{center}
   \caption{Intermediate--mass binary pulsars (IMBPs) known. The table
   provides the pulsar's rotational period ($P$), orbital period
   ($P_b$), orbital eccentricity ($e$) and a lower limit on the companion
   mass ($M_{WD}$) assuming a neutron star mass of
   1.35\,M$_\odot$.}
   \begin{tabular}{lllll}\hline
   PSR          & $P$    & $P_b$  & $e$         & $M_{\rm WD}$  \\
                & (ms)   & (days) & (10$^{-3}$) & (M$_\odot$) \\ \hline
   J0621$+$1002 & 28.8   & 8.3    & 2.5         & $>0.44$     \\
   B0655$+$64   & 195.7  & 1.0    & 0.0075      & $>0.66$     \\
   J1022$+$1001 & 16.4   & 7.8    & 0.097       & $>0.71$     \\
   J1157$-$5112 & 43.6   & 3.5    & 0.40        & $>1.18$     \\
   J1232$-$6501 & 88.2   & 1.9    & 0.11        & $>0.14$     \\ \\
   J1420$-$5625 & 34.1   & 40.3   & 3.5         & $>0.37$     \\ 
   J1435$-$6100 & 9.3    & 1.4    & 0.010       & $>0.88$     \\
   J1454$-$5846 & 45.2   & 12.4   & 1.9         & $>0.86$     \\
   J1603$-$7202 & 14.8   & 6.3    & 0.0092      & $>0.29$     \\
   J1745$-$0952 & 19.4   & 4.9    & 0.018       & $>0.11$     \\ \\
   J1757$-$5322 & 8.9    & 0.5    & 0.040       & $>0.56$     \\ 
   J1810$-$2005 & 32.8   & 15.0   & 0.025       & $>0.28$     \\
   J1904$+$0412 & 71.0   & 14.9   & 0.22        & $>0.22$     \\
   J2145$-$0750 & 16.0   & 6.8    & 0.019       & $>0.42$     \\
  \hline
  \end{tabular}\label{tb:imbp}\end{center}\end{table}

\begin{table*}
\caption{Known pulsars that lie within the Parkes multibeam survey
region that were not detected.}\setlength{\tabcolsep}{3.5pt}
\begin{tabular}{llllllll} \hline
PSR J & PSR B & $l$ & $b$ & DM & Discovery & S$_{1400}$ & Discovery \\ 
      &       & ($^\circ$) & ($^\circ$) & (cm$^{-3}$pc) & telescope & (mJy) & Reference \\ \hline
J1124$-$5916 & --- & 292.04 & $+$1.75 & 330 & Parkes  & 0.08 & Camilo et
      al. (2002b)\nocite{cmg+02} \\
J1156$-$5909 & --- & 295.91 & $+$2.97& 219 & Parkes & --- &Lyne et al. (1998) \nocite{lml+98}\\
J1617$-$5055 & --- & 332.50 & $-$0.28 & 467 & Parkes & --- & Kaspi et al. (1998)\nocite{kcm+98}\\
J1747$-$2958 & --- & 359.30 & $-$0.84 & 102 & Parkes & 0.25 & Camilo et al. (2002a)\nocite{cmgl02} \\ 
J1800$-$2343 & B1757$-$23 & 6.13 & $-$0.12 & 280 & Ooty & --- & Mohanty (1983)\nocite{moh83}\\ \\
J1817$-$2311 & B1814$-$23 & 8.49 & $-$3.27 & 240 & Ooty & --- & Mohanty (1983)\nocite{moh83}\\
J1901$+$1306 & --- & 45.79 & $+$3.68 & !75 & Arecibo & ---  & Nice, Fruchter
      \& Taylor (1995)\nocite{nft95}\\
J1907$+$1247 & B1904$+$12 & 46.10& $+$2.37 & 257 & Arecibo  & --- & Hulse \&
      Taylor (1975)\nocite{ht75b}\\
J1909$+$1450 & ---         & 48.18& $+$2.83 & 120 & Arecibo  & --- & Nice,
      Fruchter \& Taylor (1995)\nocite{nft95}\\  
J1910$+$0004 & B1908$+$00 & 35.17 &$-$4.18 & 202 & Arecibo & --- & Deich et
      al. (1993)\nocite{dma+93} \\ \\
J1918$+$1541 & ---         & 49.89 & $+$1.36 & !13 & Arecibo & --- & Nice,
      Fruchter \& Taylor (1995)\nocite{nft95}\\ \hline
\end{tabular}\label{tb:notfound}
\end{table*}

\begin{table*}\caption{Results for 281 previously known pulsars.  Catalogued and new dispersion measures and flux densities are provided along with new pulse widths.  An asterisk indicates pulsars that saturated the digitizer; for these no flux densities or pulse widths were measured. The pulsars, PSR~J0820$-$3927 and J0821$-$4217 were discovered during the Parkes high latitude survey (Burgay et al., in preparation).  The catalogued flux densities and dispersion are taken from 56 different journal articles.  Full bibliographic references may be obtained from the ATNF pulsar catalogue, version 1.13 (Manchester et al., in preparation).}\label{tb:knownPsrs}\setlength{\tabcolsep}{5pt}
\begin{tabular}{llllcrrllrlll}\hline
PSR~J & PSR~B & $l$ & $b$ & Beam & Radial & S/N & DM$^{\rm cat}$ & DM & S$_{1400}^{\rm cat}$ & S$_{1400}$ & W$_{50}$ & W$_{10}$  \\
 & & ($^\circ$) & ($^\circ$) & & Distance & & (cm$^{-3}$pc) & (cm$^{-3}$pc) & (mJy) & (mJy) & (ms) & (ms)  \\ \hline 
J0725$-$1635   & ---   & 231.47 & $-$0.33 & 11 &   1.0 &     31.3 & 98.98(3) & 98.7(4) & 0.3 & 0.33(4) & 4.1 & --- \\ 
 J0729$-$1836   & B0727$-$18   & 233.76 & $-$0.34 &  1 &   0.6 &      0.4 & 61.292(10) & 61.4(3) & 1.4 & 1.40(15) & 5.9 & 25 \\ 
 J0742$-$2822   & B0740$-$28   & 243.77 & $-$2.44 &  6 &   0.4 &   1494.7 & 73.758(8) & 73.71(17) & 15.0 & 15.0(15) & 5.4 & 8.3 \\ 
 J0820$-$3927   & ---   & 257.26 & $-$1.58 &  1 &   1.4 &     14.4 & 196.5(1) & 197.0(6) & --- & 0.32(4) & 161 & --- \\ 
 J0820$-$4114   & B0818$-$41   & 258.75 & $-$2.73 & 13 &   1.1 &      0.1 & 113.4(2) & 113.4(8) & 5.2 & 5.2(5) & 135 & --- \\ 
 \\
J0821$-$4217   & ---   & 259.77 & $-$3.18 &  8 &   2.2 &      8.1 & 266.5(1) & 266.63(19) & --- & --- & 20 & --- \\ 
 J0828$-$3417   & B0826$-$34   & 253.97 & +2.56 &  4 &   1.3 &     12.5 & 52.9(6) & 52.3(8) & 0.2 & 0.25(4) & --- & --- \\ 
 J0835$-$4510$^*$   & B0833$-$45   & 263.55 & $-$2.79 &  4 &   0.8 &   2624.9 & 67.99(1) & 67.81(8) & 1100.0 & --- & --- & --- \\ 
 J0837$-$4135   & B0835$-$41   & 260.90 & $-$0.34 &  8 &   0.9 &   1156.3 & 147.29(7) & 147.29(7) & 16.0 & 16.0(16) & 8.9 & 18 \\ 
 J0842$-$4851   & B0840$-$48   & 267.18 & $-$4.10 &  6 &   0.6 &    126.2 & 196.85(8) & 196.85(8) & 0.6 & 0.62(7) & 8.3 & --- \\ 
 \\
J0846$-$3533   & B0844$-$35   & 257.19 & +4.71 &  4 &   0.7 &    239.1 & 94.16(11) & 94.12(10) & 2.7 & 2.7(3) & 22 & 76 \\ 
 J0857$-$4424   & ---   & 265.46 & +0.82 &  1 &   0.3 &    121.5 & 184.429(4) & 184.02(17) & 0.9 & 0.88(10) & 9.9 & 17 \\ 
 J0904$-$4246   & B0903$-$42   & 265.07 & +2.86 &  5 &   0.3 &    113.1 & 145.8(5) & 145.8(5) & 0.6 & 0.60(7) & 21 & 32 \\ 
 J0905$-$4536   & ---   & 267.24 & +1.01 &  3 &   0.2 &     63.8 & 116.8(2) & 182.5(14) & 0.8 & 0.83(9) & 35 & --- \\ 
 J0905$-$5127   & ---   & 271.63 & $-$2.85 &  9 &   1.1 &     88.4 & 196.432(4) & 196.1(4) & 1.1 & 1.10(12) & 9.0 & 175 \\ 
 \\
J0907$-$5157   & B0905$-$51   & 272.15 & $-$3.03 &  8 &   0.3 &    815.7 & 103.72(6) & 103.72(6) & 9.3 & 9.3(9) & 13 & 24 \\ 
 J0908$-$4913   & B0906$-$49   & 270.27 & $-$1.02 & 13 &   0.3 &    888.9 & 180.37(4) & 180.37(4) & 10.0 & 10.0(10) & 2.8 & --- \\ 
 J0924$-$5302   & B0922$-$52   & 274.71 & $-$1.93 &  7 &   0.8 &    131.0 & 152.9(5) & 153.5(4) & 1.1 & 1.10(12) & 15 & 25 \\ 
 J0924$-$5814   & B0923$-$58   & 278.39 & $-$5.60 & 13 &   1.3 &    148.9 & 57.4(3) & 57.4(3) & 4.3 & 4.3(4) & 41 & 80 \\ 
 J0934$-$5249   & B0932$-$52   & 275.69 & $-$0.70 &  8 &   0.9 &    124.1 & 99.4(11) & 101.1(15) & 1.2 & 1.20(13) & 25 & 42 \\ 
 \\
J0941$-$5244   & ---   & 276.45 & +0.09 &  5 &   0.7 &     36.9 & 157.94(1) & 157.8(7) & 0.3 & 0.28(4) & 18 & --- \\ 
 J0942$-$5552   & B0940$-$55   & 278.57 & $-$2.23 &  6 &   1.1 &    491.3 & 180.2(5) & 179.8(3) & 10.0 & 10.0(10) & 11 & 49 \\ 
 J0942$-$5657   & B0941$-$56   & 279.35 & $-$2.99 & 13 &   0.5 &    135.8 & 159.74(12) & 159.74(12) & 0.7 & 0.72(8) & 7.0 & 16 \\ 
 J0955$-$5304   & B0953$-$52   & 278.26 & +1.16 & 12 &   1.1 &     71.3 & 156.9(2) & 156.8(3) & 0.9 & 0.94(10) & 7.8 & 33 \\ 
 J1001$-$5507   & B0959$-$54   & 280.23 & +0.08 &  3 &   0.8 &    754.1 & 130.32(17) & 130.32(17) & 6.3 & 6.3(6) & 15 & 31 \\ 
 \\
J1012$-$5857   & B1011$-$58   & 283.71 & $-$2.14 &  2 &   1.0 &    129.5 & 383.9(1) & 383.13(20) & 1.7 & 1.70(18) & 10 & 23 \\ 
 J1016$-$5345   & B1014$-$53   & 281.20 & +2.45 &  2 &   1.4 &     47.8 & 66.8(18) & 67.0(4) & 0.8 & 0.82(9) & 8.7 & 17 \\ 
 J1017$-$5621   & B1015$-$56   & 282.73 & +0.34 &  9 &   0.4 &    156.5 & 439.1(9) & 438.7(5) & 2.9 & 2.9(3) & 7.3 & --- \\ 
 J1032$-$5911   & B1030$-$58   & 285.91 & $-$0.98 &  4 &   0.3 &    130.6 & 418.20(17) & 418.20(17) & 0.9 & 0.93(10) & 7.7 & 17 \\ 
 J1038$-$5831   & B1036$-$58   & 286.28 & $-$0.02 &  2 &   0.3 &    116.4 & 72.74(3) & 72.7(4) & 0.8 & 0.79(9) & 16 & 22 \\ 
 \\
J1042$-$5521   & B1039$-$55   & 285.19 & +3.00 & 13 &   0.1 &    118.7 & 306.5(4) & 306.5(4) & 0.6 & 0.62(7) & 25 & 39 \\ 
 J1046$-$5813   & B1044$-$57   & 287.07 & +0.73 &  5 &   0.9 &     93.9 & 240.2(5) & 239.9(6) & 1.1 & 1.10(12) & 9.1 & 16 \\ 
 J1048$-$5832   & B1046$-$58   & 287.42 & +0.58 & 12 &   0.7 &    536.7 & 129.1(2) & 129.2(5) & 6.5 & 6.5(7) & 5.6 & 9.7 \\ 
 J1056$-$6258   & B1054$-$62   & 290.29 & $-$2.97 & 12 &   1.0 &    821.6 & 320.3(6) & 320.11(18) & 21.0 & 21(2) & 20 & 39 \\ 
 J1059$-$5742   & B1056$-$57   & 288.34 & +1.95 &  8 &   0.4 &    204.8 & 108.7(3) & 108.7(3) & 1.2 & 1.20(13) & 18 & 34 \\ 
 \\
J1104$-$6103   & ---   & 290.33 & $-$0.83 &  1 &   0.9 &     19.2 & 78.506(15) & 78.4(3) & 0.2 & 0.24(3) & 4.9 & --- \\ 
 J1105$-$6107   & ---   & 290.49 & $-$0.85 &  4 &   0.4 &     56.6 & 271.01(2) & 270.43(4) & 0.8 & 0.75(8) & 4.8 & --- \\ 
 J1107$-$5947   & B1105$-$59   & 290.25 & +0.52 & 12 &   1.2 &     23.7 & 158.4(11) & 158.4(11) & 0.4 & 0.43(5) & --- & --- \\ 
 J1110$-$5637   & B1107$-$56   & 289.28 & +3.53 & 13 &   0.9 &    160.2 & 262.56(6) & 262.0(3) & 1.8 & 1.80(19) & 22 & 28 \\ 
 J1112$-$6613   & B1110$-$65   & 293.19 & $-$5.23 &  8 &   1.9 &     31.4 & 249.3(10) & 249.5(5) & 2.6 & --- & 15 & --- \\ 
 \\
J1114$-$6100   & B1112$-$60   & 291.44 & $-$0.32 & 12 &   1.1 &     67.6 & 677.0(4) & 677.0(4) & 2.0 & 2.0(2) & 29 & 60 \\ 
 J1121$-$5444   & B1119$-$54   & 290.08 & +5.87 &  5 &   0.4 &    175.1 & 204.7(6) & 204.5(3) & 1.3 & 1.30(14) & 21 & 27 \\ 
 J1123$-$6259   & ---   & 293.18 & $-$1.78 &  6 &   0.6 &     61.7 & 223.26(3) & 223.14(9) & 0.6 & 0.56(7) & 6.6 & 15 \\ 
 J1123$-$6651   & ---   & 294.47 & $-$5.44 &  7 &   1.5 &     12.1 & 111.196(5) & 111.19(9) & 0.4 & 0.36(5) & --- & --- \\ 
 J1126$-$6054   & B1124$-$60   & 292.83 & +0.29 &  1 &   1.1 &     64.1 & 280.27(3) & 280.25(14) & 1.0 & 1.00(11) & 5.1 & 105 \\ 
 \\
J1133$-$6250   & B1131$-$62   & 294.21 & $-$1.30 & 12 &   0.9 &    152.9 & 567.8(5) & 568.6(11) & 2.9 & 2.9(3) & 255 & 305 \\ 
 J1137$-$6700   & ---   & 295.79 & $-$5.17 &  1 &   0.8 &     59.5 & 228.041(9) & 227.4(5) & 1.2 & 1.20(13) & 89 & --- \\ 
 J1146$-$6030   & B1143$-$60   & 294.98 & +1.34 & 10 &   0.8 &    318.0 & 111.68(7) & 111.68(7) & 3.6 & 3.6(4) & 11 & 15 \\ 
 J1157$-$6224   & B1154$-$62   & 296.71 & $-$0.20 & 12 &   0.3 &    394.8 & 325.2(5) & 324.4(3) & 5.9 & 5.9(6) & 13 & 47 \\ 
 J1202$-$5820   & B1159$-$58   & 296.53 & +3.92 &  4 &   1.2 &    169.9 & 145.41(19) & 145.41(19) & 2.0 & 2.0(2) & 10 & 16 \\ 
 \hline\end{tabular}
\end{table*}
\addtocounter{table}{-1}\begin{table*}\setlength{\tabcolsep}{5pt}
\caption{-- {\it continued}}\begin{tabular}{llllcrrllrlll}\hline
PSR~J & PSR~B & $l$ & $b$ & Beam & Radial & S/N & DM$^{\rm cat}$ & DM & S$_{1400}^{\rm cat}$ & S$_{1400}$ & W$_{50}$ & W$_{10}$  \\
 & & ($^\circ$) & ($^\circ$) & & Distance & & (cm$^{-3}$pc) & (cm$^{-3}$pc) & (mJy) & (mJy) & (ms) & (ms)  \\ \hline 
J1224$-$6407   & B1221$-$63   & 299.98 & $-$1.41 &  4 &   0.4 &    548.6 & 97.47(12) & 97.47(12) & 3.9 & 3.9(4) & 6.1 & 9.1 \\ 
 J1225$-$6408   & B1222$-$63   & 300.13 & $-$1.41 &  8 &   0.2 &     49.2 & 415.1(5) & 415.1(5) & 0.4 & 0.38(5) & 8.1 & --- \\ 
 J1239$-$6832   & B1236$-$68   & 301.88 & $-$5.69 &  7 &   0.2 &    192.5 & 94.3(3) & 94.3(3) & 1.0 & 0.96(11) & 20 & 40 \\ 
 J1243$-$6423   & B1240$-$64   & 302.05 & $-$1.53 & 13 &   0.7 &   1466.0 & 297.25(8) & 297.25(8) & 13.0 & 13.0(13) & 6.6 & 11 \\ 
 J1253$-$5820   & ---   & 303.20 & +4.53 &  7 &   1.1 &    245.7 & 100.584(4) & 100.53(12) & 3.5 & 3.5(4) & 4.4 & 13 \\ 
 \\
J1259$-$6741   & B1256$-$67   & 303.69 & $-$4.83 &  8 &   1.9 &     13.4 & 94.7(9) & 94.7(9) & 1.3 & --- & --- & --- \\ 
 J1302$-$63     & ---   & 304.11 & $-$0.90 &  5 &   1.0 &     10.7 & 875(10) & 874(17) & 0.1 & 0.10(2) & 7 & --- \\ 
 J1302$-$6350   & B1259$-$63   & 304.18 & $-$0.99 &  5 &   0.4 &    124.2 & 146.72(3) & 146.68(8) & 1.7 & 1.70(18) & 23 & --- \\ 
 J1305$-$6455   & B1302$-$64   & 304.41 & $-$2.09 & 10 &   1.0 &     86.4 & 505.0(3) & 505.0(3) & 1.6 & 1.60(17) & 19 & 50 \\ 
 J1306$-$6617   & B1303$-$66   & 304.46 & $-$3.46 & 13 &   0.9 &    130.4 & 436.9(2) & 437.6(3) & 2.5 & 2.5(3) & 23 & 59 \\ 
 \\
J1316$-$6232   & ---   & 305.85 & +0.19 &  9 &   0.9 &     33.5 & 983.3(5) & 983.3(5) & 0.7 & 0.74(8) & --- & --- \\ 
 J1319$-$6056   & B1316$-$60   & 306.31 & +1.74 &  4 &   1.0 &     94.6 & 400.94(4) & 400.7(3) & 1.2 & 1.20(13) & 6.9 & 15 \\ 
 J1326$-$5859   & B1323$-$58   & 307.50 & +3.56 & 12 &   0.6 &    867.2 & 287.30(15) & 287.30(15) & 9.9 & 9.9(10) & 7.6 & 23 \\ 
 J1326$-$6408   & B1323$-$63   & 306.75 & $-$1.53 &  3 &   0.7 &    133.0 & 502.7(4) & 502.7(4) & 1.4 & 1.40(15) & 11 & 46 \\ 
 J1326$-$6700   & B1322$-$66   & 306.31 & $-$4.37 &  3 &   0.5 &    725.5 & 209.6(3) & 209.6(3) & 11.0 & 11.0(11) & 42 & 56 \\ 
 \\
J1327$-$6222   & B1323$-$62   & 307.07 & +0.20 & 10 &   0.9 &    588.9 & 318.80(6) & 318.80(6) & 16.0 & 16.0(16) & 11 & 19 \\ 
 J1327$-$6301   & B1323$-$627  & 306.97 & $-$0.43 &  5 &   0.7 &    180.1 & 294.91(3) & 294.53(8) & 3.2 & 3.2(3) & 6.7 & 30 \\ 
 J1334$-$5839   & ---   & 308.52 & +3.75 &  1 &   0.8 &     55.4 & 119.2978(9) & 119.36(7) & 0.6 & 0.62(7) & 4.1 & --- \\ 
 J1338$-$6204   & B1334$-$61   & 308.37 & +0.31 &  9 &   0.7 &    199.8 & 640.3(7) & 640.3(7) & 3.8 & 3.8(4) & 93 & 145 \\ 
 J1340$-$6456   & B1336$-$64   & 308.05 & $-$2.56 & 12 &   1.0 &     65.7 & 76.99(13) & 76.99(13) & 1.1 & 1.10(12) & 14 & --- \\ 
 \\
J1341$-$6220   & B1338$-$62   & 308.73 & $-$0.03 &  8 &   0.8 &     78.8 & 717.3(6) & 717.9(4) & 1.9 & 1.9(2) & 12 & 41 \\ 
 J1356$-$6230   & B1353$-$62   & 310.41 & $-$0.58 & 11 &   0.3 &    535.2 & 417.3(3) & 416.8(4) & 8.7 & 8.7(9) & 20 & 42 \\ 
 J1359$-$6038   & B1356$-$60   & 311.24 & +1.13 & 11 &   0.6 &    446.5 & 293.71(14) & 293.71(14) & 7.6 & 7.6(8) & 3.9 & 7.4 \\ 
 J1401$-$6357   & B1358$-$63   & 310.57 & $-$2.14 & 10 &   1.3 &    300.7 & 98.0(5) & 97.4(3) & 6.2 & 6.2(6) & 10 & 19 \\ 
 J1413$-$6307   & B1409$-$62   & 312.05 & $-$1.72 & 11 &   1.0 &     83.7 & 121.98(6) & 121.82(18) & 0.9 & 0.90(10) & 4.1 & 9.2 \\ 
 \\
J1428$-$5530   & B1424$-$55   & 316.43 & +4.80 &  5 &   0.6 &    517.5 & 82.4(6) & 82.1(3) & 3.9 & 3.9(4) & 14 & 25 \\ 
 J1430$-$6623   & B1426$-$66   & 312.65 & $-$5.40 &  6 &   1.1 &    634.5 & 65.3(1) & 64.98(10) & 8.0 & 8.0(8) & 18 & 28 \\ 
 J1440$-$6344   & B1436$-$63   & 314.65 & $-$3.38 & 13 &   0.6 &     90.0 & 124.2(5) & 130.2(5) & 0.8 & 0.78(9) & 9.7 & 23 \\ 
 J1453$-$6413   & B1449$-$64   & 315.73 & $-$4.43 &  6 &   0.5 &   1423.8 & 71.07(2) & 71.24(8) & 14.0 & 14.0(14) & 4.4 & 9.7 \\ 
 J1512$-$5759   & B1508$-$57   & 320.77 & $-$0.11 &  3 &   0.9 &    211.2 & 628.7(2) & 627.93(10) & 6.0 & 6.0(6) & 7.8 & 16 \\ 
 \\
J1513$-$5908   & B1509$-$58   & 320.32 & $-$1.16 &  6 &   0.6 &     50.8 & 252.5(3) & 252.5(3) & 0.9 & 0.94(10) & 16 & --- \\ 
 J1522$-$5829   & B1518$-$58   & 321.63 & $-$1.22 &  7 &   1.3 &    159.0 & 199.9(2) & 199.6(8) & 4.3 & 4.3(4) & 14 & 25 \\ 
 J1527$-$5552   & B1523$-$55   & 323.64 & +0.59 &  3 &   1.3 &     43.3 & 362.7(8) & 362.7(8) & 0.8 & 0.84(9) & 17 & --- \\ 
 J1534$-$5334   & B1530$-$53   & 325.72 & +1.94 &  5 &   0.6 &    622.0 & 24.82(1) & 25.55(13) & 6.8 & 6.8(7) & 17 & 66 \\ 
 J1534$-$5405   & B1530$-$539  & 325.46 & +1.48 & 12 &   0.8 &     73.1 & 190.82(2) & 190.58(10) & 1.2 & 1.20(13) & 6.1 & 21 \\ 
 \\
J1537$-$49     & ---   & 328.74 & +5.22 & 10 &   0.5 &     24.2 & 65.0(3) & 65.0(3) & 0.3 & 0.31(4) & --- & --- \\ 
 J1539$-$5626   & B1535$-$56   & 324.62 & $-$0.81 &  5 &   0.8 &    296.6 & 175.88(6) & 175.88(6) & 4.6 & 4.6(5) & 8.8 & 17 \\ 
 J1542$-$5034   & ---   & 328.57 & +3.58 &  4 &   0.8 &     60.7 & 91.0(6) & 91.0(6) & 0.6 & 0.55(7) & 7.3 & 15 \\ 
 J1544$-$5308   & B1541$-$52   & 327.27 & +1.32 & 11 &   0.6 &    267.0 & 35.16(7) & 35.16(7) & 3.6 & 3.6(4) & 4.6 & 11 \\ 
 J1549$-$4848   & ---   & 330.49 & +4.30 &  7 &   0.5 &     64.5 & 55.983(8) & 56.0(3) & 0.5 & 0.47(6) & 6.1 & --- \\ 
 \\
J1553$-$5456   & B1550$-$54   & 327.19 & $-$0.90 &  4 &   1.0 &     52.7 & 210(7) & 232(7) & 0.8 & 0.79(9) & 22 & --- \\ 
 J1559$-$5545   & B1555$-$55   & 327.24 & $-$2.02 &  4 &   0.9 &     74.1 & 212.9(3) & 212.9(3) & 0.7 & 0.72(8) & 9.0 & 24 \\ 
 J1600$-$5044   & B1557$-$50   & 330.69 & +1.63 &  3 &   0.7 &    912.0 & 260.56(9) & 262.78(11) & 17.0 & 17.0(17) & 5.4 & 11 \\ 
 J1600$-$5751   & B1556$-$57   & 325.97 & $-$3.70 &  1 &   0.6 &    112.2 & 176.55(8) & 176.55(8) & 1.4 & 1.40(15) & 11 & 26 \\ 
 J1602$-$5100   & B1558$-$50   & 330.69 & +1.29 &  4 &   1.1 &    424.4 & 170.93(7) & 170.93(7) & 5.7 & 5.7(6) & 7.9 & 30 \\ 
 \\
J1603$-$5657   & ---   & 326.88 & $-$3.31 & 13 &   0.3 &    101.8 & 264.07(2) & 264.02(16) & 0.5 & 0.53(6) & 4.7 & 13 \\ 
 J1604$-$4909   & B1600$-$49   & 332.15 & +2.44 &  5 &   1.1 &    311.6 & 140.8(5) & 140.69(10) & 5.5 & 5.5(6) & 4.4 & 13 \\ 
 J1605$-$5257   & B1601$-$52   & 329.73 & $-$0.48 &  5 &   0.7 &    543.2 & 35.1(3) & 35.1(3) & 13.0 & 13.0(13) & 59 & 79 \\ 
 J1611$-$5209   & B1607$-$52   & 330.92 & $-$0.48 &  1 &   1.0 &     81.8 & 127.57(5) & 127.57(5) & 1.2 & 1.20(13) & 2.2 & 4.1 \\ 
 J1613$-$4714   & B1609$-$47   & 334.57 & +2.83 &  1 &   4.0 &    161.7 & 161.2(3) & 161.2(3) & 1.5 & --- & 9.7 & --- \\ 
 \hline\end{tabular}
\end{table*}
\addtocounter{table}{-1}\begin{table*}\setlength{\tabcolsep}{5pt}
\caption{-- {\it continued}}\begin{tabular}{llllcrrllrlll}\hline
PSR~J & PSR~B & $l$ & $b$ & Beam & Radial & S/N & DM$^{\rm cat}$ & DM & S$_{1400}^{\rm cat}$ & S$_{1400}$ & W$_{50}$ & W$_{10}$  \\
 & & ($^\circ$) & ($^\circ$) & & Distance & & (cm$^{-3}$pc) & (cm$^{-3}$pc) & (mJy) & (mJy) & (ms) & (ms)  \\ \hline 
J1614$-$5048   & B1610$-$50   & 332.21 & +0.17 & 12 &   0.8 &     83.8 & 582.8(3) & 582.7(3) & 2.4 & 2.4(3) & 11 & 37 \\ 
 J1615$-$5537   & B1611$-$55   & 329.04 & $-$3.46 &  9 &   0.6 &     51.1 & 124.48(8) & 124.1(5) & 0.4 & 0.44(5) & 16 & 23 \\ 
 J1617$-$4216   & ---   & 338.52 & +5.92 &  4 &   0.9 &     22.0 & 163.6(5) & 159.94(18) & 0.3 & 0.28(4) & 38 & --- \\ 
 J1622$-$4332   & ---   & 338.33 & +4.34 & 13 &   0.6 &     53.1 & 230.68(2) & 230.5(12) & 0.5 & 0.53(6) & 32 & --- \\ 
 J1623$-$4256   & B1620$-$42   & 338.89 & +4.62 &  8 &   0.5 &    122.1 & 295(5) & 286(3) & 1.3 & 1.30(14) & 13 & 29 \\ 
 \\
J1627$-$4845   & ---   & 335.14 & +0.15 &  1 &   0.5 &     29.7 & 557.8(7) & 557.8(7) & 0.5 & 0.48(6) & 36 & --- \\ 
 J1630$-$4733   & B1626$-$47   & 336.40 & +0.56 & 13 &   1.2 &     71.2 & 498(5) & 498.2(10) & 4.0 & 4.0(4) & 76 & --- \\ 
 J1633$-$4453   & B1630$-$44   & 338.73 & +1.98 &  2 &   0.4 &    165.3 & 474.1(3) & 474.1(3) & 1.9 & 1.9(2) & 10 & 25 \\ 
 J1633$-$5015   & B1629$-$50   & 334.70 & $-$1.57 & 12 &   1.1 &    252.8 & 398.41(8) & 398.69(16) & 5.7 & 5.7(6) & 8.7 & 19 \\ 
 J1637$-$4553   & B1634$-$45   & 338.48 & +0.76 &  5 &   0.7 &     73.7 & 193.23(7) & 193.11(11) & 1.1 & 1.10(12) & 3.7 & 62 \\ 
 \\
J1639$-$4604   & B1635$-$45   & 338.50 & +0.46 &  5 &   1.3 &     22.8 & 258.91(4) & 258.91(4) & 0.8 & 0.78(9) & 8.7 & --- \\ 
 J1640$-$4715   & B1636$-$47   & 337.71 & $-$0.44 &  4 &   0.8 &     41.4 & 591.7(8) & 591.7(8) & 1.2 & 1.20(13) & 23 & --- \\ 
 J1644$-$4559$^*$   & B1641$-$45   & 339.19 & $-$0.19 & 11 &   0.3 &    890.4 & 478.8(8) & 480.7(4) & 310.0 & --- & --- & --- \\ 
 J1646$-$4346   & B1643$-$43   & 341.11 & +0.97 & 13 &   1.1 &     29.4 & 490.4(3) & 490.4(3) & 1.0 & 0.98(11) & 13 & --- \\ 
 J1651$-$4246   & B1648$-$42   & 342.46 & +0.92 &  5 &   1.0 &    427.9 & 482(3) & 482(3) & 16.0 & 16.0(16) & 110 & 165 \\ 
 \\
J1651$-$5222   & B1647$-$52   & 335.01 & $-$5.17 & 12 &   1.0 &    229.9 & 179.1(6) & 178.6(3) & 2.9 & 2.9(3) & 17 & 24 \\ 
 J1653$-$3838   & B1650$-$38   & 345.88 & +3.27 & 10 &   0.8 &    107.2 & 207.2(2) & 206.8(4) & 1.3 & 1.30(14) & 4.3 & 15 \\ 
 J1700$-$3312   & ---   & 351.06 & +5.49 &  5 &   0.3 &    157.6 & 166.97(9) & 166.7(7) & 1.2 & 1.20(13) & 30 & 50 \\ 
 J1701$-$3726   & B1658$-$37   & 347.76 & +2.83 &  6 &   0.7 &    258.1 & 303.4(5) & 299.2(8) & 2.9 & 2.9(3) & 43 & 105 \\ 
 J1701$-$4533   & B1657$-$45   & 341.36 & $-$2.18 &  9 &   0.9 &    103.7 & 526.0(6) & 526.0(6) & 2.5 & 2.5(3) & 21 & 35 \\ 
 \\
J1703$-$3241   & B1700$-$32   & 351.79 & +5.39 &  2 &   1.1 &    534.6 & 110.306(14) & 110.13(3) & 7.6 & 7.6(8) & 39 & 50 \\ 
 J1703$-$4851   & ---   & 338.99 & $-$4.51 & 11 &   1.2 &     57.6 & 150.29(3) & 151.4(3) & 1.1 & 1.10(12) & 13 & 56 \\ 
 J1705$-$3423   & ---   & 350.72 & +3.98 &  7 &   0.8 &    276.9 & 146.36(10) & 146.30(7) & 4.1 & 4.1(4) & 12 & 21 \\ 
 J1707$-$4053   & B1703$-$40   & 345.72 & $-$0.20 & 11 &   0.7 &    276.1 & 360.0(2) & 357.7(5) & 7.2 & 7.2(7) & 33 & 79 \\ 
 J1708$-$3426   & ---   & 351.08 & +3.41 &  6 &   0.5 &    156.7 & 190.7(3) & 189.9(3) & 1.5 & 1.50(16) & 23 & 35 \\ 
 \\
J1709$-$4429   & B1706$-$44   & 343.10 & $-$2.69 &  9 &   0.7 &    443.3 & 75.69(5) & 75.56(17) & 7.3 & 7.3(7) & 6.0 & 13 \\ 
 J1713$-$3949   & ---   & 347.30 & $-$0.42 &  6 &   0.8 &     10.3 & 337(3) & 338.3(17) & 0.3 & 0.35(4) & 12 & --- \\ 
 J1717$-$3425   & B1714$-$34   & 352.12 & +2.02 &  2 &   0.3 &    276.1 & 585.21(6) & 585.21(6) & 3.3 & 3.3(3) & 14 & 31 \\ 
 J1717$-$4054   & B1713$-$40   & 346.82 & $-$1.73 &  8 &   2.3 &    183.2 & 308.5(5) & 308.5(5) & 54.0 & --- & 15 & 30 \\ 
 J1719$-$4006   & B1715$-$40   & 347.65 & $-$1.53 &  2 &   0.8 &     69.8 & 386.6(2) & 386.80(14) & 1.1 & 1.10(12) & 6.7 & 13 \\ 
 \\
J1720$-$2933   & B1717$-$29   & 356.50 & +4.25 &  3 &   0.9 &    160.9 & 42.64(3) & 42.52(10) & 2.1 & 2.1(2) & 17 & 33 \\ 
 J1721$-$3532   & B1718$-$35   & 351.69 & +0.67 &  1 &   0.6 &    372.8 & 496.81(16) & 496.81(16) & 11.0 & 11.0(11) & 26 & 69 \\ 
 J1722$-$3207   & B1718$-$32   & 354.56 & +2.53 &  6 &   0.7 &    309.1 & 126.064(8) & 126.2(3) & 3.4 & 3.4(4) & 10 & 20 \\ 
 J1722$-$3632   & B1718$-$36   & 350.93 & $-$0.00 &  8 &   0.4 &     92.7 & 416.2(2) & 415.8(5) & 1.6 & 1.60(17) & 22 & 37 \\ 
 J1722$-$3712   & B1719$-$37   & 350.49 & $-$0.51 &  1 &   1.0 &    215.3 & 99.50(4) & 99.44(11) & 3.2 & 3.2(3) & 4.0 & 9.0 \\ 
 \\
J1730$-$2304   & ---   & 3.14 & +6.02 & 12 &   1.2 &     82.2 & 9.611(2) & 9.608(15) & 3.7 & 3.7(4) & 1.1 & --- \\ 
 J1730$-$3350   & B1727$-$33   & 354.13 & +0.09 &  5 &   0.9 &    106.8 & 260.6(13) & 260.6(13) & 3.2 & 3.2(3) & 8.6 & 22 \\ 
 J1732$-$4128   & B1729$-$41   & 347.98 & $-$4.46 &  4 &   0.8 &     66.5 & 195.3(4) & 195.3(4) & 0.6 & 0.63(7) & 16 & 24 \\ 
 J1733$-$2228   & B1730$-$22   & 4.03 & +5.75 & 11 &   1.5 &     50.7 & 41.14(3) & 40.8(10) & 2.3 & 2.3(2) & 60 & --- \\ 
 J1733$-$3716   & B1730$-$37   & 351.58 & $-$2.28 &  2 &   0.8 &    141.5 & 153.5(3) & 153.5(3) & 3.4 & 3.4(4) & 6.2 & 55 \\ 
 \\
J1737$-$3555   & B1734$-$35   & 353.17 & $-$2.27 &  3 &   0.7 &     72.8 & 89.41(4) & 89.06(15) & 0.7 & 0.74(8) & 6.9 & 16 \\ 
 J1738$-$3211   & B1735$-$32   & 356.47 & $-$0.49 &  4 &   0.9 &     92.6 & 49.59(4) & 49.7(4) & 2.8 & 2.8(3) & 12 & 26 \\ 
 J1739$-$2903   & B1736$-$29   & 359.21 & +1.06 & 10 &   0.8 &    141.1 & 138.56(3) & 138.5(3) & 2.0 & 2.0(2) & 6.7 & --- \\ 
 J1739$-$3131   & B1736$-$31   & 357.10 & $-$0.22 &  7 &   0.3 &    240.8 & 599.5(3) & 599.5(3) & 4.9 & 4.9(5) & 22 & 63 \\ 
 J1740$-$3015   & B1737$-$30   & 358.29 & +0.24 &  3 &   0.5 &    503.7 & 152.14(20) & 152.14(20) & 6.4 & 6.4(7) & 5.2 & 12 \\ 
 \\
J1741$-$3927   & B1737$-$39   & 350.55 & $-$4.75 &  3 &   1.0 &    332.7 & 158.5(6) & 158.5(5) & 4.7 & 4.7(5) & 9.3 & 24 \\ 
 J1743$-$3150   & B1740$-$31   & 357.30 & $-$1.15 &  3 &   1.0 &    113.0 & 193.05(7) & 192.3(5) & 1.9 & 1.9(2) & 45 & 68 \\ 
 J1744$-$2335   & ---   & 4.46 & +2.94 &  7 &   2.9 &     48.3 & 96.66(2) & 98.4(11) & 0.2 & --- & 26 & --- \\ 
 J1745$-$3040   & B1742$-$30   & 358.55 & $-$0.96 &  1 &   0.7 &    586.3 & 88.373(4) & 88.46(9) & 13.0 & 13.0(13) & 7.8 & 23 \\ 
 J1748$-$2021   & B1745$-$20   & 7.73 & +3.80 &  3 &   0.8 &     19.1 & 220.4(3) & 220.0(3) & 0.4 & 0.37(5) & 87 & --- \\ 
 \hline\end{tabular}
\end{table*}
\addtocounter{table}{-1}\begin{table*}\setlength{\tabcolsep}{5pt}
\caption{-- {\it continued}}\begin{tabular}{llllcrrllrlll}\hline
PSR~J & PSR~B & $l$ & $b$ & Beam & Radial & S/N & DM$^{\rm cat}$ & DM & S$_{1400}^{\rm cat}$ & S$_{1400}$ & W$_{50}$ & W$_{10}$  \\
 & & ($^\circ$) & ($^\circ$) & & Distance & & (cm$^{-3}$pc) & (cm$^{-3}$pc) & (mJy) & (mJy) & (ms) & (ms)  \\ \hline 
J1748$-$2444   & ---   & 3.96 & +1.56 &  3 &   0.4 &     26.4 & 207.33(9) & 206.6(2) & 0.3 & 0.34(4) & 5.2 & --- \\ 
 J1748$-$2446A  & B1744$-$24A  & 3.84 & +1.70 &  6 &   1.2 &     12.9 & 242.15(4) & 242.15(4) & 0.6 & 0.61(7) & --- & --- \\ 
 J1749$-$3002   & B1746$-$30   & 359.46 & $-$1.24 &  9 &   0.3 &    146.8 & 509.4(3) & 508.7(5) & 3.7 & 3.7(4) & 45 & 88 \\ 
 J1750$-$3157   & B1747$-$31   & 357.98 & $-$2.52 &  2 &   1.0 &     56.1 & 206.34(4) & 206.6(11) & 1.2 & 1.20(13) & 10 & 84 \\ 
 J1750$-$3503   & ---   & 355.31 & $-$4.08 & 13 &   0.2 &     49.0 & 189.35(2) & 189.1(4) & 0.8 & 0.79(9) & 67 & --- \\ 
 \\
J1752$-$2806   & B1749$-$28   & 1.54 & $-$0.96 & 10 &   0.2 &   1603.4 & 50.372(8) & 50.45(12) & 18.0 & 18.0(18) & 9.1 & 15 \\ 
 J1753$-$2501   & B1750$-$24   & 4.27 & +0.51 &  6 &   1.0 &     49.1 & 676.2(7) & 676.2(7) & 2.3 & 2.3(2) & 57 & --- \\ 
 J1756$-$2435   & B1753$-$24   & 5.03 & +0.04 &  7 &   0.7 &    108.9 & 367.1(4) & 365.2(5) & 2.0 & 2.0(2) & 25 & 39 \\ 
 J1757$-$2421   & B1754$-$24   & 5.28 & +0.05 &  5 &   0.8 &    232.1 & 179.441(12) & 178.7(3) & 3.9 & 3.9(4) & 15 & 24 \\ 
 J1759$-$2205   & B1756$-$22   & 7.47 & +0.81 &  9 &   0.3 &    151.4 & 177.157(5) & 177.02(20) & 1.3 & 1.30(14) & 3.9 & 11 \\ 
 \\
J1759$-$2922   & ---   & 1.20 & $-$2.89 &  9 &   1.1 &     41.3 & 79.42(6) & 79.4(3) & 0.6 & 0.56(7) & 10.0 & --- \\ 
 J1801$-$2304   & B1758$-$23   & 6.84 & $-$0.07 &  3 &   1.1 &     40.1 & 1073.9(6) & 1073.9(6) & 2.2 & 2.2(2) & --- & --- \\ 
 J1801$-$2451   & B1757$-$24   & 5.25 & $-$0.88 &  8 &   1.0 &     28.3 & 289.01(4) & 289.01(4) & 0.8 & 0.85(9) & 9.6 & --- \\ 
 J1801$-$2920   & B1758$-$29   & 1.44 & $-$3.25 &  9 &   0.7 &    138.1 & 125.613(14) & 125.55(19) & 1.8 & 1.80(19) & 52 & 63 \\ 
 J1803$-$2137   & B1800$-$21   & 8.40 & +0.15 & 10 &   1.0 &    199.3 & 233.99(5) & 233.8(3) & 7.6 & 7.6(8) & 14 & 43 \\ 
 \\
J1803$-$2712   & B1800$-$27   & 3.49 & $-$2.53 &  1 &   0.9 &     59.8 & 165.5(3) & 165.3(3) & 1.0 & 1.00(11) & 18 & --- \\ 
 J1804$-$2717   & ---   & 3.51 & $-$2.74 &  6 &   0.9 &     28.0 & 24.674(5) & 24.666(11) & 0.8 & 0.78(9) & --- & --- \\ 
 J1806$-$1154   & B1804$-$12   & 17.14 & +4.42 & 12 &   0.6 &    203.9 & 122.41(5) & 122.0(11) & 2.6 & 2.6(3) & 32 & 41 \\ 
 J1807$-$0847   & B1804$-$08   & 20.06 & +5.59 & 12 &   0.8 &    944.3 & 112.3802(11) & 112.47(11) & 15.0 & 15.0(15) & 8.9 & 13 \\ 
 J1807$-$2459   & ---   & 5.84 & $-$2.20 & 11 &   0.7 &     10.1 & 134.0(4) & 134.0(5) & 1.1 & 1.10(12) & --- & --- \\ 
 \\
J1807$-$2715   & B1804$-$27   & 3.84 & $-$3.26 & 10 &   1.0 &     39.9 & 312.98(3) & 313.1(4) & 0.9 & 0.91(10) & 14 & --- \\ 
 J1808$-$0813   & ---   & 20.63 & +5.75 &  5 &   1.1 &    119.5 & 151.27(6) & 150.6(6) & 1.8 & 1.80(19) & 24 & 40 \\ 
 J1808$-$2057   & B1805$-$20   & 9.45 & $-$0.40 &  7 &   0.9 &    134.6 & 606.8(9) & 607(4) & 2.6 & 2.6(3) & 76 & --- \\ 
 J1809$-$0743   & ---   & 21.25 & +5.67 &  4 &   0.1 &     37.4 & 240.70(14) & 240.8(4) & 0.3 & 0.29(4) & 12 & --- \\ 
 J1809$-$2109   & B1806$-$21   & 9.41 & $-$0.72 &  5 &   0.9 &     58.7 & 381.91(5) & 381.0(4) & 0.8 & 0.84(9) & 13 & --- \\ 
 \\
J1812$-$1718   & B1809$-$173  & 13.11 & +0.54 & 10 &   1.1 &     52.0 & 254.6(4) & 254.6(4) & 1.0 & 1.00(11) & 19 & --- \\ 
 J1812$-$1733   & B1809$-$176  & 12.90 & +0.39 & 10 &   0.6 &    110.2 & 518(4) & 528(8) & 3.3 & 3.3(3) & 63 & --- \\ 
 J1816$-$1729   & B1813$-$17   & 13.43 & $-$0.42 &  1 &   0.9 &     74.1 & 525.5(7) & 526.6(3) & 1.2 & 1.20(13) & 17 & 34 \\ 
 J1816$-$2650   & B1813$-$26   & 5.22 & $-$4.91 &  2 &   0.9 &     67.8 & 128.12(3) & 128.0(5) & 1.1 & 1.10(12) & 40 & --- \\ 
 J1818$-$1422   & B1815$-$14   & 16.41 & +0.61 &  5 &   0.5 &    291.9 & 622.0(4) & 621.7(5) & 7.1 & 7.1(7) & 19 & 53 \\ 
 \\
J1820$-$0427   & B1818$-$04   & 25.46 & +4.73 &  3 &   0.9 &    562.7 & 84.435(17) & 84.39(8) & 6.1 & 6.1(6) & 11 & 20 \\ 
 J1820$-$1346   & B1817$-$13   & 17.16 & +0.48 & 12 &   0.8 &     83.7 & 776.7(17) & 779.5(6) & 2.0 & 2.0(2) & 33 & 120 \\ 
 J1820$-$1818   & B1817$-$18   & 13.20 & $-$1.72 &  4 &   0.9 &     64.0 & 436.0(12) & 436.6(3) & 1.1 & 1.10(12) & 16 & --- \\ 
 J1822$-$1400   & B1820$-$14   & 17.25 & $-$0.18 &  4 &   1.1 &     24.9 & 651.1(9) & 647.9(6) & 0.8 & 0.80(9) & 9.1 & --- \\ 
 J1822$-$2256   & B1819$-$22   & 9.35 & $-$4.37 &  8 &   0.9 &    209.2 & 121.20(4) & 119(3) & 2.4 & 2.4(3) & 46 & 81 \\ 
 \\
J1823$-$0154   & ---   & 28.08 & +5.26 & 10 &   1.0 &     78.1 & 135.87(5) & 136.18(12) & 0.8 & 0.78(9) & 8.1 & 17 \\ 
 J1823$-$1115   & B1820$-$11   & 19.77 & +0.95 &  9 &   0.4 &    118.9 & 428.59(9) & 428.6(5) & 3.2 & 3.2(3) & 26 & --- \\ 
 J1824$-$1118   & B1821$-$11   & 19.81 & +0.74 & 13 &   0.6 &     77.6 & 603(2) & 604(4) & 1.3 & 1.30(14) & 23 & --- \\ 
 J1824$-$1945   & B1821$-$19   & 12.28 & $-$3.11 &  2 &   0.8 &    492.3 & 224.649(5) & 224.452(14) & 4.9 & 4.9(5) & 2.9 & 5.5 \\ 
 J1824$-$2452   & B1821$-$24   & 7.80 & $-$5.58 & 11 &   0.9 &      7.7 & 119.857(7) & 120.1(9) & 0.2 & 0.18(3) & --- & --- \\ 
 \\
J1825$-$0935   & B1822$-$09   & 21.45 & +1.32 &  8 &   1.1 &    548.6 & 19.39(4) & 19.33(20) & 12.0 & 12.0(12) & 12 & 42 \\ 
 J1825$-$1446   & B1822$-$14   & 16.81 & $-$1.00 &  5 &   1.1 &     83.4 & 357(5) & 353(3) & 2.6 & 2.6(3) & 12 & 27 \\ 
 J1826$-$1131   & B1823$-$11   & 19.80 & +0.29 &  7 &   0.9 &     48.2 & 320.58(6) & 322(4) & 0.7 & 0.71(8) & 160 & --- \\ 
 J1826$-$1334   & B1823$-$13   & 18.00 & $-$0.69 & 11 &   1.2 &     56.5 & 231.09(8) & 231.09(8) & 2.1 & 2.1(2) & 5.8 & --- \\ 
 J1827$-$0958   & B1824$-$10   & 21.29 & +0.80 &  5 &   0.3 &     93.4 & 430.1(3) & 430.1(3) & 1.8 & 1.80(19) & 22 & --- \\ 
 \\
J1829$-$1751   & B1826$-$17   & 14.60 & $-$3.42 &  5 &   0.1 &    724.3 & 217.109(9) & 216.42(9) & 7.7 & 7.7(8) & 15 & 20 \\ 
 J1830$-$1059   & B1828$-$11   & 20.81 & $-$0.48 &  3 &   0.9 &    125.4 & 161.50(15) & 159.67(11) & 1.4 & 1.40(15) & 3.2 & 6.3 \\ 
 J1832$-$0827   & B1829$-$08   & 23.27 & +0.30 &  9 &   1.0 &    126.7 & 300.853(13) & 300.48(4) & 2.1 & 2.1(2) & 7.1 & 25 \\ 
 J1832$-$1021   & B1829$-$10   & 21.59 & $-$0.60 &  5 &   0.1 &    118.8 & 475.7(3) & 474.8(4) & 1.3 & 1.30(14) & 8.4 & 18 \\ 
 J1833$-$0338   & B1831$-$03   & 27.66 & +2.27 &  1 &   0.4 &    303.9 & 234.537(14) & 234.17(16) & 2.8 & 2.8(3) & 7.5 & 23 \\ 
 \hline\end{tabular}
\end{table*}
\addtocounter{table}{-1}\begin{table*}\setlength{\tabcolsep}{5pt}
\caption{-- {\it continued}}\begin{tabular}{llllcrrllrlll}\hline
PSR~J & PSR~B & $l$ & $b$ & Beam & Radial & S/N & DM$^{\rm cat}$ & DM & S$_{1400}^{\rm cat}$ & S$_{1400}$ & W$_{50}$ & W$_{10}$  \\
 & & ($^\circ$) & ($^\circ$) & & Distance & & (cm$^{-3}$pc) & (cm$^{-3}$pc) & (mJy) & (mJy) & (ms) & (ms)  \\ \hline 
J1833$-$0827   & B1830$-$08   & 23.39 & +0.06 &  3 &   0.8 &    113.9 & 410.925(12) & 410.925(12) & 3.6 & 3.6(4) & 4.9 & 15 \\ 
 J1834$-$0010   & B1831$-$00   & 30.81 & +3.73 & 10 &   0.5 &     18.4 & 88.65(15) & 88.2(4) & 0.3 & 0.29(4) & --- & --- \\ 
 J1834$-$0426   & B1831$-$04   & 27.04 & +1.75 &  8 &   1.0 &    166.3 & 79.308(8) & 79(4) & 5.0 & 5.0(5) & 84 & 99 \\ 
 J1835$-$0643   & B1832$-$06   & 25.09 & +0.55 & 10 &   0.2 &     68.6 & 472.9(10) & 472.7(3) & 1.3 & 1.30(14) & 32 & --- \\ 
 J1835$-$1106   & ---   & 21.22 & $-$1.51 & 13 &   1.2 &     95.0 & 132.679(3) & 132.57(10) & 2.2 & 2.2(2) & 4.7 & 9.4 \\ 
 \\
J1836$-$0436   & B1834$-$04   & 27.17 & +1.13 &  5 &   1.2 &     84.1 & 231.5(3) & 231.3(3) & 1.8 & 1.80(19) & 9.1 & 17 \\ 
 J1836$-$1008   & B1834$-$10   & 22.26 & $-$1.42 &  3 &   0.2 &    260.0 & 316.97(3) & 316.1(3) & 3.7 & 3.7(4) & 8.7 & 17 \\ 
 J1837$-$0045   & ---   & 30.67 & +2.75 &  1 &   1.1 &     32.1 & 86.98(9) & 87.4(10) & 0.6 & 0.61(7) & 9.6 & --- \\ 
 J1837$-$0653   & B1834$-$06   & 25.19 & +0.00 &  3 &   0.1 &    159.2 & 316.1(4) & 315.5(12) & 2.5 & 2.5(3) & 105 & 170 \\ 
 J1837$-$1837   & ---   & 14.81 & $-$5.50 &  3 &   0.3 &     58.3 & 100.74(13) & 100.8(3) & 0.4 & 0.36(5) & 9.0 & 18 \\ 
 \\
J1841$-$0425   & B1838$-$04   & 27.82 & +0.28 &  2 &   0.9 &    127.0 & 325.487(15) & 325.14(12) & 2.6 & 2.6(3) & 5.3 & 9.9 \\ 
 J1842$-$0359   & B1839$-$04   & 28.35 & +0.17 &  7 &   0.2 &    213.0 & 195.98(8) & 197.4(18) & 4.4 & 4.4(5) & 285 & 340 \\ 
 J1844+00     & ---   & 32.62 & +1.88 &  3 &   1.2 &    284.7 & 345.54(20) & 345.54(20) & 8.6 & 8.6(9) & 12 & 43 \\ 
 J1844$-$0244   & B1842$-$02   & 29.73 & +0.24 &  1 &   0.4 &     63.5 & 428.5(5) & 428.5(5) & 0.9 & 0.87(10) & 20 & --- \\ 
 J1844$-$0256   & ---   & 29.57 & +0.12 &  4 &   0.9 &     14.4 & 820.2(3) & 820.2(3) & 0.5 & 0.46(6) & 68 & --- \\ 
 \\
J1844$-$0433   & B1841$-$04   & 28.10 & $-$0.55 &  9 &   0.3 &    131.2 & 123.158(20) & 123.0(5) & 1.1 & 1.10(12) & 13 & 28 \\ 
 J1844$-$0538   & B1841$-$05   & 27.07 & $-$0.94 &  1 &   1.1 &     85.2 & 412.8(3) & 411.0(3) & 2.2 & 2.2(2) & 19 & --- \\ 
 J1845$-$0316   & ---   & 29.39 & $-$0.26 &  7 &   1.1 &     12.2 & 500.00(14) & 500.00(14) & 0.3 & 0.35(4) & 12 & --- \\ 
 J1845$-$0434   & B1842$-$04   & 28.19 & $-$0.78 & 12 &   0.1 &    174.0 & 230.8(17) & 230.2(17) & 1.6 & 1.60(17) & 17 & 29 \\ 
 J1847$-$0402   & B1844$-$04   & 28.88 & $-$0.94 &  1 &   0.8 &    267.9 & 141.979(5) & 141.0(5) & 4.3 & 4.3(4) & 22 & 30 \\ 
 \\
J1848+0647   & ---   & 38.70 & +3.65 &  8 &   0.6 &     17.0 & 27.9(2) & 25.0(8) & 0.2 & 0.17(3) & 13 & --- \\ 
 J1848+0826   & ---   & 40.15 & +4.44 & 10 &   0.4 &      9.5 & 90.77(7) & 90.4(3) & 0.1 & 0.11(2) & --- & --- \\ 
 J1848$-$0123   & B1845$-$01   & 31.34 & +0.04 &  3 &   0.7 &    514.1 & 159.531(8) & 159.96(12) & 8.6 & 8.6(9) & 17 & 37 \\ 
 J1849+06     & ---   & 38.12 & +3.30 &  9 &   1.1 &     14.6 & 235(3) & 235(3) & 0.3 & 0.35(4) & 89 & --- \\ 
 J1849$-$0636   & B1846$-$06   & 26.77 & $-$2.50 & 13 &   0.9 &    123.4 & 148.168(12) & 148.3(10) & 1.4 & 1.40(15) & 89 & --- \\ 
 \\
J1850+1335   & B1848+13   & 44.99 & +6.34 & 11 &   0.7 &     83.2 & 60.147(8) & 60.0(4) & 0.7 & 0.65(7) & 5.7 & 11 \\ 
 J1851+0418   & B1848+04   & 36.72 & +2.05 & 10 &   0.9 &     34.0 & 115.54(5) & 112.01(10) & 0.7 & 0.66(8) & 71 & --- \\ 
 J1851+1259   & B1848+12   & 44.51 & +5.93 &  4 &   1.5 &     43.6 & 70.615(16) & 70.8(10) & 0.8 & 0.75(8) & 11 & 21 \\ 
 J1852+0031   & B1849+00   & 33.52 & +0.02 & 10 &   0.5 &     98.6 & 798.2(16) & 798.2(16) & 2.2 & 2.2(2) & 235 & --- \\ 
 J1854+1050   & B1852+10   & 42.89 & +4.22 & 12 &   0.3 &     88.4 & 207.2(3) & 208(12) & 1.0 & 1.03(14) & 43 & --- \\ 
 \\
J1855$-$0941   & ---   & 24.72 & $-$5.24 &  8 &   0.3 &     39.5 & 151.99(14) & 151.99(14) & 0.5 & 0.48(6) & 26 & --- \\ 
 J1856+0113   & B1853+01   & 34.56 & $-$0.50 & 10 &   0.5 &     20.8 & 96.79(10) & 96.83(19) & 0.2 & 0.19(3) & 3.8 & --- \\ 
 J1857+0057   & B1854+00   & 34.42 & $-$0.81 &  3 &   0.9 &     49.8 & 82.39(11) & 83.0(7) & 0.9 & 0.92(10) & 22 & --- \\ 
 J1857+0212   & B1855+02   & 35.62 & $-$0.39 &  7 &   0.7 &    117.5 & 506.77(18) & 504.2(4) & 1.6 & 1.60(17) & 14 & 23 \\ 
 J1857+0943   & B1855+09   & 42.29 & +3.06 &  2 &   0.4 &    240.2 & 13.309(5) & 13.301(4) & 4.3 & 4.3(4) & 0.55 & 3.0 \\ 
 \\
J1859+00     & ---   & 34.40 & $-$1.59 &  7 &   1.1 &    163.9 & 420(3) & 420(3) & 4.8 & 4.8(5) & 54 & 97 \\ 
 J1901+00     & ---   & 34.47 & $-$2.05 &  8 &   0.9 &     27.8 & 345.5(11) & 345.5(11) & 0.3 & 0.35(4) & 22 & --- \\ 
 J1901+0156   & B1859+01   & 35.82 & $-$1.37 &  2 &   0.4 &     39.7 & 105.394(7) & 102.8(4) & 0.4 & 0.38(5) & 6.1 & --- \\ 
 J1901+0331   & B1859+03   & 37.21 & $-$0.64 &  7 &   0.2 &    359.1 & 402.080(12) & 400.82(10) & 4.2 & 4.2(4) & 11 & 36 \\ 
 J1901+0716   & B1859+07   & 40.57 & +1.06 &  7 &   0.7 &     74.0 & 252.81(7) & 252.5(5) & 0.9 & 0.90(10) & 11 & 34 \\ 
 \\
J1902+0556   & B1900+05   & 39.50 & +0.21 & 11 &   0.3 &    151.0 & 177.486(13) & 177.7(4) & 1.2 & 1.20(13) & 11 & 29 \\ 
 J1902+0615   & B1900+06   & 39.81 & +0.34 &  3 &   0.8 &     99.7 & 502.900(17) & 502(3) & 1.1 & 1.10(12) & 24 & --- \\ 
 J1902+0723   & ---   & 40.74 & +0.98 &  7 &   0.4 &     16.1 & 105.0(3) & 105.0(4) & 0.2 & 0.17(3) & --- & --- \\ 
 J1903+0135   & B1900+01   & 35.73 & $-$1.96 & 10 &   1.1 &    350.3 & 245.163(6) & 244.95(11) & 5.5 & 5.5(6) & 9.9 & 23 \\ 
 J1904+0004   & ---   & 34.45 & $-$2.81 &  9 &   1.0 &     96.3 & 233.61(4) & 233.6(3) & 2.1 & 2.1(2) & 7.8 & 22 \\ 
 \\
J1904+1011   & B1901+10   & 43.43 & +1.87 &  7 &   0.9 &     26.0 & 135(2) & 136.0(20) & 0.6 & 0.58(7) & 28 & --- \\ 
 J1905+0709   & B1903+07   & 40.94 & +0.07 &  9 &   1.0 &     67.3 & 245.34(10) & 247(13) & 1.8 & 1.80(19) & 39 & --- \\ 
 J1905$-$0056   & B1902$-$01   & 33.69 & $-$3.55 &  6 &   1.1 &     68.3 & 229.131(5) & 228.2(3) & 0.9 & 0.92(10) & 6.2 & 15 \\ 
 J1906+0641   & B1904+06   & 40.60 & $-$0.30 & 10 &   1.1 &     56.5 & 473.15(4) & 473.15(4) & 1.7 & 1.70(18) & 18 & --- \\ 
 J1908+0457   & ---   & 39.27 & $-$1.47 &  8 &   0.4 &     89.2 & 360(5) & 353(3) & 0.9 & 0.93(10) & 42 & 59 \\ 
 \hline\end{tabular}
\end{table*}
\addtocounter{table}{-1}\begin{table*}\setlength{\tabcolsep}{5pt}
\caption{-- {\it continued}}\begin{tabular}{llllcrrllrlll}\hline
PSR~J & PSR~B & $l$ & $b$ & Beam & Radial & S/N & DM$^{\rm cat}$ & DM & S$_{1400}^{\rm cat}$ & S$_{1400}$ & W$_{50}$ & W$_{10}$  \\
 & & ($^\circ$) & ($^\circ$) & & Distance & & (cm$^{-3}$pc) & (cm$^{-3}$pc) & (mJy) & (mJy) & (ms) & (ms)  \\ \hline 
J1908+0500   & ---   & 39.29 & $-$1.40 &  8 &   0.5 &     89.2 & 201.42(2) & 201.27(17) & 0.8 & 0.79(9) & 3.9 & 7.6 \\ 
 J1908+0734   & ---   & 41.58 & $-$0.27 &  7 &   0.7 &     30.9 & 11.104(11) & 11.09(15) & 0.5 & 0.54(6) & 2.8 & --- \\ 
 J1908+0916   & B1906+09   & 43.17 & +0.36 &  9 &   0.1 &     20.0 & 249.8(5) & 249.8(5) & 0.2 & 0.23(3) & 26 & --- \\ 
 J1909+0007   & B1907+00   & 35.12 & $-$3.98 & 11 &   0.8 &    112.2 & 112.787(6) & 112.65(12) & 0.9 & 0.87(10) & 8.0 & 31 \\ 
 J1909+0254   & B1907+02   & 37.60 & $-$2.71 &  8 &   1.3 &     38.7 & 171.734(9) & 172.4(14) & 0.6 & 0.63(7) & 11 & --- \\ 
 \\
J1909+1102   & B1907+10   & 44.83 & +0.99 &  1 &   0.8 &    182.0 & 149.982(4) & 149.74(14) & 1.9 & 1.9(2) & 4.8 & 13 \\ 
 J1910+0358   & B1907+03   & 38.61 & $-$2.34 & 12 &   0.9 &     51.8 & 82.93(10) & 78.0(8) & 1.5 & 1.50(16) & 265 & --- \\ 
 J1910+0714   & ---   & 41.52 & $-$0.87 & 11 &   0.8 &     39.4 & 124.06(5) & 125.5(15) & 0.4 & 0.36(5) & 22 & 735 \\ 
 J1910+1231   & B1907+12   & 46.20 & +1.59 &  2 &   0.1 &     46.8 & 258.64(12) & 257.7(7) & 0.3 & 0.28(4) & 18 & --- \\ 
 J1910$-$0309   & B1907$-$03   & 32.28 & $-$5.68 & 11 &   1.0 &     37.6 & 205.53(3) & 205.13(11) & 0.6 & 0.55(7) & 7.6 & --- \\ 
 \\
J1912+1036   & B1910+10   & 44.79 & +0.15 &  9 &   0.9 &     17.6 & 147.0(5) & 147.0(6) & 0.2 & 0.22(3) & --- & --- \\ 
 J1913+0936   & B1911+09   & 44.03 & $-$0.55 &  3 &   0.7 &     14.0 & 157(2) & 156.5(11) & 0.1 & 0.14(2) & --- & --- \\ 
 J1913+1400   & B1911+13   & 47.88 & +1.59 &  1 &   0.5 &    152.2 & 145.052(5) & 145.0(3) & 1.2 & 1.20(13) & 6.2 & 22 \\ 
 J1914+1122   & B1911+11   & 45.62 & +0.20 & 10 &   0.3 &     68.2 & 100(10) & 100(10) & 0.6 & 0.55(7) & 25 & --- \\ 
 J1915+07     & ---   & 42.63 & $-$1.60 &  1 &   0.6 &     36.8 & 112.5(18) & 112.5(18) & 0.2 & 0.21(3) & 23 & --- \\ 
 \\
J1915+0738   & ---   & 42.47 & $-$1.80 &  7 &   1.1 &     37.4 & 39.00(8) & 39.3(6) & 0.3 & 0.34(4) & 11 & 24 \\ 
 J1915+1009   & B1913+10   & 44.71 & $-$0.65 &  2 &   0.5 &    178.7 & 241.693(10) & 241.6(3) & 1.3 & 1.30(14) & 6.7 & 13 \\ 
 J1915+1606   & B1913+16   & 49.97 & +2.12 &  1 &   0.0 &     29.7 & 168.77(1) & 168.73(8) & 0.4 & 0.42(5) & 7.0 & --- \\ 
 J1916+07     & ---   & 42.85 & $-$2.02 &  2 &   0.8 &    159.2 & 305.1(8) & 305.1(8) & 2.8 & 2.8(3) & 115 & --- \\ 
 J1916+0951   & B1914+09   & 44.56 & $-$1.02 & 11 &   0.8 &     57.1 & 60.953(6) & 61.24(19) & 0.9 & 0.91(10) & 9.6 & --- \\ 
 \\
J1916+1030   & B1913+105  & 45.10 & $-$0.64 &  1 &   0.5 &     22.4 & 387.2(3) & 387.2(3) & 0.2 & 0.22(3) & 20 & --- \\ 
 J1916+1312   & B1914+13   & 47.58 & +0.45 & 12 &   0.2 &    162.2 & 237.016(11) & 236.7(3) & 1.2 & 1.20(13) & 6.2 & 11 \\ 
 J1917+1353   & B1915+13   & 48.26 & +0.62 &  6 &   0.6 &    204.6 & 94.538(4) & 94.51(12) & 1.9 & 1.9(2) & 4.0 & 9.1 \\ 
 J1918+08     & ---   & 43.71 & $-$2.02 &  5 &   0.8 &     32.2 & 30(1) & 29.8(15) & 0.3 & 0.31(4) & 58 & --- \\ 
 J1918+1444   & B1916+14   & 49.10 & +0.87 &  3 &   1.1 &     79.8 & 27.202(17) & 30.3(13) & 1.0 & 1.00(11) & 28 & --- \\ 
 \\
J1919+0134   & ---   & 37.58 & $-$5.56 & 10 &   1.1 &     40.9 & 191.9(4) & 192.0(7) & 0.8 & 0.77(9) & 55 & --- \\ 
 J1921+1419   & B1919+14   & 49.06 & +0.02 & 11 &   0.8 &     45.8 & 91.64(4) & 91.5(5) & 0.7 & 0.68(8) & 22 & --- \\ 
 J1926+1434   & B1924+14   & 49.92 & $-$1.04 &  8 &   0.9 &     24.2 & 211.41(8) & 211.1(8) & 0.5 & 0.48(6) & 16 & --- \\ 
 J1930+1316   & B1927+13   & 49.12 & $-$2.32 & 13 &   1.1 &     11.8 & 207.3(9) & 207.6(4) & 0.2 & 0.18(3) & --- & --- \\ 
 J1932+1059   & B1929+10   & 47.38 & $-$3.88 &  6 &   0.9 &   1697.0 & 3.181(4) & 3.3(2) & 36.0 & 36(4) & 7.4 & 14 \\ 
 \\
J1933+1304   & B1930+13   & 49.35 & $-$3.13 &  7 &   0.9 &     37.6 & 177.9(2) & 177.0(6) & 0.4 & 0.42(5) & 29 & --- \\ 
 \hline\end{tabular}
\end{table*}

\subsection{Previously known pulsars}

 The survey region contains 264 known radio pulsars not discovered by
 the multibeam survey; over half of these were found during the second
 Molonglo survey \cite{mlt+78}, the Jodrell `B' survey \cite{clj+92}
 or the Parkes 20-cm survey \cite{jlm+92}.  We have obtained folded
 pulse profiles for 249 of these pulsars using the multibeam data.
 Four pulsars (J1841$-$0345, J1842$-$0415, J1844$-$0310 and
 J1905$+$0616) were discovered independently in the multibeam survey
 and other surveys. Parameters for these pulsars were provided in
 Morris et al. (2002). The remaining 11 pulsars listed in
 Table~\ref{tb:notfound} were not detected.  This was expected for 10
 out of the 11 as these weak pulsars were originally discovered during
 long observations of supernova remnants or globular clusters or with
 the highly sensitive Arecibo telescope. As PSR~J1156$-$5909 was
 discovered during the Parkes Southern sky survey, it should have
 easily been detected in the multibeam survey. However, as reported in
 D'Amico et al. (1998)\nocite{dsb+98}, PSR~J1156$-$5909 has frequent
 and deep nulls during which the pulsar is undetectable.

 A further 32 previously known pulsars that lie outside the nominal
 survey region were detected mainly due to i) observations that were
 slightly outside of the survey region, ii) early observations that
 were made in the Galactic longitude range $220^\circ < l < 260^\circ$
 and iii) bright pulsars being observable many beam widths away from
 their actual position.  For the 281 previously known pulsars that
 were detected during the multibeam survey we list, in
 Table~\ref{tb:knownPsrs}, each pulsar's name, Galactic position, the
 beam corresponding to the highest S/N detection of this pulsar, the
 radial distance from the centre of this beam to the actual pulsar's
 position in units of beam radii and the S/N of the pulse profile in
 this detection.  If available, we also provide a previously published
 value for the pulsar's dispersion measure and flux density at
 1400\,MHz to compare with our measurements.  The final columns in
 this table give the pulse width at 50\% and 10\% of the peak height
 obtained from the multibeam data.

\subsubsection{Flux densities}

 It is notoriously difficult to obtain flux density measurements that
 agree with earlier values as (i) low dispersion measure pulsars
 scintillate, (ii) pulsar receiver systems are complex with different
 systematic biases in different systems, (iii) the received power is a
 function of telescope elevation and sky temperature and (iv) radio
 frequency interference (RFI) may significantly affect measured
 values.  The flux density measurements given in
 Table~\ref{tb:knownPsrs} were obtained in an identical way to those
 for the Parkes multibeam discoveries listed in this and previous
 papers in this series.  Full details of the method applied to obtain
 the flux densities were provided in Manchester et al. (2001). To
 summarise, the flux densities were calibrated using catalogued
 1400\,MHz flux densities for 13 pulsars that had high dispersion
 measures (to minimise variations caused by scintillation).  The
 effect of the varying sky background temperature was determined by
 scaling the values of the sky background temperature at 408\,MHz from
 the Haslam et al. (1982)\nocite{hssw82} all-sky survey to 1374\,MHz,
 assuming a spectral index of $-$2.5.  We correct for off--centre
 pointing by assuming a Gaussian beam shape of width 14.4\,arcminutes.
 Due to uncertainties in this method, we do not provide flux density
 measurements for pulsars more than 1.8\,beam widths away from the
 centre of the beam (this cutoff is chosen as the beam shape is still
 reasonably well modelled by a Gaussian up to this approximate
 distance). The mean pulse profiles at 1374\,MHz that were used in
 determining the flux densities are shown in
 Figure~\ref{fg:prf_known}.  The brightest pulsars (indicated with an
 asterisk in Table~\ref{tb:knownPsrs}) saturated the digitizer leading
 to significantly underestimated flux density measurements and poor
 pulse profiles.  No flux densities or pulse widths are given for
 these pulsars in Table~\ref{tb:knownPsrs}.

 In Figure~\ref{fg:compareS1400} we compare flux density measurements
 for all the pulsars with dispersion measures greater than
 100\,cm$^{-3}$pc that are not expected to scintillate strongly.  In
 general, there is good agreement with earlier results.  Discrepancies
 can only be explained by calibration or measurement errors in our or
 in earlier work. We note that the observing bandwidth of 288\,MHz
 used in these multibeam observations is much larger than that used
 for previous 20-cm observations.  This has the effect of averaging
 over multiple interference maxima (scintles) and hence lowers the
 uncertainty in the flux density value due to scintillation
 effects. Some previous studies have tended to select their best
 quality data when determining a flux density. This results in an
 overestimate of the mean flux density. In any case, the sample of
 flux densities provided here were obtained using identical on-line
 and off-line systems and software as the published results for the
 multibeam survey discoveries.  This leads to a sample of almost 1000
 pulsars whose flux densities have been measured in an identical way.

 For the 38 pulsars listed in Table~\ref{tb:compareS1400} we have
 obtained the first flux density measurement at 20\,cm.  As flux
 density measurements at other observing frequencies exist in the
 literature, it is possible to determine the spectral indices of these
 pulsars.  These spectral indices range from $-$3.3 for PSR~B0826$-$34
 to $-$0.3 for PSR~B1804$+$12 and lie well within the range of $-$3.4
 to 0.2 found by Lorimer et al. (1995)\nocite{lylg95}.  The mean
 spectral index of $-$1.9 is, however, slightly steeper than $-$1.6
 found in the earlier analysis.  Discrepancies do however exist.  For
 instance, Lorimer et al. (1995) measured spectral indices of $-$1.4
 and $-$0.7 for PSRs~B1813$-$26 and B1907$+$03 respectively compared
 to the steeper values of $-$2.2 and $-$2.1 obtained with our data.
 The spectra for these two pulsars steepen at higher frequencies, an
 effect commonly observed in other pulsars \cite{mkkw00}.

\begin{table*}
\caption{Flux density measurements at 1400\,MHz for those pulsars with no previously catalogued value at this observing frequency, but do have an earlier flux measurement at 400\,MHz. If a value exists in the literature at  600\,MHz then this is also provided along with the spectral index obtained from the 1400 and 400\,MHz values. References for these flux density measurements are 1. Manchester, Newton \& Cooke (1981; unpublished), 2. Qiao et al. (1995), 3. Manchester et al. (1996), 4. Lyne et al. (1998), 5. Stokes et al. (1986)  6. Lorimer et al. (1995) 7. Camilo \& Nice (1995) 8. Lorimer, Camilo \& Xilouris (2002), 9. Hulse \& Taylor (1975) and 10. Costa, McCulloch \& Hamilton (1991).  Unfortunately, not all the S$_{400}$ measurements have been published with corresponding uncertainties.  Taking a typical uncertainty of 10\% leads to an error in the spectral index determination of $\sim$\,0.3.}\label{tb:compareS1400}
\begin{tabular}{lllllllll}\hline
PSR J & PSR B & DM & $S_{1400}$ & S$_{400}$ & S$_{600}$ & SI  & Ref. & Ref. \\
      &       & (cm$^{-3}$pc) & (mJy) & (mJy) & (mJy) & & S$_{400}$ & S$_{600}$\\ \hline
J0828$-$3417  & B0826$-$34  & 52.3(8)     & 0.25(4)   & 16    & ---   & $-$3.3   & 1  & ---  \\
J0842$-$4851  & B0840$-$48  & 196.85(8)   & 0.62(7)   & 6.2   & ---   & $-$1.8   & 1  & ---  \\
J0904$-$4246  & B0903$-$42  & 145.8(5)    & 0.60(7)   & 8     & 4     & $-$2.1   & 1  & 2    \\
J0905$-$4536  & ---         & 182.5(14)   & 0.83(9)   & 13    & ---   & $-$2.2   & 3  & ---  \\
J0924$-$5814  & B0923$-$58  & 57.4(3)     & 4.3(4)    & 22    & ---   & $-$1.3   & 1  & ---  \\ \\
J1042$-$5521  & B1039$-$55  & 306.5(4)    & 0.62(7)   & 14    & ---   & $-$2.5   & 1  & ---  \\
J1112$-$6613  & B1110$-$65  & 249.5(5)    & 2.6(3)    & 19    & 11    & $-$1.6   & 1  & 10   \\
J1121$-$5444  & B1119$-$54  & 204.5(3)    & 1.30(14)  & 24    & ---   & $-$2.3   & 1  & ---  \\
J1123$-$6259  & ---         & 223.14(9)   & 0.56(7)   & 11    & ---   & $-$2.4   & 4  & ---  \\
J1239$-$6832  & B1236$-$68  & 94.3(3)     & 0.96(11)  & 6.5   & ---   & $-$1.5   & 1  & ---  \\ \\
J1259$-$6741  & B1256$-$67  & 94.7(9)     & 1.30(14)  & 4.5   & ---   & $-$1.0   & 1  & ---  \\
J1326$-$6700  & B1322$-$66  & 209.6(3)    & 11.0(11)  & 28    & ---   & $-$0.8   & 1  & ---  \\
J1603$-$5657  & ---         & 264.02(16)  & 0.53(6)   & 8     & ---   & $-$2.2   & 4  & ---  \\
J1622$-$4332  & ---         & 230.5(12)   & 0.53(6)   & 16    & ---   & $-$2.7   & 4  & ---  \\
J1700$-$3312  & ---         & 166.7(7)    & 1.20(13)  & 21    & ---   & $-$2.3   & 3  & ---  \\ \\
J1703$-$4851  & ---         & 151.4(3)    & 1.10(12)  & 22    & ---   & $-$2.4   & 3  & ---  \\
J1705$-$3423  & ---         & 146.30(7)   & 4.1(4)    & 31    & ---   & $-$1.6   & 3  & ---  \\
J1732$-$4128  & B1729$-$41  & 195.3(4)    & 0.63(7)   & 9     & ---   & $-$2.1   & 1  & ---  \\
J1806$-$1154  & B1804$-$12  & 122.0(11)   & 2.6(3)    & 4     & ---   & $-$0.3   & 5  & ---  \\
J1816$-$2650  & B1813$-$26  & 128.0(5)    & 1.10(12)  & 18    & 10.5  & $-$2.2   & 6  & 6    \\ \\
J1834$-$0010  & B1831$-$00  & 88.2(4)     & 0.29(4)   & 5.1   & 2.8   & $-$2.3   & 6  & 6    \\
J1848+0826    & ---         & 90.4(3)     & 0.11(2)   & 2.8   & ---   & $-$2.6   & 7  & ---  \\
J1854+1050    & B1852+10    & 208(12)     & 1.03(14)  & 11    & ---   & $-$1.9   & 5  & ---  \\
J1901+0156    & B1859+01    & 102.8(4)    & 0.38(5)   & 13.7  & 4.2   & $-$2.9   & 6  & 6    \\
J1902+0723    & ---         & 105.0(4)    & 0.17(3)   & 0.6   & ---   & $-$1.0   & 7  & ---  \\ \\
J1904+1011    & B1901+10    & 136.0(20)   & 0.58(7)   & 4.4   & ---   & $-$1.6   & 8  & ---  \\
J1908+0500    & ---         & 201.27(17)  & 0.79(9)   & 6.1   & ---   & $-$1.6   & 7  & ---  \\
J1908+0734    & ---         & 11.09(15)   & 0.54(6)   & 3.5   & ---   & $-$1.5   & 7  & ---  \\
J1908+0916    & B1906+09    & 249.8(5)    & 0.23(3)   & 5     & ---   & $-$2.5   & 9  & ---  \\
J1910+0358    & B1907+03    & 78.0(8)     & 1.50(16)  & 21    & 16    & $-$2.1   & 6  & 6    \\ \\
J1910+0714    & ---         & 125.5(15)   & 0.36(5)   & 5.4   & ---   & $-$2.2   & 7  & ---  \\
J1910+1231    & B1907+12    & 257.7(7)    & 0.28(4)   & 5     & ---   & $-$2.3   & 9  & ---  \\
J1912+1036    & B1910+10    & 147.0(6)    & 0.22(3)   & 1.6   & ---   & $-$1.6   & 8  & ---  \\
J1913+0936    & B1911+09    & 156.5(11)   & 0.14(2)   & 0.8   & ---   & $-$1.4   & 8  & ---  \\
J1914+1122    & B1911+11    & 100(10)     & 0.55(7)   & 1.1   & 0.9   & $-$0.6   & 6  & 6    \\ \\
J1915+0738    & ---         & 39.3(6)     & 0.34(4)   & 1.9   & ---   & $-$1.4   & 7  & ---  \\
J1930+1316    & B1927+13    & 207.6(4)    & 0.18(3)   & 5     & ---   & $-$2.6   & 9  & ---  \\
J1933+1304    & B1930+13    & 177.0(6)    & 0.42(5)   & 2.0   & ---   & $-$1.2   & 8  & ---  \\
\hline \end{tabular}
\end{table*}
\nocite{mnc81}\nocite{qmlg95}\nocite{mld+96}\nocite{lml+98}\nocite{sstd86}\nocite{lylg95}\nocite{cn95}\nocite{lcx02}\nocite{ht75b}\nocite{cmh91}

\subsubsection{Dispersion Measures}

 For the pulsars listed in Table~\ref{tb:betterDM} we have obtained
 dispersion measure measurements more than an order of
 magnitude more precise than earlier results.  Large discrepancies
 exist between the measured and previously determined dispersion
 measures for the three pulsars listed in Table~\ref{tb:badDM}.
 Small, but significant, changes in the absolute value of the
 dispersion measure may be accounted for by dispersion measure
 variations (see, for example, Hobbs et al. submitted to MNRAS).  The
 largest discrepancy in Table~\ref{tb:badDM} exists for
 PSR~J0905$-$4536.  Earlier archived observations of this pulsar from
 the Parkes telescope also suggest a much higher dispersion measure
 value than that obtained by D'Amico et al. (1998)\nocite{dsb+98}.  We
 therefore believe that this earlier result was in error.

\begin{table}
\caption{Dispersion measure values that have been measured more than an order-of-magnitude more precisely than in earlier studies. The earlier dispersion measures (DM$_{\rm cat}$) and the new measurements (DM) have been obtained from Table~\ref{tb:knownPsrs}.}
\begin{tabular}{llll}\hline
PSR J & PSR B & DM$_{\rm cat}$ & DM \\
      &       & (cm$^{-3}$pc) & (cm$^{-3}$pc) \\ \hline
J0842$-$4851  & B0840$-$48   & 197.0(10)   & 196.85(8)     \\
J0907$-$5157  & B0905$-$51   & 104.0(7)    & 103.72(6)     \\
J1001$-$5507  & B0959$-$54   & 130(2)      & 130.32(17)    \\
J1032$-$5911  & B1030$-$58   & 419(5)      & 418.20(17)    \\
J1107$-$5947  & B1105$-$59   & 159(19)     & 158.4(11)     \\ \\
J1239$-$6832  & B1236$-$68   & 96(5)       & 94.3(3)       \\
J1326$-$6408  & B1323$-$63   & 505(5)      & 502.7(4)      \\
J1327$-$6222  & B1323$-$62   & 318.4(9)    & 318.80(6)     \\
J1340$-$6456  & B1336$-$64   & 77(2)       & 76.99(13)     \\
J1537$-$49    & ---   & 65(4)       & 65.0(3)       \\ \\
J1602$-$5100  & B1558$-$50   & 172(1)      & 170.93(7)     \\
J1639$-$4604  & B1635$-$45   & 259(2)      & 258.91(4)     \\
J1646$-$4346  & B1643$-$43   & 490(5)      & 490.4(3)      \\
J1717$-$3425  & B1714$-$34   & 587.7(7)    & 585.21(6)     \\
J1717$-$4054  & B1713$-$40   & 317(9)      & 308.5(5)      \\ \\
J1732$-$4128  & B1729$-$41   & 195(5)      & 195.3(4)      \\
J1801$-$2451  & B1757$-$24   & 289.0(10)   & 289.01(4)     \\
J1826$-$1334  & B1823$-$13   & 231.0(10)   & 231.09(8)     \\
J1827$-$0958  & B1824$-$10   & 430(4)      & 430.1(3)      \\
J1833$-$0827  & B1830$-$08   & 411(2)      & 410.925(12)   \\ \\
J1844+00    & ---   & 335(67)     & 345.54(20)    \\
J1844$-$0310  & ---   & 836(7)      & 836.1(5)      \\
J1845$-$0316  & ---   & 500(5)      & 500.00(14)    \\
J1852+0031  & B1849+00   & 787(17)     & 798.2(16)     \\
J1859+00    & ---   & 412(82)     & 420(3)        \\ \\
J1901+00    & ---   & 346(69)     & 345.5(11)     \\
J1905+0616  & ---   & 259(7)      & 257.9(6)      \\
J1908+0916  & B1906+09   & 250(20)     & 249.8(5)      \\
J1916+07    & ---   & 305(30)     & 305.1(8)      \\
J1916+1030  & B1913+105  & 387(10)     & 387.2(3)      \\
\hline \end{tabular}
\label{tb:betterDM}
\end{table}

\begin{table}
\caption{DM measurements that are different from earlier work.  Pulsars are included
in this table if they are more than 3$\sigma$ discrepant and have a difference in DM greater than 5cm$^{-3}$pc. References are 1. D'Amico et al. (1998) and 2. Newton, Manchester \& Cooke (1981).}
\begin{tabular}{lllll}\hline
PSR J & PSR B & DM$_{\rm cat}$ & DM  & Ref. \\
      &       & (cm$^{-3}$pc) & (cm$^{-3}$pc) & \\ \hline
J0905$-$4536  & ---  & 116.8(2)  & 182.5(14)   & 1    \\
J1440$-$6344  & B1436$-$63  & 124.2(5)  & 130.2(5)    & 2     \\
J1651$-$4246  & B1648$-$42  & 525(8)    & 482(3)      & 2     \\
\hline \end{tabular}
\label{tb:badDM}
\end{table}
\nocite{nmc81}

\section{CONCLUSION}

 Observations for the Parkes multibeam pulsar survey have been
 completed. Processing of the data has so far led to over 700 new
 pulsar discoveries.  Combining the new discoveries with redetections
 of previously known pulsars results in a sample of almost 1000
 pulsars in the Galactic plane that have been analysed in a similar
 fashion.  When we have completed processing the data from the
 multibeam survey, a well-defined sample of pulsars in the Galactic
 plane will exist with flux densities and dispersion measures all
 acquired in an identical manner.  This sample will be used to update
 earlier studies of the pulsar population such as determining the
 pulsar birthrate and the total number of active pulsars in the
 Galaxy.

\section*{Acknowledgements} 

 We gratefully acknowledge the technical assistance with hardware and
 software provided by Jodrell Bank Observatory, CSIRO ATNF,
 Osservatorio Astronomico di Bologna and the Swinburne centre for
 Astrophysics and Supercomputing.  The Parkes radio telescope is part
 of the Australia Telescope which is funded by the Commonwealth of
 Australia for operation as a National Facility managed by CSIRO.  The
 Arecibo Observatory, a facility of the National Astronomy and
 Ionosphere Center, is operated by Cornell University under a
 cooperative agreement with the U.S. National Science Foundation.  IHS
 holds an NSERC UFA and is supported by a Discovery Grant. DRL is a
 University Research Fellow funded by the Royal Society.  FC
 acknowledges support from NSF grant AST-02-05853 and a NRAO travel
 grant.  VMK is a Canada Research Chair and is supported by an NSERC
 Discovery Grant and Steacie Supplement, by NATEQ, CIAR and NASA. NDA,
 AP and MB received support from the Italian Ministry of University
 and Research (MIUR) under the national program {\it Cofin 2002}.

\begin{figure*} 
\centerline{\psfig{file=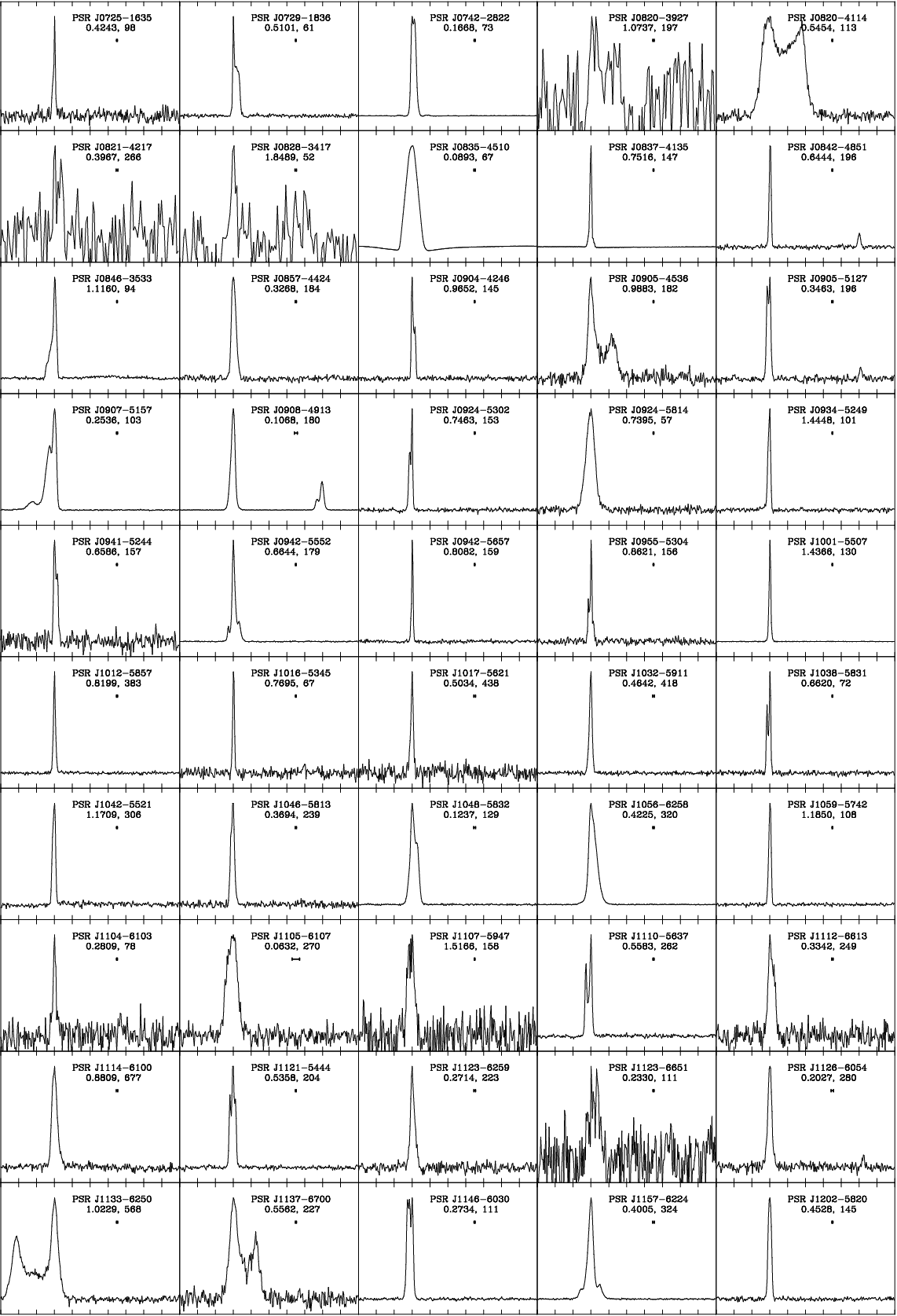,width=150mm}} 
\caption{Mean 1374\,MHz pulse profiles for 281 pulsars redetected in the
Parkes multibeam survey. The highest point in the profile is placed at phase
0.3. For each profile, the pulsar Jname, pulse period (s) and dispersion
measure (cm$^{-3}$pc) are given. The
small horizontal bar under the period indicates the effective resolution of
the profile by adding the bin size to the effects of interstellar dispersion
in quadrature.}
\label{fg:prf_known}
\end{figure*}
\addtocounter{figure}{-1}

\begin{figure*} 
\centerline{\psfig{file=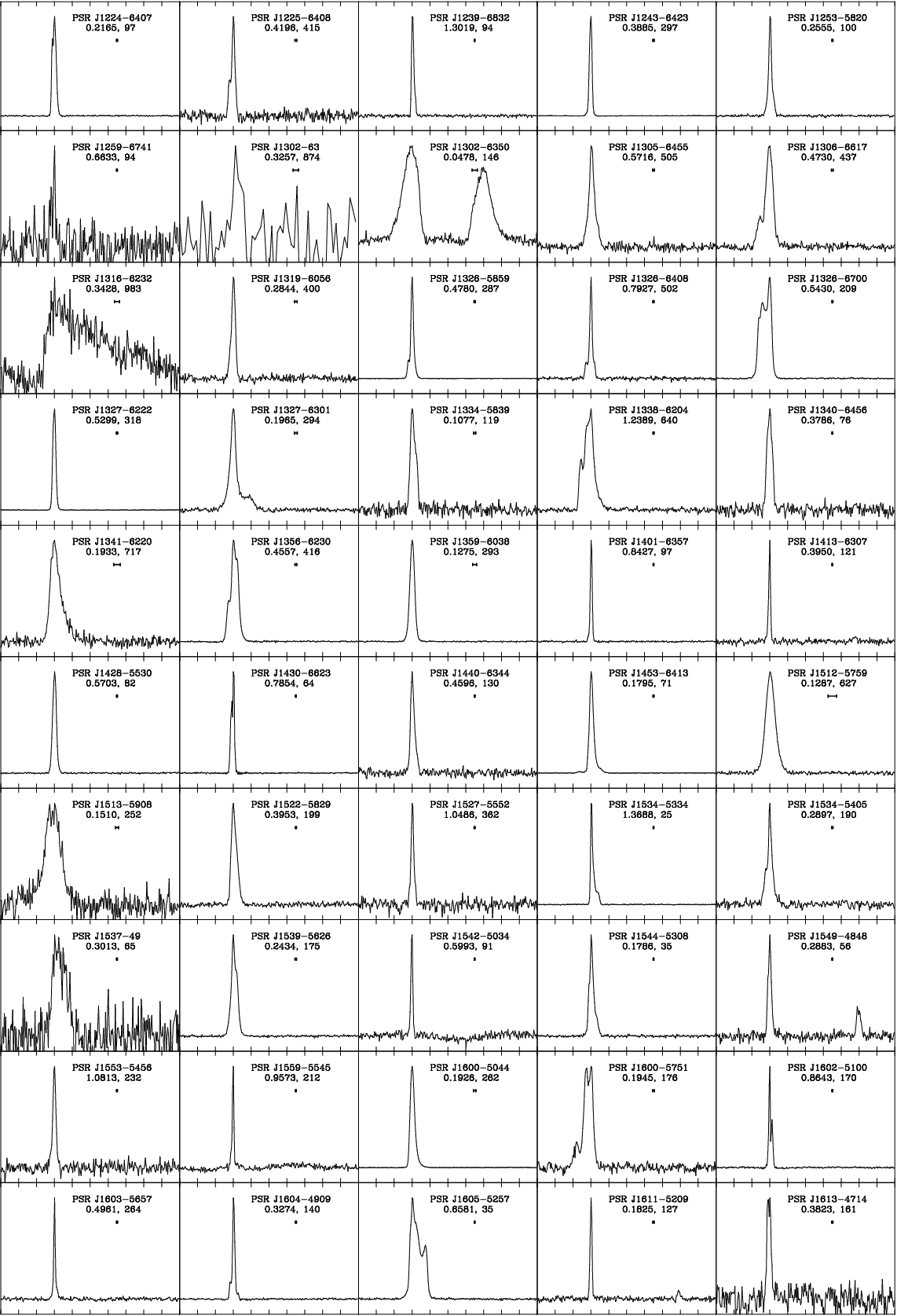,width=150mm}} 
\caption{-- {\it continued}}
\end{figure*}
\addtocounter{figure}{-1}

\begin{figure*} 
\centerline{\psfig{file=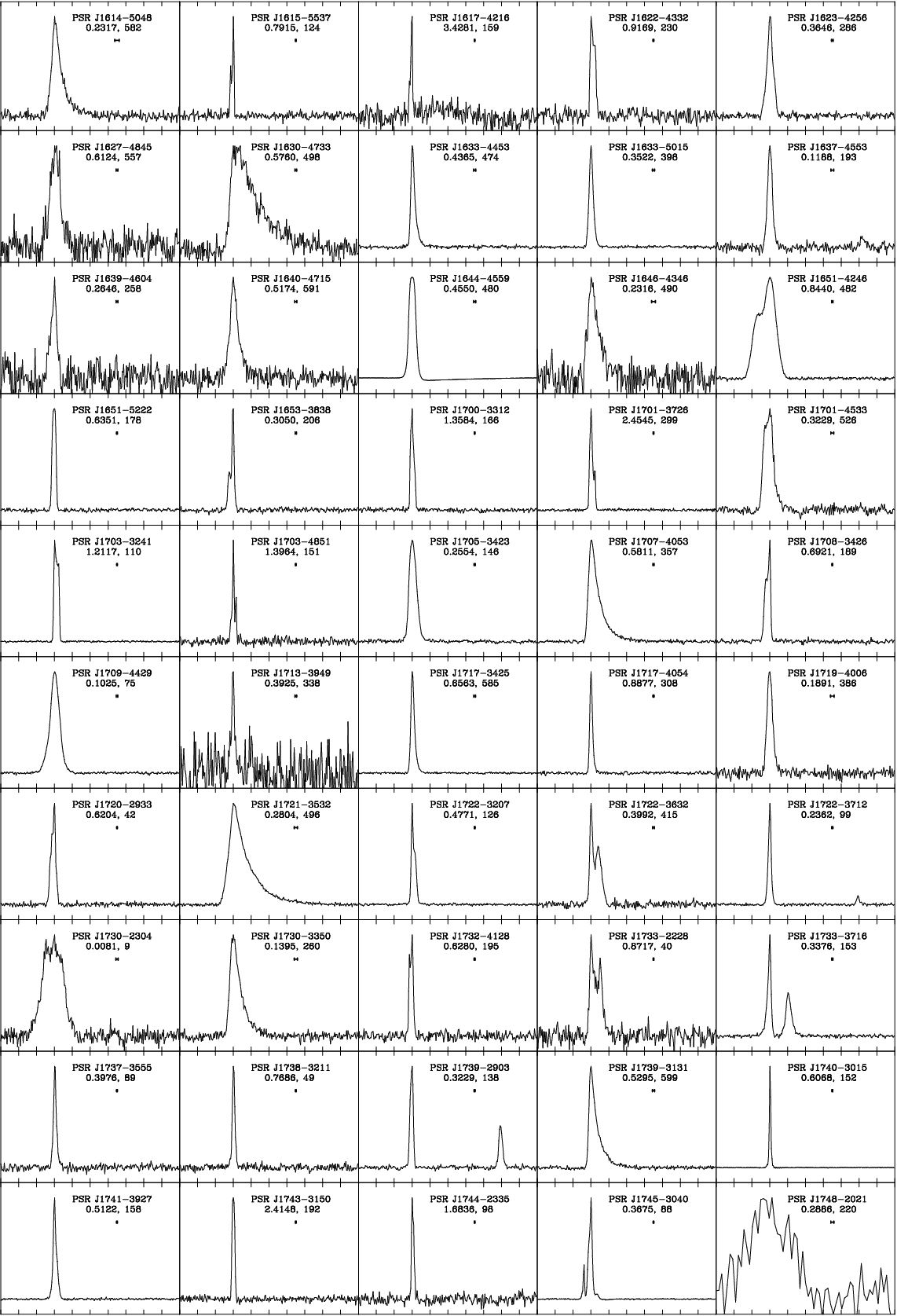,width=150mm}} 
\caption{-- {\it continued}}
\end{figure*}
\addtocounter{figure}{-1}

\begin{figure*} 
\centerline{\psfig{file=known4.ps,width=150mm}} 
\caption{-- {\it continued}}
\end{figure*}
\addtocounter{figure}{-1}

\begin{figure*} 
\centerline{\psfig{file=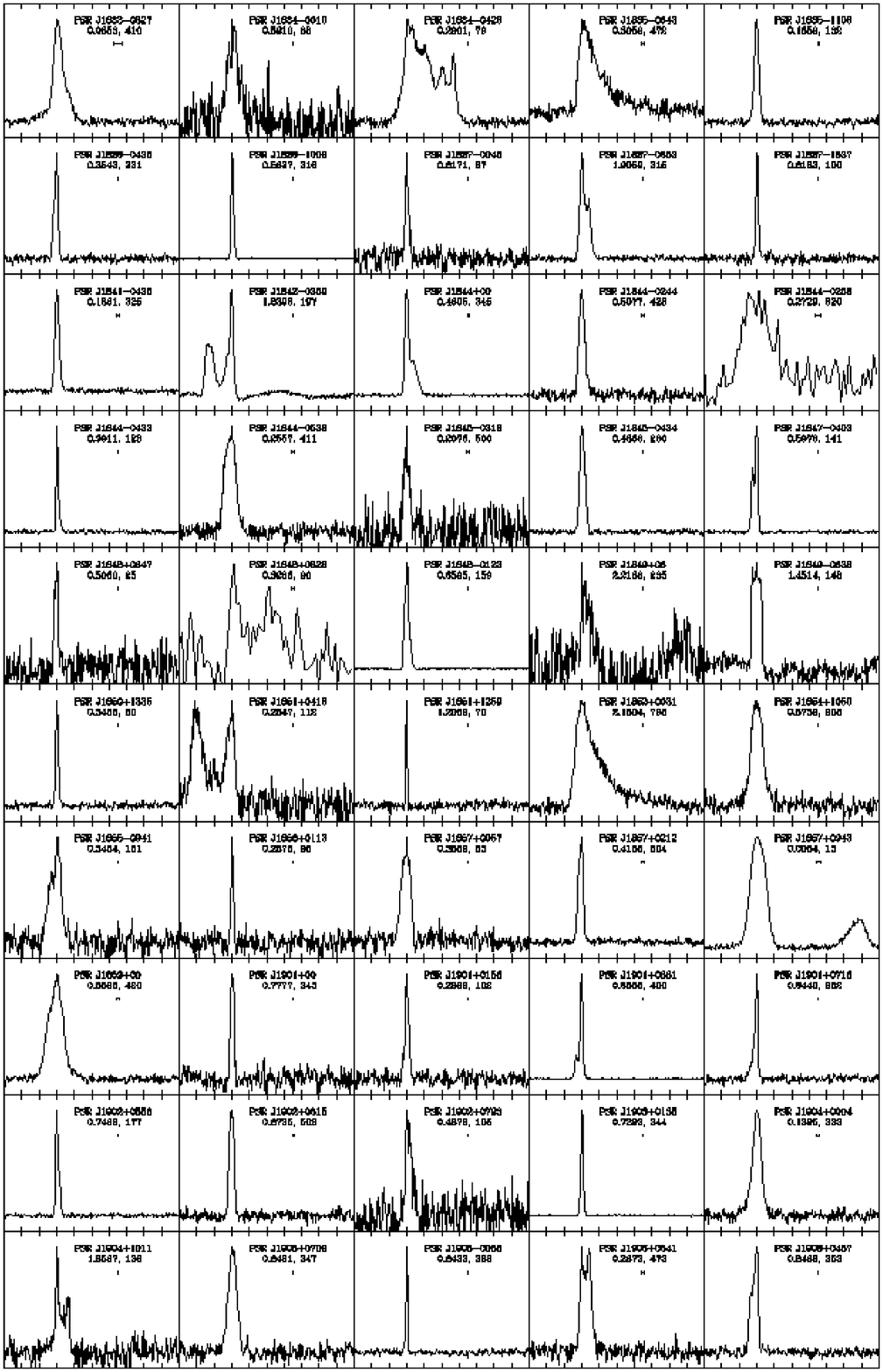,width=150mm}} 
\caption{-- {\it continued}}
\end{figure*}
\addtocounter{figure}{-1}

\begin{figure*} 
\centerline{\psfig{file=known6.ps,width=150mm}} 
\caption{-- {\it continued}}
\end{figure*}

\begin{figure}
\psfig{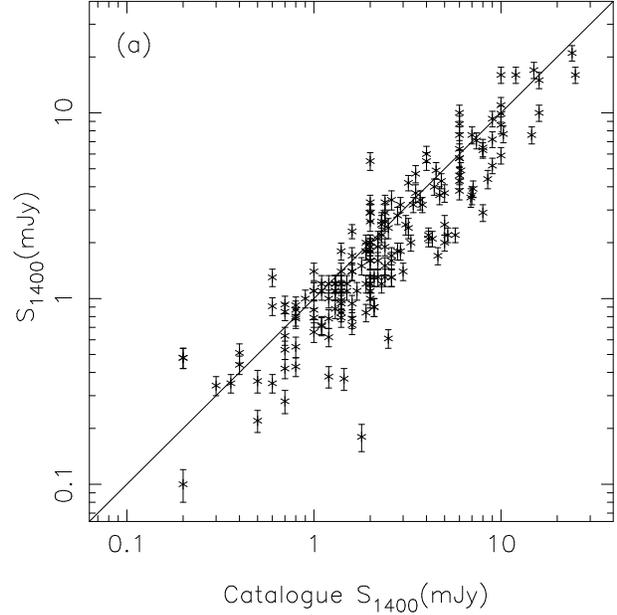}
\caption{Flux density measurement comparisons.   The new flux
  densities and previously published flux densities listed 
  in Table~\ref{tb:knownPsrs} are compared.  The solid lines indicate equality between the
  catalogued and measured flux values.}\label{fg:compareS1400}
\end{figure}

\bibliographystyle{mn}
\bibliography{journals,modrefs,psrrefs,crossrefs}

\end{document}